\documentclass[aps,preprint,showpacs,floatfix,superscriptaddress]{revtex4}

\usepackage{amsmath}
\usepackage{amssymb}
\usepackage{graphicx}

\usepackage{afterpage}

\def\G{\mathcal{G}}
\def\G{{G}}

\begin{document}

\title{Towards a nonequilibrium Green's function description of nuclear reactions: one-dimensional mean-field dynamics}

\date{\today}

\author{Arnau Rios}
\email[]{A.Rios@surrey.ac.uk}
\affiliation{National Superconducting Cyclotron Laboratory and
Department of Physics and Astronomy,
Michigan State University, East Lansing, MI 48824-1321, USA}
\affiliation{Department of Physics,
University of Surrey,
Guildford, Surrey, GU2 7XH,
United Kingdom}
\author{Brent Barker}
\affiliation{National Superconducting Cyclotron Laboratory and
Department of Physics and Astronomy,
Michigan State University, East Lansing, MI 48824-1321, USA}
\author{Mark Buchler}
\altaffiliation{Current address: 121 Argyle St., Howell, MI 48843, USA}
\affiliation{National Superconducting Cyclotron Laboratory and
Department of Physics and Astronomy,
Michigan State University, East Lansing, MI 48824-1321, USA}
\author{Pawel Danielewicz}
\affiliation{National Superconducting Cyclotron Laboratory and
Department of Physics and Astronomy,
Michigan State University, East Lansing, MI 48824-1321, USA}

\begin{abstract}
Nonequilibrium Green's function methods allow for an intrinsically consistent description of the evolution of quantal many-body body systems, with inclusion of different types of correlations. In~this paper, we focus on the practical developments needed to build a Green's function methodology for nuclear reactions. We start out by considering symmetric collisions of slabs in one dimension within the mean-field approximation. We concentrate on two issues of importance for actual reaction simulations. First, the preparation of the initial state within the same methodology as for the reaction dynamics is demonstrated by an adiabatic switching on of the mean-field interaction, which leads to the mean-field ground state. Second, the importance of the Green's function matrix-elements far away from the spatial diagonal is analyzed by a~suitable suppression process that does not significantly affect the evolution of the elements close to the diagonal.  The relative lack of importance of the far-away elements is tied to system expansion.  We also examine the evolution of the Wigner function and verify quantitatively that erasing of the off-diagonal elements corresponds to averaging out of the momentum-space details in the Wigner function.
\end{abstract}


\maketitle


\section{Introduction}

Nonequilibrium Green's function (NGF) techniques \cite{kadanoff,danielewicz84a,botermans90,kohler95} represent a powerful tool to describe the evolution of correlated quantum many-body systems.
While direct applications of those techniques have produced many interesting results in other fields \cite{haug,gasenzer-2005-72,negf07,dahlen07}, within nuclear physics they have been mostly used for derivations, as in  \cite{botermans90,Danielewicz:1991dh,Ivanov:1999tj,Effenberger:1999uv,Cassing:1999wx}, rather than exploited directly \cite{danielewicz84b,tohyama87,kohler95}.  Moreover, the direct applications of NGFs have either pertained to uniform matter, see {\em e.g.}\ \cite{PhysRevE.53.3145,PhysRevC.56.1452,PhysRevD.69.025006,Lindner:2007am}, or to the lowest-energy range of nuclear reactions \cite{tohyama87}.
By~contrast, application of the static limit of Green's function theory to stationary nuclear states has been advanced much farther, accommodating different types of many-body correlations in the description \cite{dickhoff}.
The situation with the nonequilibrium theory in reactions may be, in~part, attributed to the serious numerical issues faced by that theory.  The immediate obstacles, however, will become less of an issue as numerical capabilities increase.  Ultimately aiming at a direct application of the NGFs to the reactions, we consider here strategies for handling the challenges ahead by considering reactions of nuclear slabs in one dimension. The examples that we shall discuss concern the mean-field approximation of the NGF formalism. In the future, we shall explore the extension to correlated dynamics and higher dimensions, relevant for a realistic description of nuclear dynamics. This first study also provides insights into the dynamics of density matrices, coarse-graining and time-reversibility in quantum mechanics.

Even with correlations incorporated into the Green's function approach, that approach is not going to be free of important limitations.  Thus, the primary quantities within the Green's function theory \cite{kadanoff,danielewicz84a} are single-particle functions.  Given the general practical difficulties, it is not likely that an approach that relies on more-body functions as independent quantities can be soon developed for nuclear reactions. With this, the effects of correlations on the dynamics of single-particle functions may be accounted for on the average, assuming that the effects of correlations can be themselves expressed in terms of single-particle functions.  Specifically, within the Green's function theory \cite{kadanoff,danielewicz84a}, the single-particle functions satisfy Dyson-type equations in terms of single-particle self-energies.  In the differential form, those equations are known as Kadanoff-Baym (KB) equations~\cite{kadanoff}.  When the self-energies in those equations are approximated in terms of the single-particle functions, the equations become closed. Theoretically consistent simulations of the dynamics of correlated many-body systems can, in principle, be obtained from these equations.  The~approximations may account for the effects of different types of correlations, but only on an average, single-particle level.  Not surprisingly, in the semiclassical limit the KB equations yield the Boltzmann equation~(BE).

With the inherent averaging over the more-body effects, the Green's function approach is likely to be more suitable for central nuclear reactions than for peripheral ones.  In the former kind of reactions, many particles participate and interactions between those particles are repeated over and over. An averaging over many-body effects takes place physically.  The~description of central reactions could, thus, naturally benefit from a practical implementation of NGF techniques. These developments might potentially improve our understanding
of reaction processes such as fusion and deep-inelastic collisions at low energies, multifragmentation at intermediate and high energies, and vaporization of the participant zone at high energies.  Interestingly, the effect of fluctuations might eventually be included in the Green's function description by using stochastic methods \cite{Reinhard199298,greiner98}.

Historically, just a handful of methods have been developed to describe central nuclear reactions, that simultaneously exhibited some generality and could also be employed in generating practical predictions. The time-dependent Hartree-Fock (TDHF) method has been exhaustively employed in describing low-energy reactions \cite{bonche76,tdhf,Umar:2006ub}.  The semiclassical BE has been extensively used to analyze reactions at intermediate and high-energies \cite{Bertsch:1988ik,bonasera94}. Moreover, molecular approaches, sharing elements of both TDHF and BE, have also been successfully applied for various nuclear reaction purposes \cite{Aichelin:1986wa,Ono:1991uz}.

The semiclassical BE and its related approaches have emerged as a way of dealing with the growth of complexity of reactions with increasing incident energy, while exploiting the~shortening of the particle de Broglie wavelengths.  Because of their semiclassical nature, however, these descriptions have remained genuinely disconnected from the methods employed for nuclear structure, peripheral reactions or giant nuclear excitations, all with quantal underpinnings~\cite{ring}. More importantly, there is no systematic way of improving upon the BE-type approaches.  Their accuracy has remained elusive, since direct comparisons to data necessarily involve adjustable parameters.  At times, cross-comparisons between different approaches, such as molecular dynamics and TDHF, have been attempted~\cite{wong82}.

TDHF \cite{tdhf} (and some of its extensions \cite{Reinhard199298,lacroix04}) is an approach to central collisions that emerges from first principles, without {\em ad hoc} assumptions put in.  The nuclear system is described in terms of a product wave-function, and the TDHF trajectory is found from a~variational principle.  Individual wave-functions are seen to satisfy Schr\"odinger-like single-particle equations of motion, with potential mean-fields evaluated self-consistently.  Within TDHF, the only allowed nuclear excitations are those describable in terms of the evolution of single-particle wave-functions.  The validity of TDHF requires a negligible role played by correlations in the dynamics \cite{tohyama87}.  In a fermion system, antisymmetrization of the many-body wave-function can suppress the correlation admixtures brought in by the interparticle interactions, beyond what can be absorbed into the renormalization of those interactions.  However, as the incident energy increases, the effects of Pauli principle weaken.  Correlations can then lead to a fast thermalization of the occupation of single-particle states and to enhanced stopping, compared to what can be found in TDHF.  Specifically aiming at increasing incident energies, it is therefore important to develop a quantal approach to central nuclear reactions that, besides the mean-field effects, can also incorporate correlations.

Aside from reactions, time-dependent descriptions can lead to improvements in the understanding of nuclear structure~\cite{tdhf}.  Thus, the response properties of nuclei can be studied by exciting the ground state with an external field and simulating the subsequent system evolution \cite{simenel03}.  The inclusion of correlations, particularly if they give rise to dissipation, can introduce substantial modifications in the structure of the response.  Obviously then the correlations can significantly affect the properties of giant resonances and of other collective excitations and may be, in this context, considered from a time-dependent perspective \cite{lacroix04}.

The study of 1D nuclear systems (nuclear slabs), as undertaken here, is likely to appear academic. However, slabs offer an excellent initial testing ground for the ideas involved in the time evolution, whether for correlated or uncorrelated systems. 1D calculations are straightforward to implement numerically with nearly no computational compromise and, thus, serve as an optimal starting point for the time-dependent studies.  Historically, the~TDHF approach has followed such a path and the pioneering work on nuclear slabs of Ref.~\cite{bonche76} became a benchmark for the large number of later studies in 2D and 3D~\cite{flocard78}.  One-dimensional collisions in TDHF were subsequently discussed from the semiclassical point of view and the dynamics were directly compared to that in the Vlasov description~\cite{tang81,wong82}.  Different attempts to introduce the effects of correlations into 1D nuclear dynamics have been described in the literature, mostly in the context of symmetric collisions of slabs.  Specific efforts involved extending TDHF along the lines of the relaxation-time approximation \cite{kohler80,kohler84,grange81} and further the extension motivated by NGF theory, in Ref.~\cite{tohyama85}.  Moreover, in Ref.~\cite{kohler79} collisions of a slab of a finite extension with semi-infinite matter were considered.

The NGF approach that we try to develop has a chance to generalize both the TDHF and BE approaches. The method can also remove some of the serious limitations of those two approaches.  Unfortunately, even at the single-particle level, NGF techniques require handling vast amounts of information that can easily overwhelm the capabilities of computing systems, rendering this approach impractical.  One of our primary motivations for this study is finding whether all the information in the Green's function is equally important for the reaction dynamics.  Another motivation is finding out whether separate numerical infrastructure needs to be developed for preparing the initial states of nuclei and for following the reaction dynamics.

In the next section, we discuss the KB equations, orienting the discussion around 1D applications. After this, in Sec.~\ref{sec:MF}, we consider the mean-field approximation for the KB equations as well as the basic conceptual and numerical details of the procedure we employ in solving those equations. The preparation of the initial state by adiabatic switching on of the interactions is described in Sec.~\ref{sec:initial}.  After the initial-state slabs have been prepared, the collisions of slabs may be simulated.  The phenomenology of slab collisions is reviewed in Sec.~\ref{sec:collisions}. The suppression of off-diagonal elements in the density matrix is of relevance for the practical adaptation of NGF techniques to nuclear reactions.  A suppression scheme and some of its consequences for slab collisions are presented in Sec.~\ref{sec:suppression}. In addition to the practical consequences, we discuss the loss of time reversibility, induced by the introduced off-diagonal cuts, in Sec.~\ref{sec:rev}.  The evolution of the colliding system in terms of Wigner function and impact of the cuts on the Wigner function is discussed in Sec.~\ref{sec:wigner}.  Finally, in~Sec.~\ref{sec:conclusions}, we sum up the paper and draw some conclusions, based on the results found, for future developments of correlated NGF approaches to nuclear collisions.  Two appendices are dedicated to some practical aspects of 1D calculations.  In Appendix \ref{appendix1}, we show how to interpret 1D results in the 3D manner and how to directly adapt 3D mean-field parameterizations for 1D calculations.  In Appendix \ref{appendix2}, we discuss the choice of frequency for the harmonic-oscillator potential that may be used at the start of adiabatic conversion of the single-particle Hamiltonian, when aiming at the mean-field ground-state.

\section{Kadanoff-Baym equations}
\label{sec:KB}

The KB equations describe the evolution, under rather liberal assumptions, of the expectation values of products of two single-particle annihilation and creation operators at different time arguments, \emph{i.e.}\ 2-point Wightman functions. In the context of nonequilibrium theory, these are related to the time-ordered single-particle Green's functions and give access to the most important observables of the reaction. Within a NGF picture, the initial state of the system is specified in terms of an $A$-body density matrix, $\hat \rho_i$, at time $t_0$. Such an~initial state is required to be of some simplified structure. Here, we shall assume that the initial state is uncorrelated, \emph{i.e.}\ completely describable in terms of one-body density matrices. This approximation can eventually be relaxed, leading to a modification of the KB equations, which we shall not treat here \cite{danielewicz84a}. Under the mentioned constraints, the Wightman functions are defined as expectation values with respect to the initial density matrix, $\hat \rho_i$, of products of Heisenberg-picture creation, $\hat a^\dagger(x,t)$, and destruction, $\hat a(x,t)$, operators:
\begin{align}
\label{eq:G<}
\G^<(x_1,t_1; x_2,t_2) &=
i \left\langle \hat a^\dagger(x_2,t_2) \, \hat a(x_1,t_1) \right\rangle \, , \\
\G^>(x_1,t_1; x_2,t_2) &=
- i \left\langle \hat a(x_1,t_1) \, \hat a^\dagger(x_2,t_2)  \right\rangle  \, ,
\end{align}
where discrete quantum numbers are suppressed, fermions are assumed, and
$\left\langle \cdot \right\rangle \equiv \textrm{Tr} \left\{\hat \rho_i \, \cdot \right\}$.
Up to a factor, the time-diagonal (\emph{i.e.}~$t=t'$) Green's function $\G^<$ reduces to the one-body density matrix of the system.

The KB equations for 1D systems, governing the evolution of Green's functions in their arguments,
\begin{align}
	\left\{ i \hbar \, \frac{\partial}{\partial t_1} + \frac{ \hbar^2}{2m} \, \frac{\partial^2} {\partial x_1^2}  \right\} \, \G^{\lessgtr} (\mathbf{11'})
	&= \int \!\! \textrm{d} x_{\bar{1}} \, \Sigma_{HF}( \mathbf{1} \bar{\mathbf{1}} ) \, \G^{\lessgtr} ( \bar{\mathbf{1}} \mathbf{1}' ) \nonumber \\
	&+ \int_{t_0}^{t_1} \!\!\! \textrm{d} \bar{\mathbf{1}} \, \Sigma^{+}( \mathbf{1} \bar{\mathbf{1}} ) \, \G^{\lessgtr} ( \bar{\mathbf{1}} \mathbf{1}' )
	+ \int_{t_0}^{t_{1'}} \!\!\! \textrm{d} \bar{\mathbf{1}} \, \Sigma^{\lessgtr}( \mathbf{1} \bar{\mathbf{1}} ) \, \G^{-} ( \bar{\mathbf{1}} \mathbf{1}' ) \, ,
        \label{eq:kb1} \\
	\left\{- i  \hbar \, \frac{\partial}{\partial t_{1'}} +  \frac{ \hbar ^2}{2m} \, \frac{\partial^2}{\partial x_{1'}^2} \right\} \, \G^{\lessgtr} (\mathbf{11'})
	& = \int \!\! \textrm{d} x_{\bar{1}} \, \G^{\lessgtr} ( \mathbf{1} \bar{\mathbf{1}} ) \, \Sigma_{HF}( \bar{\mathbf{1}} \mathbf{1}') \nonumber \\
	&+ \int_{t_0}^{t_{1}} \!\!\! \textrm{d} \bar{\mathbf{1}} \, \G^{+} ( \mathbf{1} \bar{\mathbf{1}} ) \, \Sigma^{\lessgtr}( \bar{\mathbf{1}} \mathbf{1}' )
	 + \int_{t_0}^{t_{1'}} \!\!\! \textrm{d} \bar{\mathbf{1}} \, \G^{\lessgtr} (\mathbf{1}  \bar{\mathbf{1}} ) \, \Sigma^{-}( \bar{\mathbf{1}}  \mathbf{1}' ) \, ,
         \label{eq:kb1p}
\end{align}
follow from considerations of the equations of motion for creation and destruction operators~\cite{kadanoff}. The simplified notation $\mathbf{1}=(x_1,t_1,\sigma_1,\ldots)$ has been introduced and the retarded~(+) and advanced (-) functions are defined according to:
\begin{align}
F^{\pm}(\mathbf{1},\mathbf{2}) = F^\delta (\mathbf{1},\mathbf{2})
\pm \Theta\left[ \pm (t_1-t_2)\right] \left[ F^>(\mathbf{1},\mathbf{2}) - F^<(\mathbf{1},\mathbf{2}) \right] \, ,
\end{align}
with $F^\delta$ standing for a possible singular contribution at $t_1=t_2$. The generalized self-energy $\Sigma(\mathbf{1},\mathbf{2})$ introduces interaction effects on the time evolution and also describes excitation processes within the system \cite{kadanoff,danielewicz84a}. The self-energy in the previous equations has been separated in two different components. The first one involves the Hartree-Fock (HF) contribution, $\Sigma_{HF}(\mathbf{1},\mathbf{2})$, which accounts for the instantaneous, one-body interaction of the considered particle with the mean-field produced by the other particles of the system. The term involving $\Sigma^\lessgtr$ describes time-dependent excitation processes, beyond the mean-field changes. Such terms account for the effect of correlations on the time evolution and need to be included for a complete description of nuclear reactions.

While we orient our discussion around the intended 1D applications, the extension of the Green's functions and of the KB equations to 3D is obvious.  Otherwise, the complex integro-differential KB equations have to be solved in a self-consistent manner, since the self-energies are functionals of the Green's functions that are being solved for~\cite{baym62}. For certain intrinsically-consistent many-body approximations to the self-energy, one can show that the time evolution induced by the self-consistent KB equations preserves the conservation laws obeyed by the system as a whole~\cite{baym62,danielewicz84a}. This is not a trivial issue, since the
conservation laws can be broken by many-body approximations that, on the face, have quite sensible appearance.  On the other hand, when obeyed, the conservation laws can be extremely useful in practice, as {\em e.g.} in testing the numerical implementation of the equations.

An attractive feature of the KB equations is their generality. The time evolution induced by Eqs.~\eqref{eq:kb1} and \eqref{eq:kb1p} can easily incorporate various types of correlations, described by different approximations to the self-energy~\cite{kadanoff}. In other words, the NGF framework is able to include systematically more and more complicated processes in the self-energy without spoiling the conservation properties. The KB equations are nominally derived without any particular assumption on the physical system under consideration. Consequently, they have been used to study the time evolution of a number of many-body problems. In the nuclear context, only the time evolution of uniform nuclear systems has been discussed so far~\cite{danielewicz84b,kohler95}. We~are not aware of any attempt to generalize these works to finite nuclei. Elsewhere, the KB equations have been used to study the uniform electron gas~\cite{kwong98}, semiconductor systems \cite{haug}, nanostructures \cite{negf07}, inhomogeneous atomic and molecular systems~\cite{dahlen07}, or quantum dots \cite{balzer09}.

The KB equations account for retardation effects if terms beyond the mean-field are included in the time-evolution of the system. From Eq.~\eqref{eq:kb1}, one can easily see that the Green's function $\G^<$ at times $t_1$ and $t_{1'}$ depends on the Green's functions and self-energies at all the previous times t, $t_0<t<t_1$ and $t_0<t<t_{1'}$, via the time integrals on the~r.h.s. Consequently, to find a solution of the KB equations, one must keep track of all the previous time-steps. This might become a major concern in the numerical implementation of the KB equations \cite{kohler99}.
In the mean-field approximation, under which the effects of interactions are local in time, this is not a practical hindrance.

Finally, it is important to note that the KB equations generally respect the non-local features of quantum mechanics. The matrix structure of the single-particle functions, in~space and time, codes information about single-particle positions and momenta, as well as about energies.  In nonstationary and nonuniform systems, neither energies nor momenta are well defined.  In the self-energy terms in the equations, both the nonlocalities associated with the finite range of interactions are accounted for, as well as those of de Broglie type~\cite{danielewicz84a}.

The matrix structure, effectively doubling the number of variables compared to the TDHF approach~\cite{ring}, puts the NGF approach at a computational disadvantage.  Let us for example just consider the difficulties in storing density matrices rather than wavefunctions, in $D$ dimensions.  A uniform mesh of size $N_x$ in each direction will yield $N_x^D$ locations for which wavefunction values need to be stored.  With $N_s$ occupied single-particle states, the wavefunction information for TDHF can be stored within a $N_x^D \times N_s$ matrix.   On the other hand, the density matrix needs to be stored in a generally much larger $N_x^{2D}$ matrix. The difference in storage, between NGF and TDHF, is, though, not yet large in 1D.  Even for fairly accurate meshes with $N_x \sim 100$, the numerical implementation of NGF is not computationally demanding in 1D and can be carried out straightforwardly.  In fact, we expect that even the correlated calculations may be carried out in 1D without any truncation compromises.

Inclusion of correlations in NGF requires, further, augmenting the dimension of the storage matrix to account for the time variables $(t_1,t_{1'})$~\cite{kohler99}.  Overall, because of the high-power growth with the number of dimensions $D$, the storage becomes a serious issue for $D>1$.  The calculations at $D>1$ likely need to involve truncations in space-time meshes.  To optimize these truncation schemes, a better understanding of the role played and the structure of off-diagonal elements in the Green's functions is essential (see Sec.~\ref{sec:offdiag} for further details).

\section{Mean-field approximation}
\label{sec:MF}

The KB equations simplify substantially when the correlation effects, described in terms of $\Sigma^\lessgtr$, are neglected.  In that case, the evolution equations for $\G^<$ and $\G^>$ can be decoupled.  Since the single-particle observables are more straightforwardly expressed in terms of $\G^<$, we concentrate on the evolution of the latter.  The evolution equations simplify even more when assuming a negligible range for the nucleon-nucleon interactions.  The self-energy then takes the form
\begin{align}
\label{eq:sigHF}
\Sigma_{HF}( \mathbf{1} \mathbf{1'} ) = \delta(\mathbf{1} - \mathbf{1'}) \,  U(\mathbf{1}) \, ,
\end{align}
where $U$ depends on local densities.  With Eq.~\eqref{eq:sigHF}, the KB equations for $\G^<$ become
\begin{align}
	i \hbar \frac{\partial}{\partial t_1} \, \G^<(x_1,t_1;x_{1'},t_{1'})
        = \left[ - \frac{ \hbar^2}{2m} \, \frac{\partial^2}{\partial x_1^2}
	+ U( x_1, t_1 ) \right] \, \G^<(x_1,t_1;x_{1'},t_{1'}) \, ,
        \label{eq:kbmf1} \\
	-i \hbar \frac{\partial}{\partial t_{1'}} \, \G^<(x_1,t_1;x_{1'},t_{1'})
        = \left[ - \frac{ \hbar^2}{2m} \, \frac{\partial^2}{\partial x_{1'}^2}
	+ U( x_{1'}, t_{1'} ) \right] \, \G^<(x_1,t_1;x_{1'},t_{1'}) \, .
	\label{eq:kbmf2}
\end{align}
Thanks to the instantaneous nature of $\Sigma_{HF}$, the set of equations for the time-diagonal elements of the Wightman function, $t_1 = t_{1'}$, can also be closed.  Up to a factor, those functions are identical to the single-particle density matrix $\rho$,
\begin{align}
\rho(x_1,x_{1'};t) \equiv -i \G^<(x_1,t;x_{1'},t) \, .
\label{eq:rhoG}
\end{align}
A combination of
Eqs.~(\ref{eq:kbmf1}) and (\ref{eq:kbmf2}) yields:
\begin{align}
	i \hbar \frac{\partial}{\partial t} \, \rho(x_1,x_{1'};t)
        = \left[ - \frac{ \hbar^2}{2m} \,
	\left\{ \frac{\partial^2}{\partial x_1^2} - \frac{\partial^2}{\partial x_{1'}^2} \right\}
	+ U( x_1,t ) - U( x_{1'},t) \right] \, \rho(x_1,x_{1'};t) \, ,
        \label{eq:denmf}
\end{align}
which describes the time evolution of the density matrix in the mean-field approximation.

The connection between the equations presented here, \emph{i.e.} the mean-field approximation of the NGF approach, and TDHF, is established by decomposing the Wightman function in terms of $N_s$ single-particle states:
\begin{align}
  -i \G^<(x_1,t_1;x_{1'},t_{1'}) = \sum_{\alpha=1}^{N_s} \phi_\alpha(x_1,t_1) \, \phi^*_\alpha(x_{1'},t_{1'}) \, .
\label{eq:Gwf}
\end{align}
A substitution of \eqref{eq:Gwf} into \eqref{eq:kbmf1} then yields a set of $N_s$ TDHF single-particle equations:
\begin{align}
	i \hbar \frac{\partial}{\partial t} \phi_\alpha(x_1,t)
        = \left[ - \frac{ \hbar^2}{2m} \frac{\partial^2}{\partial x_1^2}
	+ U( x_1, t ) \right] \phi_\alpha(x_1,t) \, .
        \label{eq:tdhf}
\end{align}
At equal time-arguments for the Wightman function, the decomposition \eqref{eq:Gwf} reduces to that for the mean-field density matrix \eqref{eq:rhoG},
\begin{align}
  \rho (x_1, x_{1'} ; t) = \sum_{\alpha=1}^{N_s} \phi_\alpha(x_1,t) \, \phi^*_\alpha(x_{1'},t ) \, .
\label{eq:rhowf}
\end{align}
The equivalence between the two approaches implies that, for the same initial conditions and mean-field interactions, both the mean-field NGF and the TDHF equations will drive exactly the same time evolutions.  Notably, the decomposition \eqref{eq:Gwf} is possible, in terms of a~finite number of terms, with $\phi$ normalized to unity, when the decomposition of the density matrix \eqref{eq:rhowf} holds at some stage of the system evolution, such as for the combination of ground states prior to a collision, and the system otherwise follows mean-field dynamics that does not change the normalization of the states.

In the following, we shall assume a spin-isospin saturated system.  The Wightman functions, and the density matrix in particular, are then diagonal in the spin and isospin indices (which are still suppressed). Further, the diagonal values of the functions are independent of those indices.  In the strict 1D interpretation of the 1D calculations, the density of nucleons is
\begin{align}
n_1 (x,t) = \nu \, \rho(x,x;t) \, ,
\end{align}
where $\nu=4$ represents the spin-isospin degeneracy of the system, and the nucleon number is $A=\nu N_s$.  A 3D interpretation is further possible where the matter in 3D is assumed uniform in two directions and nonuniform in the third $x$-direction.  The $y$ and $z$ directions are described in terms of a set of plane-wave wavefunctions, independent of the $\alpha$ state or of time.  With this, the density in the 3D interpretation, $n_3(x,t) \equiv n(x,t)$, becomes proportional to the 1D density,
\begin{align}
n(x,t) = \xi \, n_1(x,t) \, .
\end{align}
The advantage of developing a 3D interpretation is that it allows to employ well-known 3D mean-field parameterizations for 1D calculations.

In Appendix \ref{appendix1}, we arrive at the following expression for the scaling factor:
\begin{align}
\xi = \sqrt{\frac{5}{3}} \left( \frac{\pi \, n_0^2}{6 \, \nu^2} \right)^{1/3}  \, ,
\end{align}
by demanding that, within the 3D interpretation, the system energy minimizes at the normal nuclear density $n_0$.  We choose a simple mean-field parametrization for our calculations:
\begin{align}
U = \frac{3}{4} \, t_0 \, n(x, t) + \frac{2+\sigma}{16} \, t_3 \, \left[ n(x, t) \right]^{1+\sigma} \, ,
\label{eq:meanf}
\end{align}
with the parameters $t_0=-2150.1 \, \text{MeV} \cdot \text{fm}^3$, $t_3=14562 \, \text{MeV} \cdot \text{fm}^{3 + 3 \sigma}$ and $\sigma=0.257$ fitted to the saturation properties of nuclear matter: $n_0=0.16 \, \text{fm}^{-3}$, energy per nucleon of $-16 \, \text{MeV}$, and incompressibility of $220 \, \text{MeV}$.

With the potential $U$ dependent on density in \eqref{eq:meanf}, the equation of motion \eqref{eq:denmf} for the density matrix becomes nonlinear.  However, the nonlinearity nominally concerns just the diagonal $x=x'$ elements of the matrix.  Any features of the matrix that are far away from the diagonal, with relatively compact support, will evolve over finite times in such a manner as if the equation were linear.  Even for more realistic mean fields, dependent on off-diagonal matrix elements, the region of that mean-field dependence would be still limited to the immediate vicinity of the diagonal, making the equation in practice linear for features of the matrix far away from the diagonal.

For a local mean-field, the time evolution of the density matrix can be implemented numerically in a rather straightforward way following the so-called Split Operator Method (SOM) \cite{feit82}.  Over a time-step $\Delta t$, the time evolution of the density matrix described by Eq.~(\ref{eq:denmf}) is formally given in terms of the single-particle kinetic $\hat K$ and mean-field $\hat U$ operators:
\begin{align}
\hat{\rho}(t+\Delta t) = T^c\left\lbrace \text{e}^{-\frac{i}{\hbar}\int_t^{t+\Delta t} \text{d}t' \, \left[ \hat K_1 + \hat U_1 \right]} \right\rbrace \, \hat{\rho}(t) \, T^a \left\lbrace
 \text{e}^{\frac{i}{\hbar}\int_t^{t+\Delta t} \text{d}t' \, \left[ \hat K_{1'} + \hat U_{1'} \right]} \right\rbrace \, ,
\label{eq:SOM}
\end{align}
where $T^c$ and $T^a$ are chronologically and antichronologically ordering operators, respectively. One problem in applying the evolution operators is that the kinetic-energy and mean-field operators do not commute.  However, an expansion of the functions in the short time step $\Delta t$ leads to the identity
\begin{align}
\label{eq:BCH}
T^c\left\lbrace \text{e}^{-\frac{i}{\hbar}\int_t^{t+\Delta t} \text{d}t' \, \left[ \hat K + \hat U \right]} \right\rbrace = \text{e}^{-\frac{i}{\hbar}\int_{t \Delta t/2}^{t+\Delta t} \text{d}t' \, \hat U } \,  \text{e}^{-\frac{i}{\hbar} \hat K \, \Delta t } \, \text{e}^{-\frac{i}{\hbar}\int_{t}^{t+\Delta t/2} \text{d}t' \, \hat U }  + {\mathcal O} (\Delta t^3) \, ,
\end{align}
that helps to circumvent the issue of commutation over short time steps $\Delta t$.
In the above expression, the evolution operator has been factored into two mean-field and one kinetic energy factors. Within the spatial representation of the density matrix, the mean-field factor reduces to a $c$-factor.  On the other hand, within the momentum or wavevector representation of the matrix, the kinetic-energy factor becomes a $c$-factor.  In each representation, those factors are just phase factors and, for small $\Delta t$, they can be trivially applied to the density matrix.  The only serious practical issue that remains is that of the switching between configuration-space and wavevector representations.  This switching can be optimally accomplished using Fast Fourier Transforms (FFT) \cite{feit82}.  Note that, in applying the evolution factors from \eqref{eq:BCH} in (\ref{eq:SOM}), the diagonal elements of the density matrix remain unaffected either in the configuration or wavevector space.  However, the changes in the off-diagonal elements, together with the Fourier transforms, combine to induce modifications in the diagonal elements as well.

At each time step of evolution, the density matrix provides full single-particle information on the system, including density values, in either interpretation, 1D~or~3D. The total particle number in the 1D interpretation is equal to
\begin{align}
A(t) = \nu \int \! \textrm{d}x \, \rho(x,x;t) \, .
\label{eq:totden}
\end{align}
Upon rescaling with $\xi$, that number provides further the number of nucleons per transverse area in the 3D interpretation.  An evolution algorithm based on Eq.~\eqref{eq:BCH} ensures numerical conservation of the particle number.  The kinetic energy may be either computed from the density matrix in the spatial or wavevector representation:
\begin{align}
K(t) = \frac{\hbar^2}{2m} \, \nu \int \! \textrm{d}x \, \frac{\partial^2}{\partial x \, \partial x'} \left. \rho(x,x';t) \right|_{x'=x}
= \frac{\hbar^2}{2m} \, \nu \int \! \textrm{d}k \, k^2 \, \rho(k,k;t) \, .
\label{eq:ekin}
\end{align}
On the other hand, the potential energy is obtained from the density, according to
\begin{align}
V(t) = \frac{\nu}{2} \int \! \textrm{d}x \, \left[
\frac{3}{4} t_0 \left[ n(x) \right]^2 + \frac{1}{8} t_3 \left[ n(x) \right]^{2+\sigma} \right] \, .
\end{align}
The net energy, $E(t)=K(t)+V(t)$, is conserved by the mean-field evolution equations, even though $K$ and $V$ may individually change with time.

Our numerical code, that implements the time evolution for 1D slabs  following the SOM, employs a constant time step and an evenly spaced mesh in space.  The typical step combination has been $\Delta t=0.5 \, \text{fm}/c$ and $\Delta x = 0.25 \, \text{fm}$.  For FFT, it is necessary to adopt periodic boundary conditions at the extremes of the computational mesh.  Typically, we~have employed a mesh spanning the interval $-L \le x \le L$, where $L=25 \, \text{fm}$.  The~code has been tested in different ways, such as in reproducing the analytic evolution of free Gaussian wavepackets.  For any initial conditions, the code conserves particle number with an accuracy that is only limited by machine precision.  The accuracy of energy conservation depends on $\Delta t$ and will be discussed in the context of manipulations of the evolution, see Sec.~\ref{sec:offdiag}.  Some further testing of the code has involved verification of time reversibility, where the evolution was followed until a reaction has been largely completed and then ran back in time to reach practically the same initial conditions.  Some information pertinent to that testing will be, again, presented in the context of manipulations of the evolution in Sec.~\ref{sec:rev}.  Finally, we have tested the NGF code against a traditional TDHF code, arriving at consistent results, within the accuracy of those codes.

\section{Preparation of the initial state}
\label{sec:initial}

The initial state for a reaction needs to be constructed out of the ground states of the two nuclear systems entering the reaction.  Observables for a ground state should, in principle, not change with time.  However, if inconsistent approximations are employed for preparing the initial state of a system and for advancing its time evolution, spurious changes are likely to appear for observables that should have been static originally.  Such spurious changes are likely to impair the proper theoretical understanding of the nuclear reaction dynamics.

One way of preparing the initial system in a consistent manner within the NGF evolution is through an imaginary-time evolution \cite{danielewicz84b,kohler95}, employing analogous approximations for the self-energy within the imaginary- and real-time domains.  The complication, though, is that the boundary conditions for the imaginary evolution tie extremal instances in the evolution.  Because of this, the evolution has to be carried out in passes, until consistency is reached over a time interval that may possibly need to expand with the progress of iterations. As~a~consequence, the imaginary-time evolution requires a development of a numerical code separate to that for the real-time evolution, of likely greater complexity than the real-time code.  Naturally, it may be worthwhile to explore alternative ways of preparing the initial state.  When trying to avert the development of a separate code for finding such an~initial state, an obvious option is that of an adiabatic switching from a precursor, simple Hamiltonian to the more realistic Hamiltonian of interest.  Here, we explore that possibility, switching between an external potential and self-consistent mean-field.

At time $t \rightarrow - \infty$, whether ultimately aiming at a correlated or uncorrelated state, the~system may be assumed to represent the ground state for an external potential $U_0$, such as a Woods-Saxon or Harmonic Oscillator (HO).  Here, we choose the latter.  Provided that the mean-field interaction is switched on slowly enough, while the external potential is extinguished, the system may be evolved to the interacting ground state \cite{messiah}.  For the sake of minimizing the changes in the system, the precursor potential $U_0$ should be chosen such that the system's geometric characteristics do not evolve significantly, while the interactions are switched on.

With the adiabatic switching on, the single-particle potential acquires an explicit time dependence, and we adopt
\begin{align}
\label{eq:Ut}
U_t =  F(t - \tau_0) \, U_0 + [1 - F(t - \tau_0)] \, U \, ,
\end{align}
where $U$ is our mean-field parametrization and, further,
\begin{align}
U_0 =  \frac{1}{2} \, m \, \Omega^2 \, x^2
\end{align}
is the precursor HO potential. A choice of the frequency $\Omega$ to minimize the evolution of geometric characteristics is discussed in Appendix~\ref{appendix2}. The switching function $F$ above is equal to $1$ at some initial argument $t_i$ ($<0$) and to $0$ at some final argument $t_f$ ($>0$) when the switching over is completed.  This switching function is constructed in terms of a~monotonically {\em decreasing} function $f$ (such as $f(t)=-t$ in the linear case):
\begin{align}
\label{eq:Fft}
F(t) = \frac{f(t) - f(t_f)}{f(t_i) - f(t_f)} \, ,
\end{align}
for the argument values $t_i \le t \le t_f$.
In the switching function for Eq.~\eqref{eq:Ut}, we have most often employed
\begin{align}
f(t) = \frac{1}{1 + e^{{t}/{\tau} } } \,
\label{eq:ft}
\end{align}
within Eq.~\eqref{eq:Fft}. Here, $\tau$ represents a transition time that, for the sake of adiabaticity of the switching, should be longer than any characteristic time of the system.
Notably, whenever $f(t_i) \simeq 1$ and $f(t_f) \simeq 0$, such as for $|t_{i,f}| \gg \tau$ in the case of \eqref{eq:Fft}, then the two functions $f$ and $F$ practically coincide,
\begin{align}
F(t) \simeq f(t) \, .
\end{align}

\begin{figure}
  \begin{center}
    \includegraphics[width=.5\textwidth]{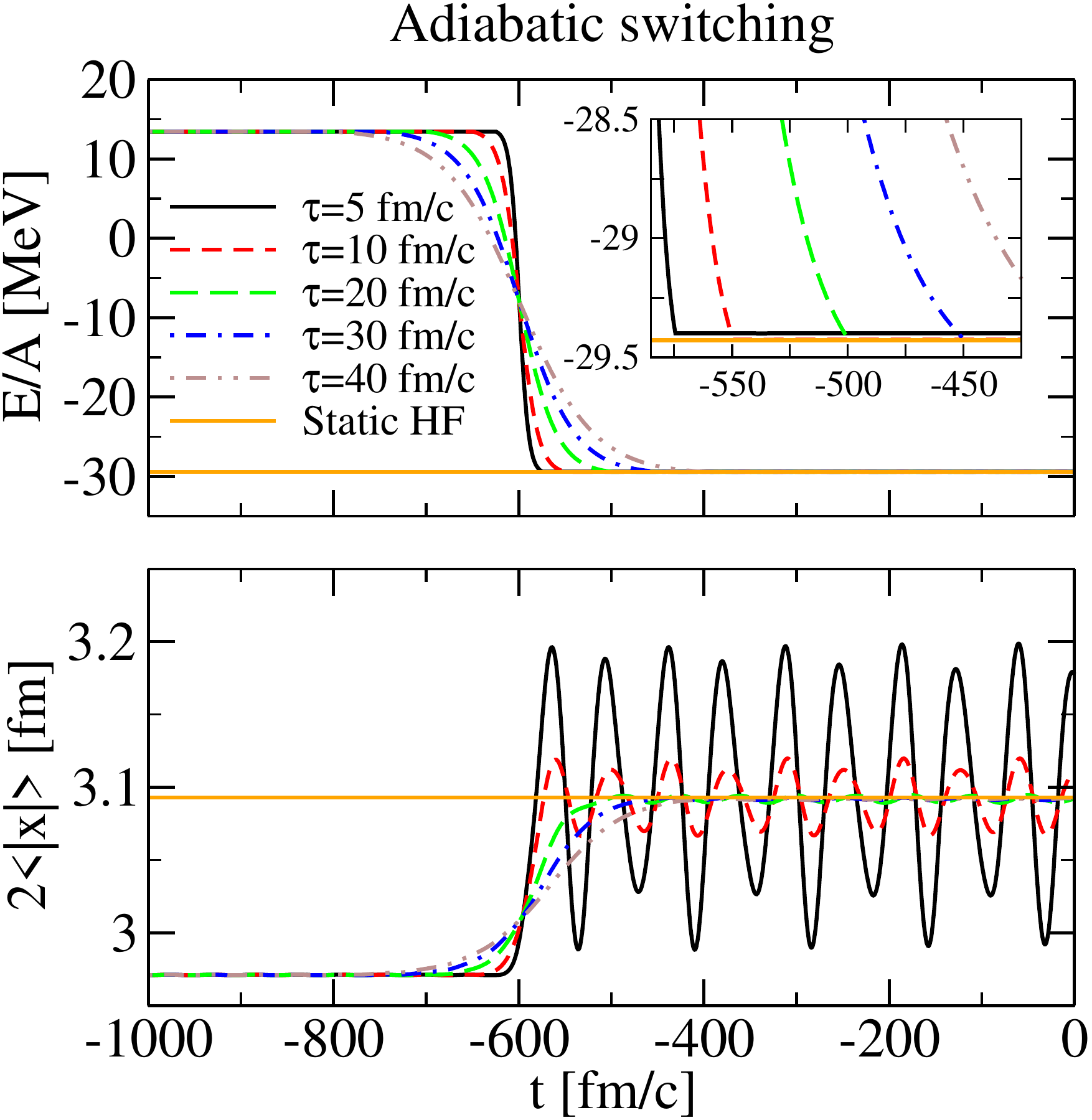}
    \caption{(Color online) Time evolution of the energy per particle (upper panel) and the size of the slab (lower panel) when starting from an HO configuration with $N_s=2$ filled shells (or $A=8$ in 1D interpretation) at time $t=-1000\, \text{fm}/c$ and transforming the single-particle potential to the mean-field form, according to Eq.~\eqref{eq:Ut} with Eq.~\eqref{eq:Fft} and~\eqref{eq:ft}. Different values of the transition time $\tau$ are considered.  Inset into the top panel shows a magnified portion of the time evolution of the energy.  For reference, the mean-field results from static Hartree-Fock solution are shown as straight solid lines.}
    \label{fig:ad_enerwidth}
  \end{center}
\end{figure}

The slower the process of switching over is, the better an approximation to the mean-field ground-state the final state of the adiabatic evolution is likely to be.  In the top panel of Fig.~\ref{fig:ad_enerwidth}, we show the evolution of the energy per nucleon for a slab initiated at time $t=-1000 \, \text{fm}/c$ with $N_s = 2$ HO shells filled (\emph{i.e.}\ $A=8$ nucleons in the 1D interpretation). We consider different transition times, $\tau$, with a fixed latent time $\tau_0 = -600 \, \text{fm}/c$.  For any of the employed transition times $\tau \ge 5 \, \text{fm}/c$, the energy evolves to a value very close to that for the static Hartree-Fock solution.  Between the employed values of $\tau = 5$ and $40 \, \text{fm}/c$, the energy per nucleon obtained at $t=0$ changes just by mere $0.04 \, \text{MeV}$!  Judging the quality of the approximation on the basis of the energy alone, though, can be treacherous.  This is because the energy is quadratic in the deviation of wavefunction from the ground state, around the energy minimum.
As a consequence, final states for the adiabatic evolution might be found, with a wavefunction poorly approximating that of the mean-field ground-state, but giving an energy close to the ground-state value. On the other hand, density $n$ is linear in the deviation of the wavefunction from the ground-state, so it may provide a better measure of the the wavefunction quality.  With this in mind, in the bottom panel of Fig.~\ref{fig:ad_enerwidth}, we show the evolution of the size of the slab, defined as
\begin{align}
D = 2\langle | x | \rangle = 2 \left. \int \textrm{d} x \, |x| \, n(x,t) \right/ \int \textrm{d} x \, n(x,t) \, .
\label{eq:width}
\end{align}
Indeed, for $\tau=5$ and $10 \, \text{fm}/c$, slab sizes exhibit significant oscillations in the final state, indicating that the ground state has not been yet satisfactorily reached.  For $\tau \gtrsim 30 \, \text{fm}/c$, as expected from the adiabatic theorem, the oscillations become insignificant, with the slab sizes practically coinciding with that for the static solution. Note also, in the context of Fig.~\ref{fig:ad_enerwidth}, that the emerging total thickness of the self-consistent slab, $2 D \sim 6.2$ fm, is quite close to that expected on the basis of Eq.~\eqref{eq:lA}, $\ell \sim 6.4$ fm.

To supplement the above results, we show in Fig.~\ref{fig:ad_den} the evolution of density from the predecessor HO form to the self-consistent one, for the transition time $\tau=40 \, \text{fm}/c$.  The densities shown here and throughout this paper will be those of the 3D interpretation. Already at $t \sim -400 \, \text{fm}/c$, the asymptotic form of the density is reached.  Notable within the density is the dip at the center of the slab.  This is a reflection of the node in the second asymmetric orbital.  As such, the dip represents a shell effect that gets to be somewhat more pronounced for the mean-field than HO potential.  The average density across the slab center ends up not too far from the normal density of $n_0 = 0.16 \, \text{fm}^{-3}$. Overall, during the adiabatic switching on of the mean-field, the changes in the density are quite modest, attesting the utility of Eq.~\eqref{eq:w0} in choosing the HO frequency, $\Omega$.

\begin{figure}
  \begin{center}
    \includegraphics[width=.75\textwidth]{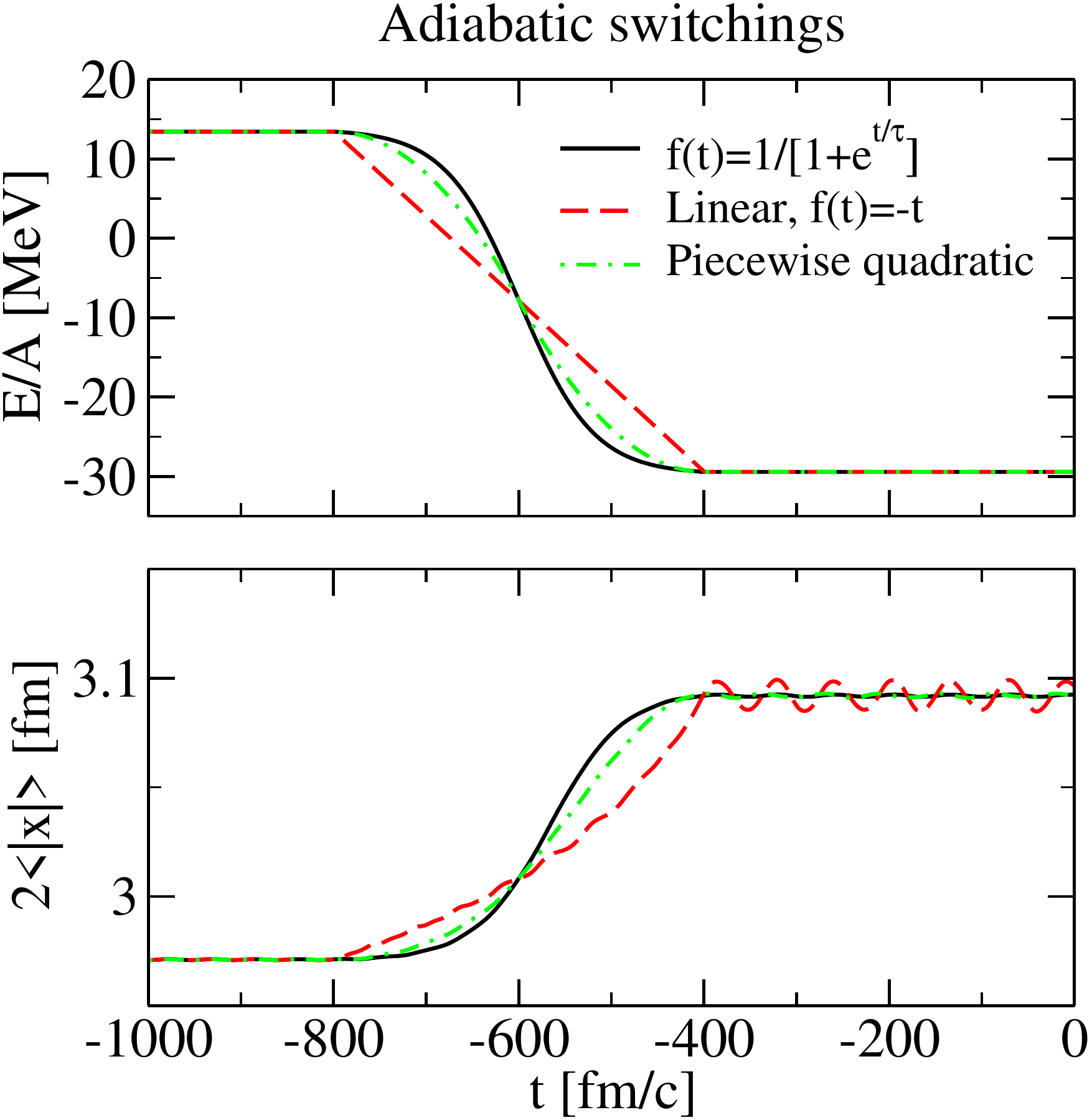}
    \caption{(Color online) Time evolution of the energy per particle (upper panel) and the slab width (lower panel), when starting from an HO configuration with $N_s=2$ filled shells at time $t=-1000\, \text{fm}/c$ and transforming the single-particle potential from the HO to the mean-field form, following different switching functions: of the Fermi-Dirac form [see Eq.~\eqref{eq:ft}], linear and piecewise quadratic.}
    \label{fig:ad_func}
  \end{center}
\end{figure}

Let us also mention that the switching function $F$ with a Fermi-Dirac $f$ as shown in Eq.~\eqref{eq:ft} is in practice optimal. This particular switching function leads to a close approximation of the ground state within the shortest possible time, from among tested functions. To illustrate that point, we show in the top panel of Fig.~\ref{fig:ad_func} the evolution of the energy per nucleon and the width for the slab considered previously, initiated with $N_s = 2$ HO shells filled at $-1000 \, \text{fm}/c$. Different switching functions are considered: one of the Fermi-Dirac type as in Eq.~\eqref{eq:ft}, with $\tau=40 \, \text{fm}/c$, and, further, a linear and a piecewise quadratic function with suitably chosen characteristic transition times.  In each case, the bulk of the changes occurs over the same period of $\sim 400 \, \text{fm}/c$ and is centered at $\tau_0=-600 \, \text{fm}/c$.  A~first striking observation concerning Fig.~\ref{fig:ad_func} is how closely the energy per nucleon follows the shape of the switching function.
This can be attributed to the relative adiabaticity of the switching, with the deviation of the wavefunction being small from an instantaneous ground state, and to the quadratic dependence of the instantaneous energy upon that deviation.  As evident in the bottom panel of Fig.~\ref{fig:ad_func}, for the Fermi-Dirac $F$ also the width of the slab follows closely the shape of the switching function, remaining in particular stable at late stages of the transition.  On the other hand, the width of the slab for the linear $F$, in the same panel, oscillates both during and after the transition, underscoring a~poorer adiabaticity of the transition for that $F$.  Inferior adiabaticity for the linear $F$ may be surprising, as the slope of the linear $f$ is lower than the slope of the Fermi-Dirac~$f$ at every $t$.  Even in the case of the piece-wise quadratic $F$, oscillations in the width may be discerned for the final state of the transition, although they are of significantly lower amplitude than for the linear switching function. This points towards the importance of using smooth functions for adiabatic transitions.

We hope that, in a similar manner as with the switching on of the mean-field, the retarded self-energies $\Sigma^\lessgtr$, representing correlations, can be switched on within the KB equations, with parallel changes taking place in the instantaneous self-energy.  The goal would be, as here, for the system to stay close to the ground state for a given instant of evolution.  With correlations, however, the switching would need to be  slow compared to characteristic correlation times~\cite{morawetz99}.  The success of the switching procedure could be, in particular, judged again with the degree to which the final state is stationary.

After the density matrix for an interacting ground state has been constructed, it needs to be boosted in order to be incorporated within the initial state of a collision.  The boost is accomplished through a simple multiplication by momentum phase factors~\cite{thouless62}:
\begin{align}
\rho_1 (x_1,x_{1'}) = \text{e}^{i P x_1/\hbar } \, \rho_0(x_1,x_{1'}) \, \text{e}^{-i P x_{1'}/\hbar} \, .
\label{eq:boost}
\end{align}
Here, $\rho_1$ is the density matrix for a slab moving at momentum per nucleon $P$ while $\rho_0$ is the matrix for an idle slab.  As a consequence of the boost, the kinetic energy for the slab increases by the amount ${ \Delta K}/{A}= {P^2}/{2 m}$.  The net density matrix is constructed as a sum of the density matrices for two countermoving slabs:
\begin{align}
\rho (x_1,x_{1'}) = \rho_1 (x_1,x_{1'}) + \rho_2 (x_1,x_{1'}) \, .
\label{eq:net_rho}
\end{align}
In this paper, we present symmetric collisions, with the second slab being a symmetric reflection of the first.  With this, the density matrix of the second slab is related to the matrix of the first by
\begin{align}
\rho_2 (x_1,x_{1'}) = \rho_1 (-x_1,-x_{1'}) \, .
\end{align}
Under the reflection symmetry, the center-of-mass (CM) energy of the collision becomes simply $E_{CM}/A = {P^2}/{2 m}$.

The net density matrix as a sum of the two individual density matrices \eqref{eq:net_rho} is equivalent to the Slater determinant for the two nuclei being built up in an independent manner~\cite{simenel08}.  With this, no coherence is assumed between the initial, far-away states of the two nuclei.

\section{Collisions of slabs}
\label{sec:collisions}

The relatively simple 1D model of nuclear collisions discussed here demonstrates a surprisingly rich range of phenomena~\cite{bonche76,kohler84}.  Qualitatively different physical processes are observed within the model when changing the CM energy for the reactions.  At low collision energies ($E_{CM}/A \sim 0.1-0.5 \, \text{MeV}$), the nuclear slabs fuse into one compound slab that remains excited for long times.   For intermediate energies ($E_{CM}/A \sim 0.5-15 \, \text{MeV}$), a fusion process is observed, followed by a break-up into a number of smaller slabs.  Higher reaction energies ($E_{CM}/A > 15 \, \text{MeV}$) yield a pile-up of density at the system center, followed by a~violent break-up phase with the formation of a fragmented low-density neck.  The process is reminiscent of multifragmentation in nuclear reactions.  Since analyzing phenomena in slab collisions is not the principal goal of our paper, the discussions of those different phenomena will be quite limited here.

\begin{figure}
  \includegraphics[height=.4\textheight]{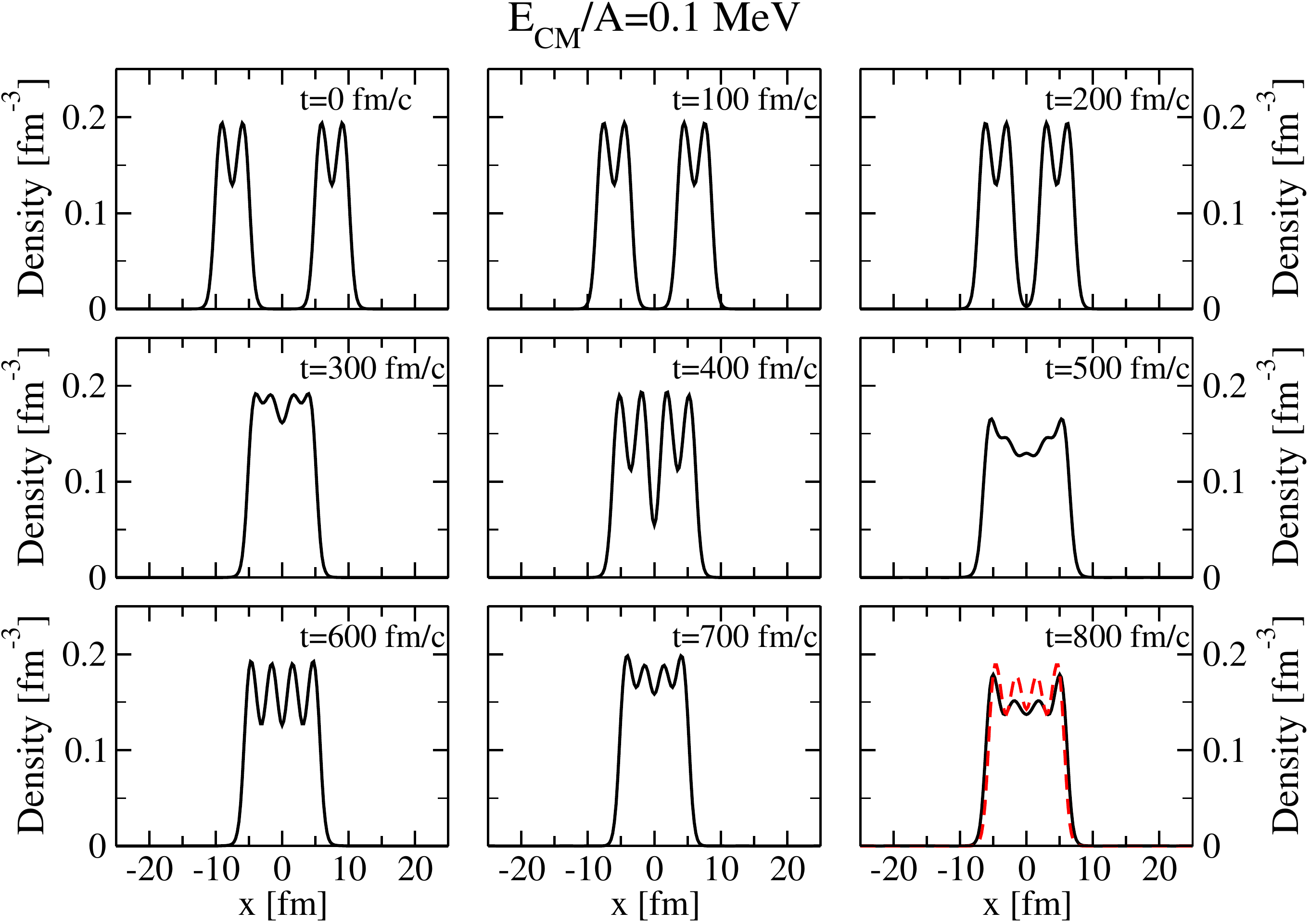}
  \caption{(Color online) Evolution of the CM density profile, represented by solid line, for a collision of two $A=8$ slabs at the collision energy of $E_{CM}/A = 0.1 \, \text{MeV}$.  The dashed line in the last panel represents the density profile for the ground state of an $A=16$ slab, provided for comparison.}
  \label{fig:denmat_ea01}
\end{figure}

At low collision energies, the left and right slabs approach each other slowly until they get in contact and start to overlap. In our simplified model, without Coulomb effects, fusion can take place no matter how low the collision energy. The formed compound slab acquires a~total mass of $2A$ and an excitation energy that exceeds the collision energy.  As~an~example, Fig.~\ref{fig:denmat_ea01} displays the evolution of the density for two $A=8$ slabs colliding at the rather low collision energy of $E_{CM}/A = 0.1 \, \text{MeV}$.  At $t=0$, the slab centroids are separated by $15 \, \text{fm}$.  Following the contact at $t \sim 250 \, \text{fm}/c$, a compound slab is formed.  The~energy associated with the translational motion of the slabs is converted into the energy of collective oscillations for the compound slab.  Those oscillations are reflected in temporal changes of different characteristics of the system. In particular, the density profile is observed to oscillate, in the later stages of the collision, around the shape of the Hartree-Fock ground state of the $A=16$ system, shown for reference in the last panel of Fig.~\ref{fig:denmat_ea01}.  The oscillations cannot easily damp out in this model~\cite{kohler80}, because their energy cannot be transferred to the transverse degrees of freedom and because of the constrained values of the occupations of the evolved single-particle states, which inhibit the transfer of energy from longer to shorter wavelengths.

\begin{figure}
  \includegraphics[width=.6\textwidth]{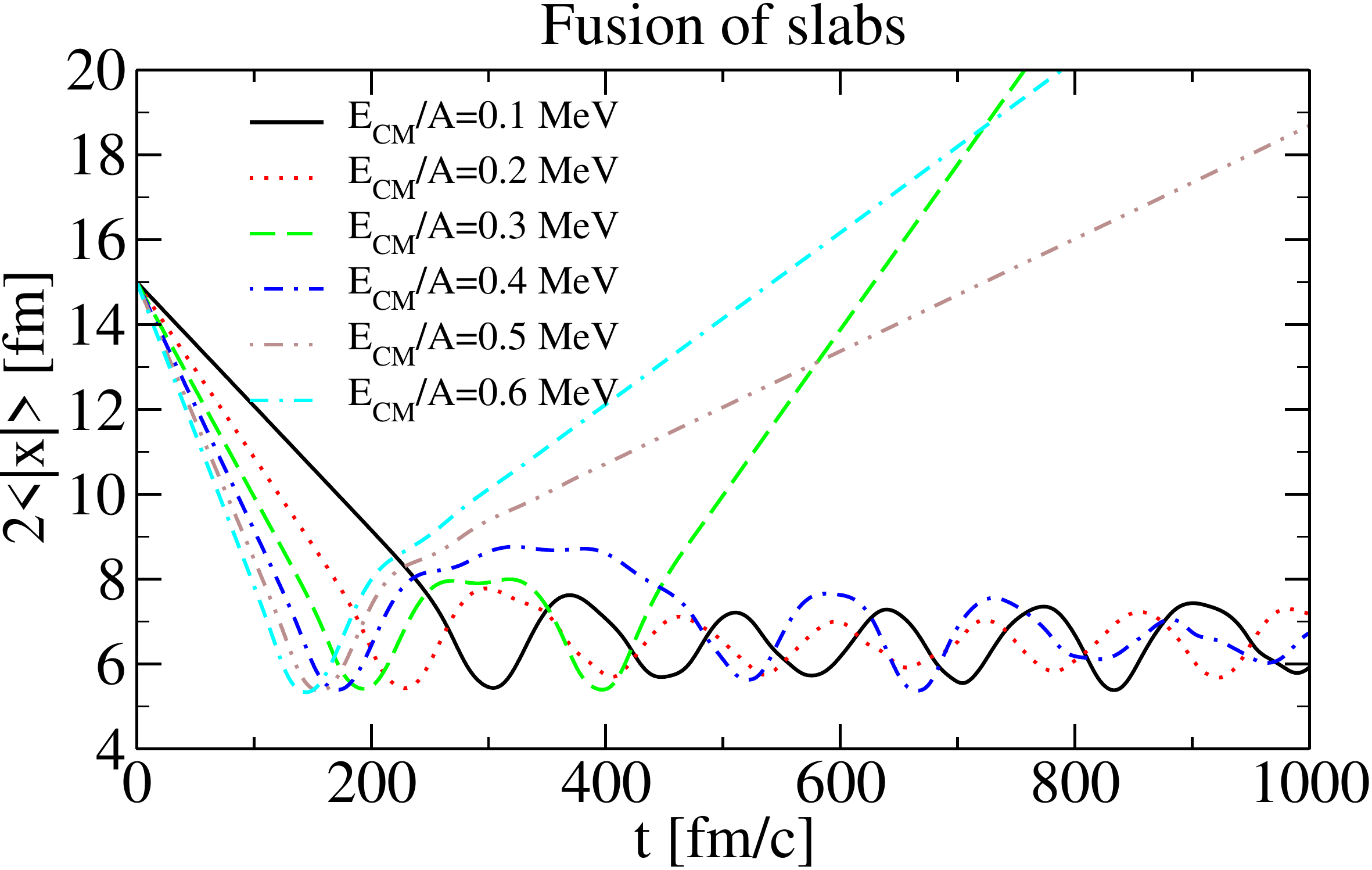}
  \caption{(Color online) Time evolution of system size for a collision of two $A=8$ slabs for different indicated collision energies. Oscillating sizes in the late stages of the evolution represent fused systems.  Sizes that grow monotonically correspond to systems that did not ultimately fuse.}
  \label{fig:fusion}
\end{figure}

To provide insight into the process of slab fusion as a function of collision energy, we~show in Fig.~\ref{fig:fusion} the evolution of the system size, as given in Eq.~\eqref{eq:width}, for different collision energies. Before contact, at early times, the slabs approach each other at a constant relative velocity equal to the slope of the size.  With the slabs starting out always $15 \, \text{fm}$ apart, higher collision energies involve earlier contact times, signaled by a sudden change in the slope of the indicated system size.  As the collision energy is changed, qualitatively different outcomes are observed for the collision final states.  For the lowest collision energies, at~times $t > 200 \, \text{fm}/c$, the system size oscillates around a central value.  Boundedness of the size signals the formation of a compound slab of mass $2A = 16$.  Following Eq.~\eqref{eq:lA}, the~expected size for a ground-state slab with that mass number is $\ell/2 = 6.4 \, \text{fm}$ and this appears to be about the value around which the oscillations occur at late times in Fig.~\ref{fig:fusion}.  For collision energies in excess of about $E_{CM}/A = 0.45 \, \text{MeV}$, the two slabs interpenetrate but then separate back into two $A=8$ fragments that move apart at a lower relative velocity than initially, as indicated by the reduced slope of the late time size in Fig.~\ref{fig:fusion}.  The reduced velocity points to internal excitations for the separating fragments.  It should be noted that, counterintuitively, the transition between the fusion and fusion-fission regimes is neither gradual nor monotonic.  Thus, while at $E_{CM}/A = 0.2 \, \text{MeV}$ the slabs fuse in Fig.~\ref{fig:fusion}, the slabs separate at $0.3 \, \text{MeV}$ and then fuse again at $0.4 \, \text{MeV}$.  The slabs that separate at $E_{CM}/A = 0.3 \, \text{MeV}$, after one full oscillation of the system, move apart faster than either at $0.5 \, \text{MeV}$ or $0.6 \, \text{MeV}$.  These observations point to underlying resonant phenomena in the transfer of energy between translational and internal degrees of freedom of the slabs~\cite{bonche76}.  Even when TDHF is supplemented with a collision term in the relaxation time approximation, the late-time separation back into original fragments persists over a certain range of energies~\cite{grange81}.

As the collision energy increases further, the final states of the collisions undergo additional changes.  Thus, at energies $E_{CM}/A \gtrsim 1.5 \, \text{MeV}$, the system subdivides into three fragments.  At the center of the system, a slab representing a mass of about $A=8$ forms and stays at rest in the system CM.  Besides, two residual slabs of mass of about $A=4$ form, moving symmetrically forward and backward in the CM.  An example of such a collision, at the energy of $E_{CM}/A = 4 \, \text{MeV}$, is shown in Fig.~\ref{fig:denmat_ea4}.  Following the fusion of the original slabs at $t \sim 50 \, \text{fm}/c$, peaks in density, not far from $n_0$, begin to emerge symmetrically at the edges of the matter at $t \sim 100 \, \text{fm}/c$.  By~$t \sim 125 \, \text{fm}/c$, those peaks separate from the central fragment. At first, this remainder central region seems to fragment further. However, while the leading peaks in the density manage to separate from the central slab, the two more central peaks lack sufficient energy to separate from each other.  In the end, the central region recontracts and oscillates, representing, in this, a highly excited state of an $A \sim 8$ system.  The $A \sim 4$ fragments remain excited too, as evidenced in the small changes with time of their central density.

\begin{figure}
  \includegraphics[height=.4\textheight]{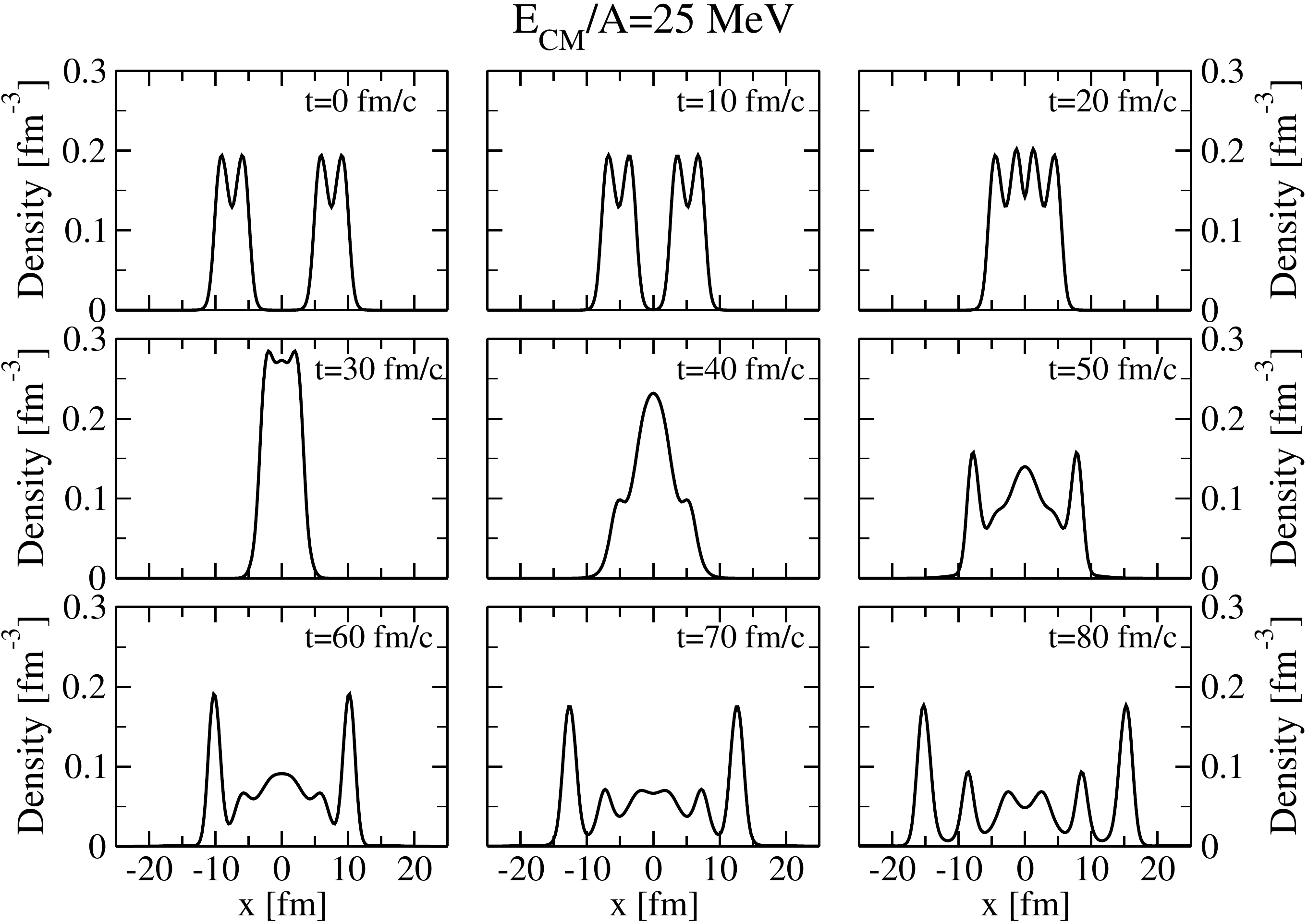}
  \caption{Evolution of the CM density profile for a collision of two $A=8$ slabs at the collision energy of $E_{CM}/A = 25\, \text{MeV}$.}
  \label{fig:denmat_ea25}
\end{figure}

Generally, as the collision energy increases, the maximal density reached at the system center rises.  The time during which the density stays substantial, on the other hand, decreases.  The decompression of the central region becomes more and more violent for higher energies, thus enhancing the number of produced fragments.  As an example, Fig.~\ref{fig:denmat_ea25} shows a collision of two $A=8$ slabs at $E_{CM}/A = 25 \, \text{MeV}$.  At the maximal compression point, $t \sim 30 \, \text{fm}/c$, the central density exceeds by more than 70\% that in normal matter.  By~$t \sim 50 \, \text{fm}/c$, leading density peaks begin to form at the edges of matter, not far from $n_0$, similarly to the situation at $E_{CM}/A = 4 \, \text{MeV}$.  While these peaks continue to separate, the~central region undergoes structural changes.  In contrast to the situation at $4 \, \text{MeV}$, rather than attempting to fragment into pieces at normal density, the central region undergoes a~rapid decompression and a low-density neck region forms between the leading residual fragments, see panels for $t = 60$ and $70 \, \text{fm}/c$ in Fig.~\ref{fig:denmat_ea25}.  The neck subsequently begins to recontract into separate fragments. The fate of those low-density structures is difficult to predict without increasing the size of the calculational box.

\section{Off-diagonal structure of the density matrix}
\label{sec:offdiag}

At equal arguments in a specific representation, such as the configuration or the momentum representation, the density matrix or, more generally, the function $-iG^<$ of Eq.~\eqref{eq:G<}, yields the single-particle density in that representation.  At different arguments, the off-diagonal matrix elements reflect existing correlations between magnitudes and phases of single-particle wavefunctions, cf.\ Eq.~\eqref{eq:rhowf}.  As we have discussed earlier, the task of following all the elements in 3D is likely to overwhelm present computer storage capabilities.  Nevertheless, the quantities of most direct physical interest, including densities, tend to be associated with either diagonal values of the functions or values close to the diagonal~\cite{kadanoff}.  Hopefully, the matrix elements which lie sufficiently far from the diagonal may be ignored and realistic NGF calculations may actually be implemented numerically.  Note, however, that the temporal evolution couples different matrix elements, cf.\ Eqs.\  \eqref{eq:kb1p} and \eqref{eq:denmf}.  Moreover, off-diagonal elements in one representation are needed for the transformation to another representation and thus for obtaining even the diagonal elements there.  Any discarding of the elements will have to be carefully tied to a specific representation and justified within that representation.  Impact on other representations of interest will need to be understood.

\begin{figure}
  \includegraphics[height=.4\textheight]{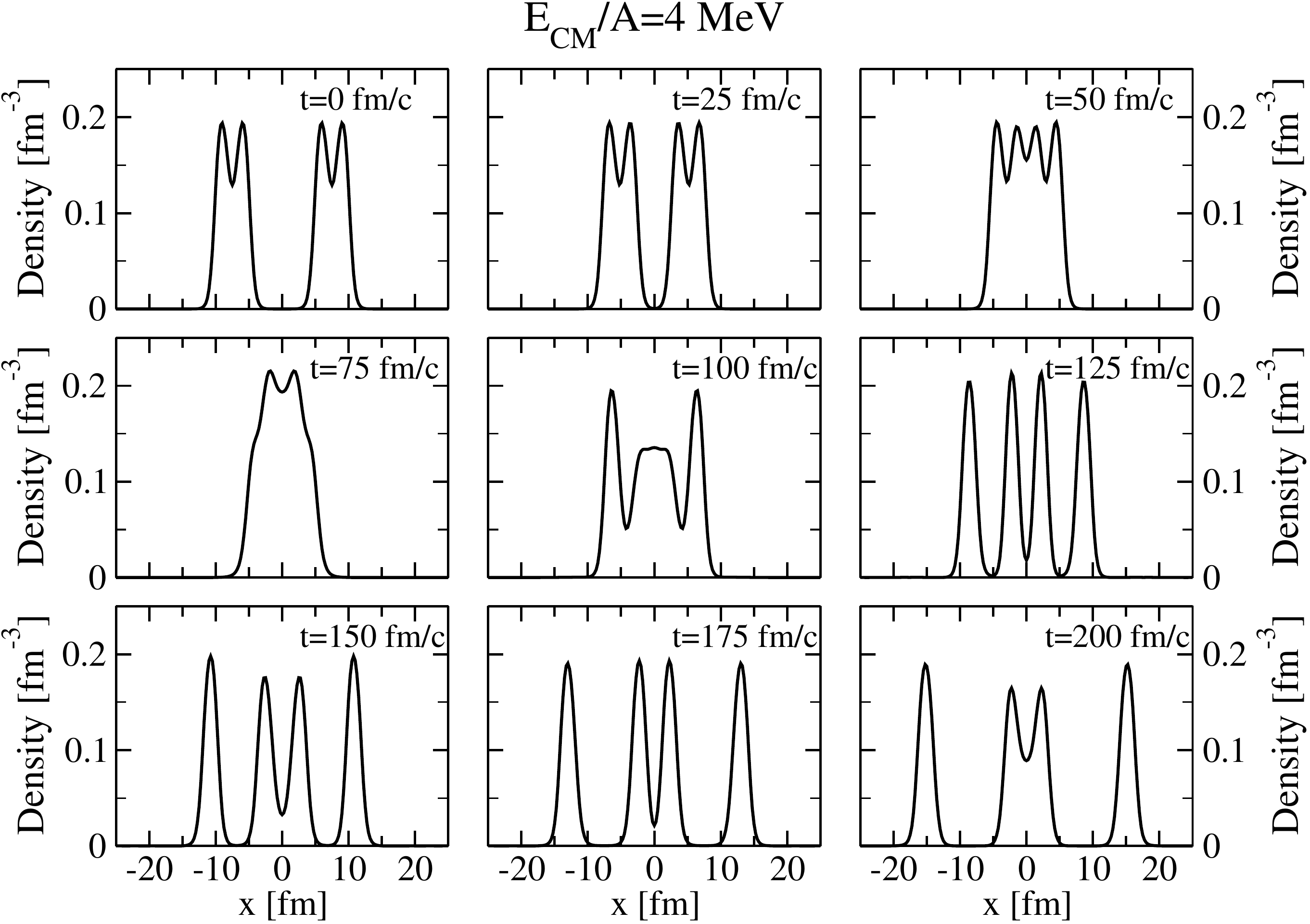}
  \caption{Evolution of the CM density profile for a collision of two $A=8$ slabs at the collision energy of $E_{CM}/A = 4 \, \text{MeV}$.}
  \label{fig:denmat_ea4}
\end{figure}

\begin{figure}
  \includegraphics[height=.4\textwidth]{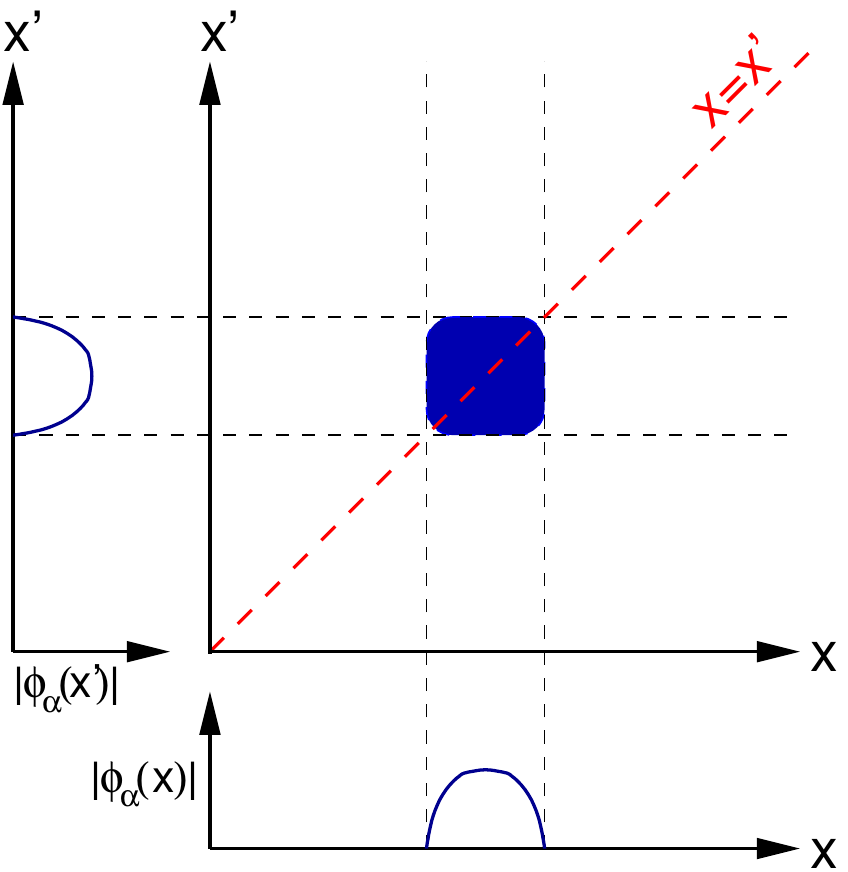} \hskip.5cm
  \includegraphics[height=.4\textwidth]{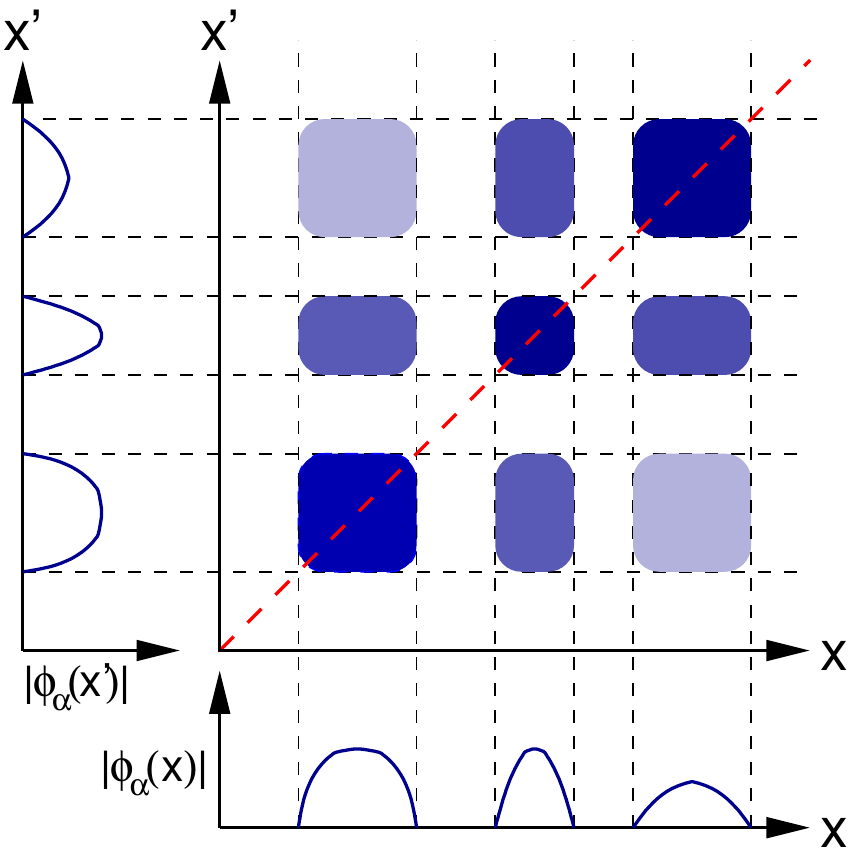}
  \caption{(Color online) Schematic illustration showing the origin of the characteristic features in the off-diagonal structure of a single-particle density matrix.  In the case of localized single-particle states (left panel), the density matrix is confined to a square-like region.  Fragmented single-particle states (right panel) give rise to a patch structure for the density matrix, with patches extending arbitrarily far away from the diagonal.}
  \label{fig:rhopl}
\end{figure}

In the following, we shall examine the off-diagonal structure of the density matrix within the spatial representation for colliding 1D slabs.  Figure~\ref{fig:rhopl} illustrates what may be generally expected, as far as the off-diagonal structure of the density matrix is concerned.  The~wavefunctions of the initial slabs are confined to within the size of the respective slab.  For the density matrix (see the left panel of Fig.~\ref{fig:rhopl}), this implies a compact support limited to the size of the slab in all directions. In particular, a square-like region in the variables $x$ and~$x'$ will develop around the $x=x'$ axis, cf.\ Eq.~\eqref{eq:rhowf} for equal time-arguments.  In the late stages of a higher-energy collision, such as in Fig.~\ref{fig:denmat_ea4}, the single-particle wavefunctions are shared between different fragments, reflecting the fact that the individual original nucleons had been probabilistically distributed within the different fragments.  Correspondingly, the~density matrix is likely to develop a patch-like structure extending far away from the diagonal (see the right panel of Fig.~\ref{fig:rhopl}).  As far as magnitudes are concerned, the contributions to the {\em diagonal} elements from different wavefunctions are all real and positive, as evidenced in Eq.~\eqref{eq:rhowf}.  On the other hand, phase differences between wavefunction values at different locations will make the contributions that arise from those individual wavefunctions come with different phases to {\em off-diagonal} elements, plausibly suppressing those elements compared to the diagonal.  Larger differences in position and larger energies will generally increase such phase differences and one might expect corresponding greater suppression of off-diagonal elements.

Physics beyond the mean-field dynamics (short-range correlations, particle and gamma decays, etc.) are likely to introduce additional decoherence between the separating fragments, beyond that stemming only from different mean-field orbitals.  Correspondingly, higher values for the off-diagonal elements are likely to persist more in a mean-field approach than in any simulation with correlations.  As a corollary, if one finds out that the far off-diagonal elements may be disregarded within the mean-field dynamics, then it should be even more possible to disregard such elements in more realistic approaches.

\begin{figure}
  \includegraphics[width=.9\textwidth]{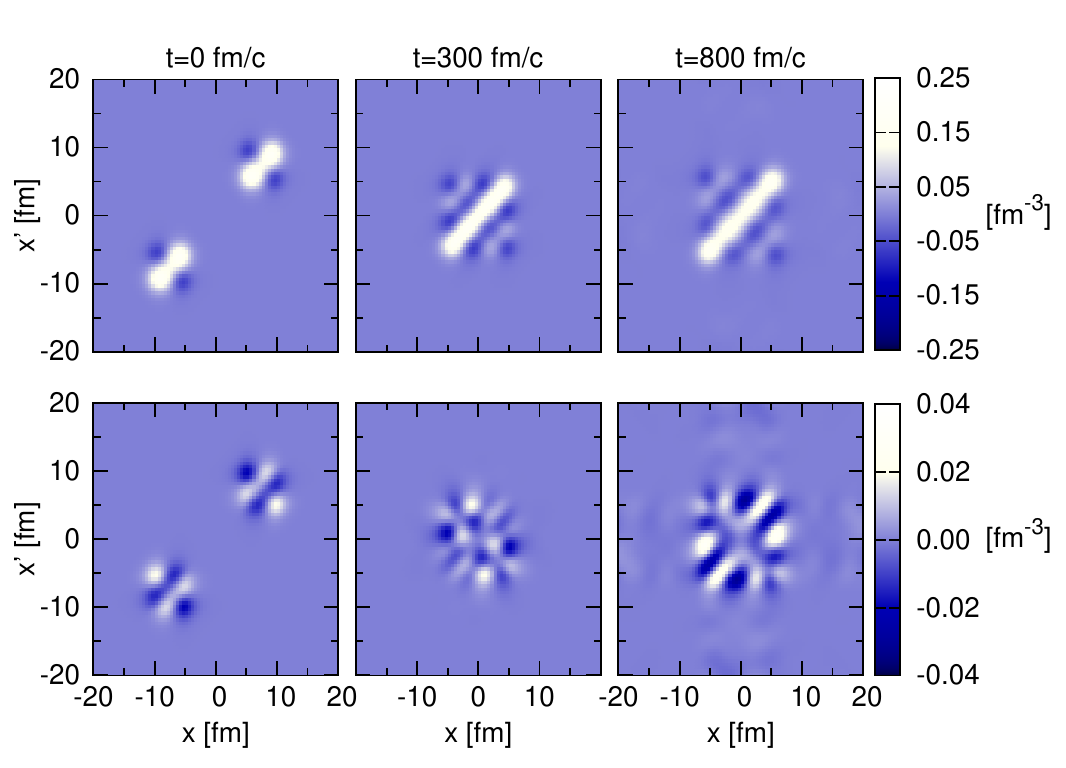}
  \caption{(Color online) Intensity plots representing the real (upper panels) and the imaginary part (lower panels) of the scaled density matrix, $\xi \nu \, \rho(x,x';t)$, for a collision at $E_{CM}/A=0.1 \, \text{MeV}$.  The~left, central and right panels show the initial, overlap and late stages of the collision, respectively.  The~scaling aims at making the values along the diagonal coincide with the 3D density shown in~Fig.~\ref{fig:denmat_ea01}. }
  \label{fig:2dfus}
\end{figure}

\begin{figure}
  \includegraphics[width=.9\textwidth]{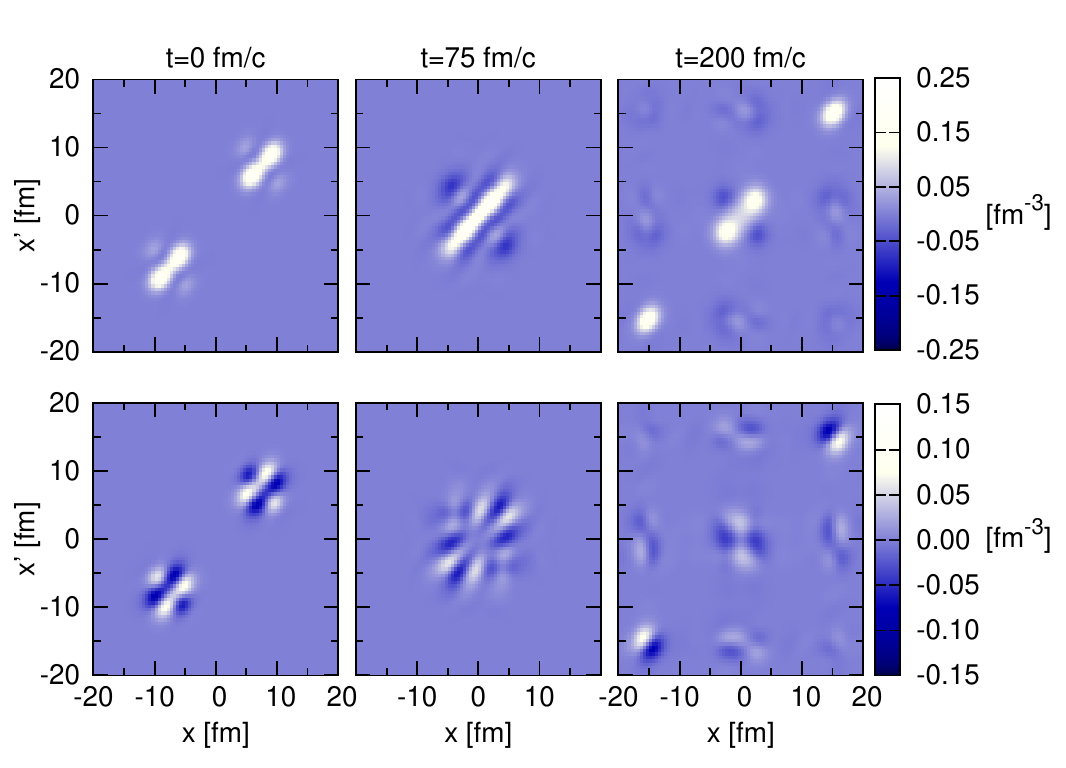}
  \caption{(Color online) Same as Fig.~\ref{fig:2dfus} but for a collision at $E_{CM}/A=4$ MeV.}
  \label{fig:2dfis}
\end{figure}

\begin{figure}
  \includegraphics[width=0.9\textwidth]{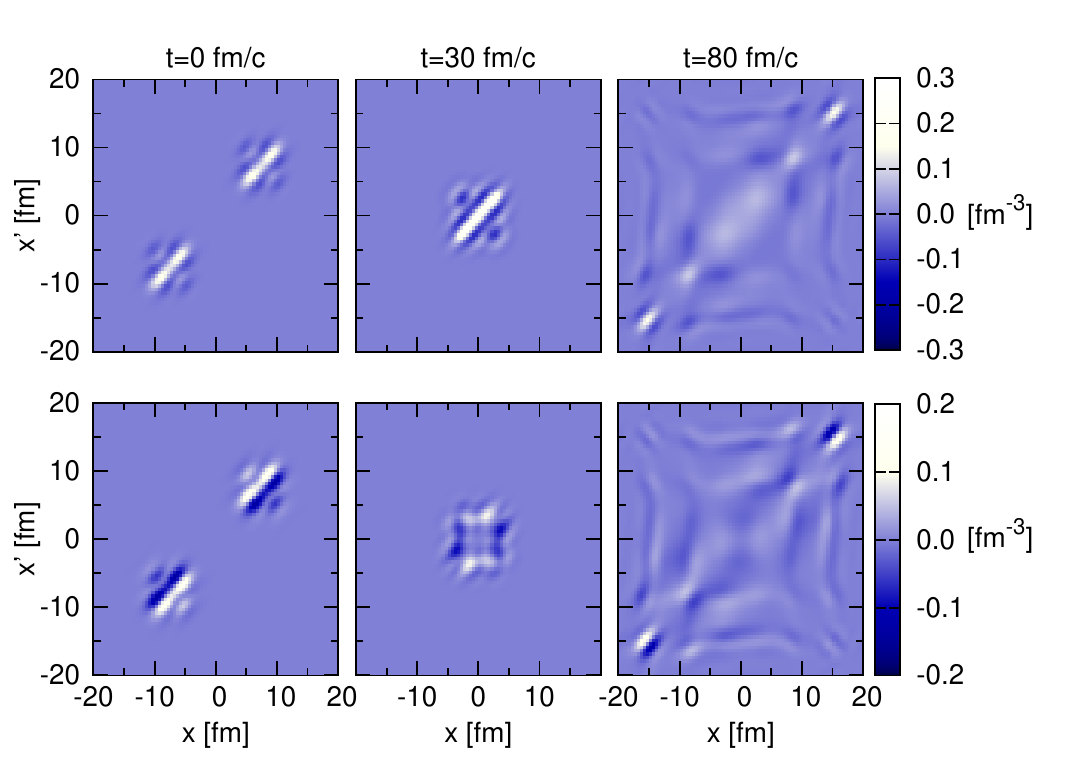}
  \caption{(Color online) Same as Fig.~\ref{fig:2dfus} but for a collision at $E_{CM}/A=25$ MeV. }
  \label{fig:2dmult}
\end{figure}

The 1D nature of the dynamics discussed here allows for a straightforward visualization of the density matrix in terms of intensity plots.  Figures~\ref{fig:2dfus}, \ref{fig:2dfis} and \ref{fig:2dmult} show such intensity plots for $\rho(x,x';t)$ at different times $t$ for the $E_{CM}/A = 0.1$, 4, and $25 \, \text{MeV}$ collisions of $A=8$ slabs.  The top and bottom panels exhibit, respectively, real and imaginary parts of the matrix.  Three sample times have been chosen to represent the initial, overlap and late instants within each collision.  In addition, Figs.~\ref{fig:denmat_ea01}, \ref{fig:denmat_ea4} and \ref{fig:denmat_ea25} show values of the density matrix along given lines perpendicular to the $x=x'$ line for the same sample times. All of the mentioned figures reflect the hermiticity and/or positive definiteness of the density matrix.  With the matrix being hermitian, its real part is a symmetric function of its two spatial arguments, while its imaginary part is antisymmetric.  In particular, the imaginary part vanishes along the $x=x'$ diagonal.  In addition, when studying symmetric collisions, we encounter the $(x,x') \rightarrow (-x,-x')$ symmetry for the matrix in the $(x,x')$ plane.

\begin{figure}
  \includegraphics[width=.9\textwidth]{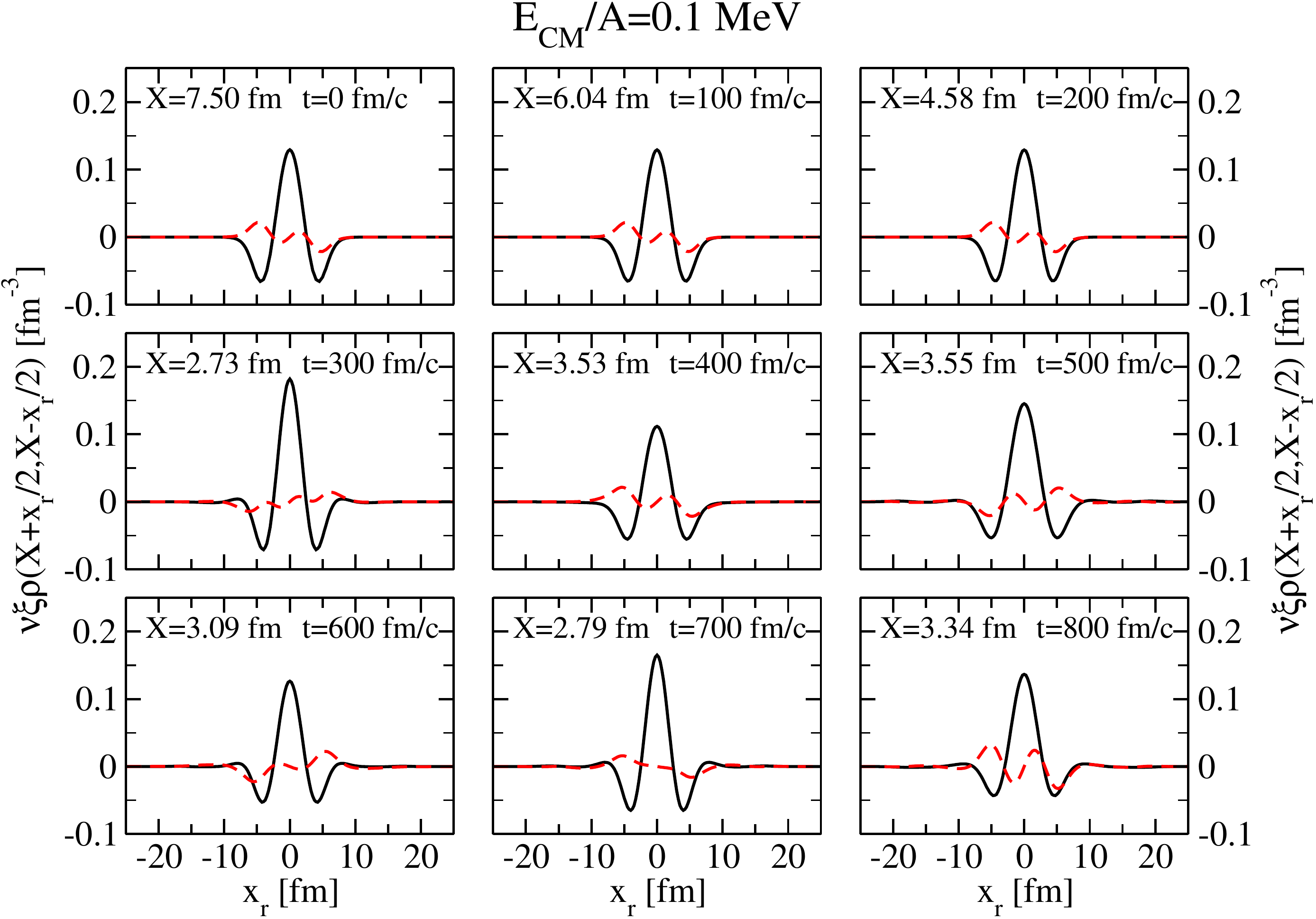}
  \caption{(Color online) Scaled density matrix, $\xi \nu \, \rho(x,x';t)$, for the collision at $E_{CM}/A = 0.1 \, \text{MeV}$, along a line $x + x' = 2X = const$, perpendicular to the diagonal $x=x'$ in the $(x,x')$ plane.  Real (solid lines) and imaginary (dashed) values are shown against the coordinate difference $x-x' = x_r$, at sample values of $X$ and $t$.}
  \label{fig:offdenfus}
\end{figure}

In the following, we discuss, in sequence, the structure of the density matrices in configuration space for the collisions at the three energies specified previously.  We start out with the matrix for the initial state of the low-energy slab fusion, illustrated in the left panels of~Fig.~\ref{fig:2dfus}.  In addition, the top left panel in Fig.~\ref{fig:offdenfus} shows a section across the $t=0$ density matrix, along a line perpendicular to the diagonal of the matrix.  The single-particle orbitals, used for the construction of the ground-state, are real and centered around $\pm 7.5 \, \text{fm}$ for the two slabs.  As it is apparent in Fig.~\ref{fig:2dfus}, the compact orbitals yield compact supports for the individual contributions of the two slabs, to the total density matrix. These contributions occupy square-like regions in the $(x,x')$ plane, centered around the points $x=x' = \pm 7.5 \, \text{fm}$.  The density matrix peaks along the diagonal, $x=x'$.  With the orbitals constructed independently for the two slabs, coherence patches at $x \simeq -x'$, as suggested by the right panel in Fig.~\ref{fig:rhopl},  are absent in Fig.~\ref{fig:2dfus}. At $t=0$, positive peaks in the real part of the matrix along the diagonal, are followed by negative values farther away from the axis, see Fig.~\ref{fig:2dfus} and also~\ref{fig:offdenfus}.  For this reaction, the imaginary part is generally much smaller in magnitude than the real part, both at $t=0$ and later.

If there were no boosts involved, the matrix out of real orbitals would have been, in fact, purely real.  Moreover, if only the lowest, $\alpha=1$, purely positive orbitals were filled, the matrix would have been positive.  The negative values of the real part of the matrix can therefore be attributed, in the initial state, to the filling of the second orbital, odd with respect to the origin of the slab.  With the filling of just those two orbitals, the matrix in the central region of the slab can be already successfully compared to that expected in the ground-state of nuclear matter:
\begin{align}
\rho_0 (x,x') = \frac{1}{2 \pi \hbar} \int_{-p_{F1}}^{p_{F1}} \text{d}p
\, \text{e}^{i p (x - x')/\hbar} = \frac{1}{\pi (x - x')}
\sin{\left[ \frac{p_{F1} (x-x')}{\hbar} \right]} \,
\label{eq:rhonm}
\end{align}
where $p_{F1}$ is the 1D Fermi momentum, cf.\ Eq.~\eqref{eq:rhowf} and Appendix \ref{appendix1}.  The function on the r.h.s.\ of Eq.~\eqref{eq:rhonm} is, in particular, familiar from the context of slit diffraction.  However, in contrast to slit diffraction, a band of momenta rather than positions is permitted here.  The~result of Eq.~\eqref{eq:rhonm} predicts alternating regions of positive and negative values in the~direction perpendicular to the diagonal as well as a central positive region twice as wide as the others.  In particular, the first zeros in the off-diagonal direction are anticipated at
\begin{align}
|x - x'| = \frac{\hbar \pi}{p_{F1}} \simeq 3.2 \, \text{fm} \, ,
\label{eq:3.2}
\end{align}
which agrees reasonably well with what may be observed in the first panel of Fig.~\ref{fig:offdenfus}.  In the estimate, we have used Eqs.~\eqref{eq:pF13} and \eqref{eq:npF3} with $n_0 = 0.16 \, \text{fm}^{-3}$, yielding $p_{F1}/\hbar = 1.0 \, \text{fm}^{-1}$.  Furthermore, when a slab is boosted to an average momentum per nucleon $P$, the ground-state density matrix gets multiplied by a phase factor, cf.\ Eq.~\eqref{eq:boost},
\begin{align}
\rho_1(x,x') = \rho_0(x,x') \left\lbrace \cos{\left[ \frac{P (x - x')}{\hbar}  \right]}        + i \sin{\left[ \frac{P (x - x')}{\hbar}  \right]}
\right\rbrace \, .
\label{eq:rho1P}
\end{align}
With a real ground-state matrix $\rho_0$, a simple relation follows for the boosted matrix, $\text{Im} \, \rho_1(x,x') = \tan{[P(x-x')/\hbar]} \, \text{Re} \, \rho_1(x,x')$.  Whenever the initial momentum per nucleon is low, \emph{i.e.} $P/\hbar < 2/\ell$, with $\ell$ the total thickness of the slab, the imaginary part of the matrix will be small compared to its real part.  In addition, in this limit, the real part becomes practically identical to the ground-state matrix, justifying our analysis in terms of the ground-state matrix for a collision at $E_{CM} = 0.1~\text{MeV}$ ($P/\hbar \sim 0.07 \, \text{fm}^{-1}$).  With the result of Eq.~\eqref{eq:rhonm} for nuclear matter, we can further expect in the central region of a ground-state slab:
\begin{align}
\text{Re} \, \left[ \rho_1(x,x') \right] & = \frac{1}{\pi (x - x')}
\sin{\left[ \frac{p_{F1} (x-x')}{\hbar} \right]}  \cos{\left[ \frac{P (x - x')}{\hbar}  \right]} \notag \\[.2ex] & \simeq \frac{1}{\pi (x - x')}
\sin{\left[ \frac{p_{F1} (x-x')}{\hbar} \right]} \, , \label{eq:Rerha} \\[.5ex]
\text{Im} \, \left[ \rho_1(x,x') \right] & = \frac{1}{\pi (x - x')}
\sin{\left[ \frac{p_{F1} (x-x')}{\hbar} \right]}  \sin{\left[ \frac{P (x - x')}{\hbar}  \right]} \notag \\[.2ex] & \simeq \frac{P}{\pi \hbar} \sin{\left[ \frac{p_{F1} (x-x')}{\hbar} \right]} \, . \label{eq:Imrha}
\end{align}
In general, we see in the central expressions of Eqs.~\eqref{eq:Rerha} and \eqref{eq:Imrha} that the off-diagonal oscillations in an initial density matrix, associated with its Fermi momentum $p_{F1}$, will generally compete with the oscillations associated with the initial momentum per nucleon $P$.  The~approximations on the r.h.s.\ of Eqs.\ \eqref{eq:Rerha} and \eqref{eq:Imrha} are in the limit of $P/\hbar \ll 2/\ell$.  The~expectations from the r.h.s.\ of \eqref{eq:Rerha} and \eqref{eq:Imrha} qualitatively agree with what is displayed for the initial slabs in Figs.~\ref{fig:2dfus} and~\ref{fig:offdenfus}, over the supports associated with the slab size.  Note that for the right-hand slab the initial momentum is negative, so its imaginary part is of the opposite sign, for a given location relative to the center of the support, compared to the matrix of the left-hand side slab.

As time progresses, the slabs at $0.1 \, \text{MeV}$ get into contact and fuse. The density matrix of the compound slab acquires a compact support and the single-particle orbitals now spread out over the extent of the $A=16$ system.  The spreading of the orbitals is reflected in the fact that, upon fusion at $t = 300$ and $800 \, \text{fm}/c$ in Fig.~\ref{fig:2dfus}, the net density matrix becomes square-like, as for the individual slabs, but with a larger spread.  Given the extremely low collision energy, the $A=16$ slab is, obviously, only moderately excited.  Accordingly, in~the~direction perpendicular to the $x=x'$ axis, the structure of oscillations in the density matrix becomes very similar to that for the initial ground state slabs, see Figs.~\ref{fig:2dfus} and~\ref{fig:offdenfus}. This is consistent with the fact that the oscillations are governed by the Fermi momentum $p_{F1}$ of normal matter and qualitatively described by Eq.~\eqref{eq:rhonm}.  In~fact, for the $A=16$ slab, additional oscillations, also controlled by $p_{F1}$, are seen in the direction perpendicular to $x=x'$, beyond those observed for the initial $A=8$ slabs.  The persistence of the imaginary part of the matrix, at late stages, may be understood in terms of collective motion with the matter flowing back and forth.  In fact, a finite net local flux $j(x)$ is conditioned on the imaginary part of the density matrix being finite next to the $x=x'$ axis:
\begin{align}
j(x) = \frac{1}{2 m \, i} \, \left(  \frac{\partial}{\partial x} - \frac{\partial}{\partial x'} \right)
\, \rho(x,x') \Big|_{x'=x} = \frac{1}{m} \, \frac{ \partial}{\partial x} \, \text{Im} \, \rho(x,x') \Big|_{x'=x}  \, .
\end{align}
When moving along the $x=x'$ axis, up to four regions are observed with alternating sign of the flux, see the bottom $t=300 \, \text{fm}/c$ panel in Fig.~\ref{fig:2dfus}.

\begin{figure}
  \includegraphics[width=.9\textwidth]{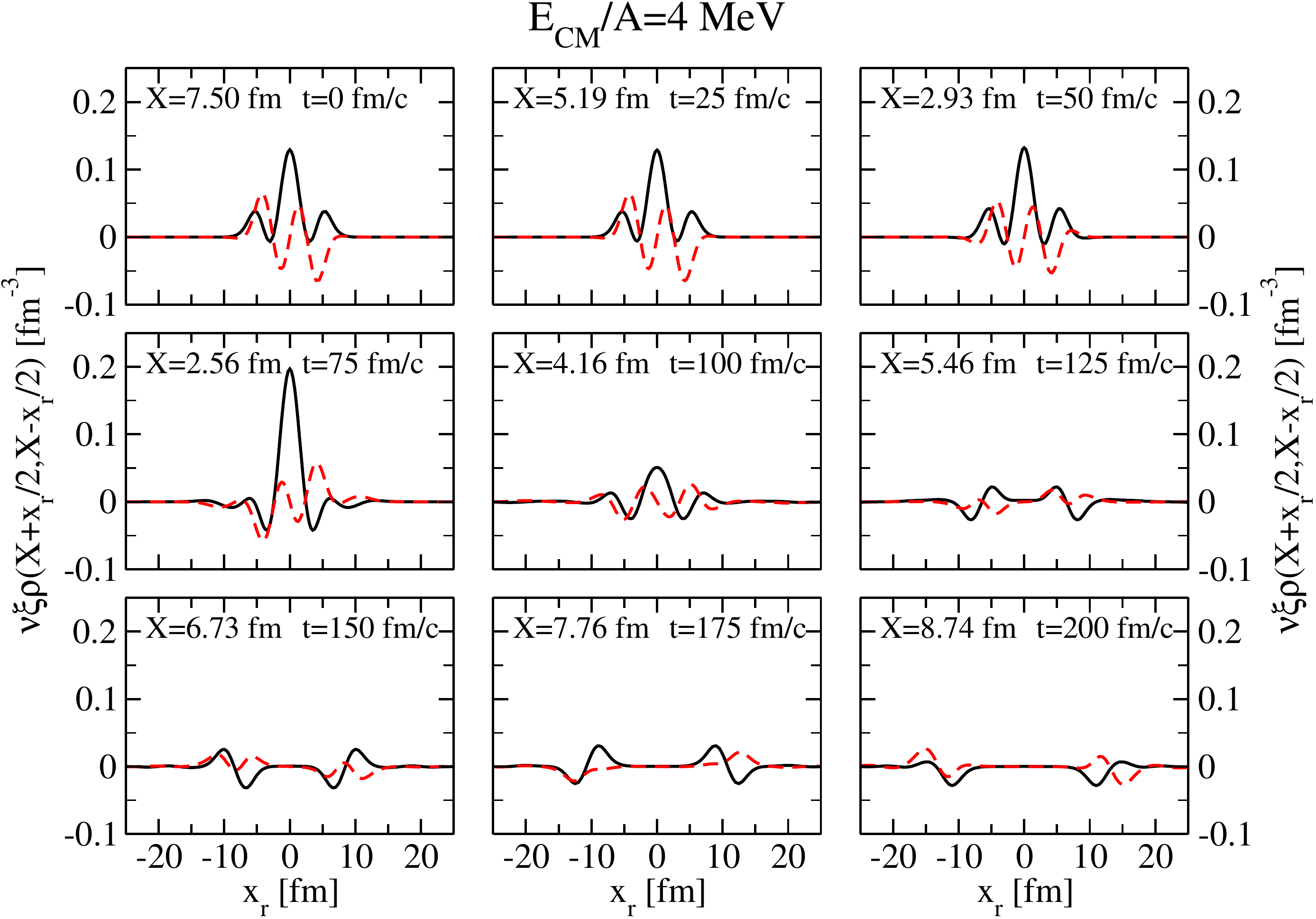}
  \caption{(Color online) Same as Fig.~\ref{fig:offdenfus} but a the collision at $E_{CM}/A=4$ MeV.}
  \label{fig:offdenfis}
\end{figure}

As the collision energy increases to $E_{CM}/A = 4 \, \text{MeV}$, some qualitative changes are observed both for the initial and later-stage density matrices in Figs.~\ref{fig:2dfis} and \ref{fig:offdenfis}, compared to the lowest collision energies.  To begin with, the initial momentum, at $P/\hbar = 0.44 \, \text{fm}^{-1}$, begins now to compete with the Fermi momentum, $p_{F1}/\hbar = 1.00 \, \text{fm}^{-1}$, in controlling the oscillations of the initial density matrix as a function of $(x-x')$, cf.~Eqs.~\eqref{eq:rho1P} and \eqref{eq:Rerha}.  In~fact, the modulation associated with $P$ basically eliminates the regions with negative values for the real part of the initial matrix in Figs.~\ref{fig:2dfis} and \ref{fig:offdenfis}.  In addition, the significant $P$-value enhances the imaginary values for the density matrix compared to the previous case, cf.~Eqs.~\eqref{eq:rho1P} and \eqref{eq:Imrha}.

At this energy, once the compound slab is formed, it remains in contact for a too short time for the single-particle orbitals to spread out evenly over the size of the compound system.  In consequence, the square-like support for the density matrix of the compound system, such as at low energy, fails to develop.  On the other hand, in contrast to the low energies, the system subsequently disintegrates into three fragments.  Each of these fragment separately achieves a final stationary state, with approximately square-like support around the diagonal, see right panels of Fig.~\ref{fig:2dfis}. In addition, cross-correlations develop between individual fragments, as signaled by the patches of significant values of the density matrix far away from the $x=x'$ diagonal. As an example, the structures at $x \sim 0$ and $x' \sim 15 \, \text{fm}$ for $t=200 \, \text{fm}/c$ in Fig.~\ref{fig:2dfis}, both for the real and imaginary parts of the matrix, signal the overlap of single-particle states between the central and the outgoing fragments.  Fainter correlation patches may be observed between the two fast fragments, see the region at $x \sim -15 \, \text{fm}$ and $x' \sim 15 \, \text{fm}$ there.  A~further insight into the features of the density matrix at the late stages of the collision, away from the $x=x'$ axis, is provided in Fig.~\ref{fig:offdenfis}, where regions of significant values far away from the axis are observed, particularly at times $t=(150$--$200) \, \text{fm}/c$.

The discussed correlation patches may be understood in terms of the fragmentation of single-particle orbitals, which we have already mentioned in the context of Fig.~\ref{fig:rhopl}.  During the breakup, the nucleons from the original orbitals have finite probabilities of ending up in different fragments.  The amplitudes for those possibilities maintain a phase relationship leading to correlations in the density matrix.  The entanglement of the internal wavefunctions is the sole reason for the persistence of the far-away off-diagonal patches in the density matrix.  The real and imaginary parts for those structures are comparable in magnitude, see Figs.~\ref{fig:2dfis} and \ref{fig:offdenfis}. This points to the involvement of significant relative phases, as expected from the significant difference in momentum and position.  Note that these entanglement correlations persist for fragments that are $30 \, \text{fm}$ apart for the $t=200 \, \text{fm}/c$ panels of Figs.~\ref{fig:2dfis} and \ref{fig:offdenfis}!

\begin{figure}
  \includegraphics[width=.9\textwidth]{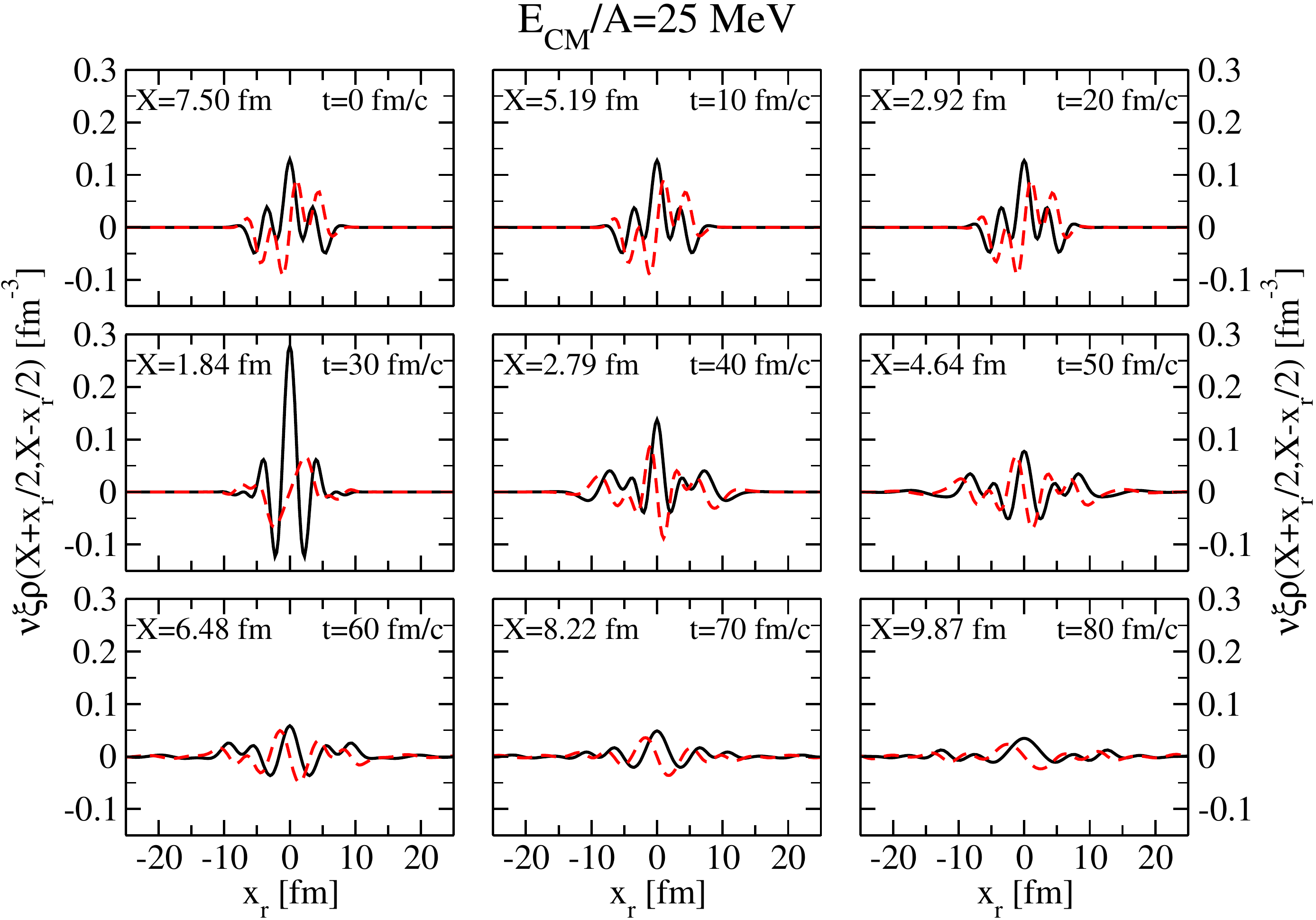}
  \caption{(Color online) Same as Fig.~\ref{fig:offdenfus} but for a collision at $E_{CM}/A=25$ MeV.}
  \label{fig:offdenmult}
\end{figure}

In the collision at $E_{CM}/A = 25 \, \text{MeV}$, the initial momentum, $P/\hbar = 1.10 \, \text{fm}^{-1}$, is already similar to the Fermi momentum.  With this, the collision can serve as an illustration of how the initial density matrix is controlled by two similar momenta, see Eq.~\eqref{eq:rho1P} and central expressions of Eqs.~\eqref{eq:Rerha} and \eqref{eq:Imrha}.  The initial momentum generally controls the relative magnitude of the real and imaginary parts of the matrix.  When the initial momentum is large, the imaginary and real parts become comparable, cf.\ Figs.~\ref{fig:2dmult} and~\ref{fig:offdenmult}.  In addition, when the initial momentum becomes comparable or exceeds the Fermi momentum, it begins to control the oscillations of the matrix in $(x - x')$.  In fact, in Figs.~\ref{fig:2dmult} and~\ref{fig:offdenmult}, the matrix values oscillate about twice as often as at low energies, in the direction perpendicular to the $x=x'$ axis, because of the initial and Fermi momentum being about the same.  On~the other hand, from the two momenta, it is the Fermi momentum that controls the extent of the peak of large magnitude of the matrix around the $x=x'$ axis, compare the left panels of Fig.~\ref{fig:2dmult} and~\ref{fig:2dfus}.  As time progresses within the $E_{CM}/A = 25 \, \text{MeV}$ collision, after the slabs come into contact, they again remain in this contact for a time that is too short for the square-like support to develop in the compound system.  With further progress of the reaction, qualitatively new developments take place at this energy, not only for the density, but also for the off-diagonal structure.  Thus, as the leading fragments recede and a low-density region emerges in-between those fragments, correlations develop all across the calculational region, involving regions over $30 \, \text{fm} \times 30 \, \text{fm}$ in size at $t = 80 \, \text{fm}/c$.  Clearly, the single-particle orbitals spread out quite substantially across the calculational region, with few spikes developing, such as those representing the leading fragments.  The spikes in the wavefunctions give rise to different structures within the off-diagonal elements of the density matrix, such as ridges.

Otherwise, for the multiply fragmented system, the general expectation is that Hubble-like correlations develop between the position and the average momentum per nucleon around that position.  Consistently with such an expectation, for an orbital expanding in the vacuum~\cite{Geza97}, the oscillations in the direction perpendicular to the diagonal of the density matrix are governed, at late times, by the factor
\begin{align}
\rho(x,x') \propto \exp \left( \frac{i \, m \, X \, x_r}{t \, \hbar} \right) \, ,
\label{eq:eXxr}
\end{align}
where $x_r = x - x'$ and $X= (x+x')/2$.  Comparing the above factor to that in Eq.~\eqref{eq:rho1P}, we see that momentum $P$ is replaced by $m X/t$. The expectation from \eqref{eq:eXxr} is that of a pattern of hyperbolas for the zeros of the real and imaginary parts of the density matrix, corresponding to constant values of the product $ X x_r = \frac{x^2 - x'^2}{2}$. More generally, the oscillations of the matrix will also follow a hyperbolic pattern.  Such a pattern indeed begins to emerge in the $t=80 \, \text{fm}/c$ panels of Fig.~\ref{fig:2dmult}, superimposed onto the structures associated with specific fragments.  The asymptotes for those hyperbolas are the two diagonals, $x=x'$ and $x=-x'$.  The late-time profiles of the matrix, exhibited in the last two panels in Fig.~\ref{fig:offdenmult}, are also consistent with the expectations based on Eq.~\eqref{eq:eXxr}.

To summarize, the significant values of the far off-diagonal elements of the density matrix are associated with phase coherence within far spread-out and fragmented orbitals.  Such a~phase relationship might be of importance if the fragmented slabs were ever to recombine.  However, at the late stages of a nuclear collision, the system tends to undergo expansion and progresses through a series of decays rather than through recombinations.  On the other hand, the phase relations may be important within individual emerging fragments, with nucleons moving back and forth between the fragment boundaries.  However, even in this case the correlations within the density matrix are characterized by a specific range, of $\sim 3.2 \, \text{fm}$, cf.\ Eq.~\eqref{eq:3.2}.  Monitoring of the far off-diagonal elements of a density matrix, substantially beyond this natural range, might be unacceptably costly within a true 3D system.  In~the following section, we examine the degree to which far-away elements can be disregarded, by suppressing such elements in 1D and observing the impact on the observables for the slab reactions.

\section{Suppression of far off-diagonal structure}
\label{sec:suppression}

As we have discussed, the significant values of far off-diagonal elements of a density matrix can be interpreted in terms of highly delocalized single-particle orbitals.  In some physical situations, {\em e.g.}~those that are relevant for nuclear reactions, such off-diagonal elements might not be important.  Consequently, one could expect to be able to disregard such elements without affecting core features of the reaction dynamics.  We are going to test this here in practice. There are two major potential benefits of suppressing off-diagonal elements.  The~first is that the suppression procedure can provide an understanding of the practical role of the far-away elements in the dynamics. The second is that the procedure can introduce significant savings at the numerical level for the anticipated effort in realistic calculations of reactions.  With the inclusion of correlations, and without any savings, a~NGF calculation in $D$ dimensions would actually require following a self-consistent evolution of $N_x^{2D} \times N_t^2$ values of the Green's functions.  Here, $N_x$ and $N_t$ represent the number of discretization steps in space and time, respectively.  The ability of reducing the number of matrix elements of the relative position, down to a nominal fraction within each dimension, is critical to carry through realistic NGF calculations.

In the correlated case, it would be additionally beneficial if it were possible to consider only short relative times, $t_r = t-t'$, in the functions, \emph{i.e.}~minimize the memory timescales.  In~fact, with inclusion of relative time in the consideration, the situation gets generally more complicated for the {\em spatial} arguments, than in the mean-field case with $t_r = 0$ only.  Thus, for $t_r \ne 0$, a~superposition of the products of orbitals, such as in Eq.~\eqref{eq:Gwf}, tends not to maximize exclusively around the relative position $x_r = (x - x') = 0$, but rather over a~trajectory within $(x_r, t_r)$ space, that ends up at $x_r =0$ when $t_r=0$.  Depending on how strong the effect of correlations is, the values for the superposition may persist as a function of $t_r$ along that trajectory.  Consistently, the cases of $t_r=0$ and $t_r \ne 0$ can be handled within a Wigner representation for the Wightman functions, which will be discussed in this paper in Section~\ref{sec:wigner}.  On the other hand, consideration of short memory times only~\cite{kohler99} may be warranted by physically short transition times through binary interaction regions, eliminating, in practice, the need to cope with the mentioned trajectories in relative space-time.

In a nuclear reaction, fragments separate from each other in configuration space. It is therefore natural to seek a suppression of the far-away elements in the configuration rather than any other space where the density matrix could be represented.  (Note that the density matrix would maximize around the diagonal in other representations as well, {\em e.g.} around the $p=p'$ momentum axis). As physics observables tend to be obtained from elements of the density matrix close to the diagonal, any satisfactory procedure of suppressing far-away elements needs to retain the near-diagonal region nearly intact.  Conservation laws, for one, rely on the near-diagonal region and can be used for testing of the suppression procedure.  Otherwise, the results near or on the axis may be directly compared to the dynamics with and without the suppression of the far-away elements.

Since only elements away from the axis need to be suppressed, $x \ne x'$, the suppression procedure needs to be in some way nonlocal.  The object on which the operation of suppression needs to be carried out, \emph{i.e.}\ the density matrix, is an operator itself rather than a wavefunction.  Operators that act on operators are termed superoperators within the quantum-mechanical context~\cite{frensley90}.  These have proven useful in the studies of decoherence and quantum transport phenomena~\cite{breuer}.   Thus, for example we can represent the equation of motion~\eqref{eq:denmf} for the density matrix as
\begin{align}
i \hbar \, \frac{\partial}{\partial t} \, \hat \rho = \hat{\mathcal L} \, \hat \rho \, ,
\label{eq:rhoL}
\end{align}
where $\hat{\mathcal L}$ is the Liouvillian superoperator,
\begin{align}
\hat{\mathcal L} = \hat{K}_1 + \hat{U}_1 - \hat{K}_{1'} - \hat{U}_{1'} \, .
\end{align}
To achieve our goal of suppressing the off-diagonal elements, we supplement the Liouvillian superoperator with an absorptive superpotential ${W}$:
\begin{align}
\hat{\mathcal L} \longrightarrow \hat{\mathcal L}_s = \hat{K}_1 + \hat{U}_1 + {W}(x_1, x_{1'};t) - \hat{K}_{1'} - \hat{U}_{1'} - {W}^*(x_{1'}, x_{1};t) \, .
\label{eq:Ls}
\end{align}
The dependence on both spatial arguments in $W$ is needed in order to suppress exclusively those elements that are away from the diagonal.  The particular combination of the superpotential and its complex conjugate is enforced by the requirement that the density matrix stay hermitian.
With the modified Liouvillian, the evolution of the density matrix over a~time step may be represented as
\begin{align}
\hat \rho(t+\Delta t) =  T^c\left\lbrace \text{e}^{-\frac{i}{\hbar}\int_t^{t+\Delta t} \text{d}t' \, \left[ \hat K_1 + \hat U_1 + \hat W \right]} \right\rbrace \, \hat \rho(t) \, T^a \left\lbrace
 \text{e}^{\frac{i}{\hbar}\int_t^{t+\Delta t} \text{d}t' \, \left[ \hat K_{1'} + \hat U_{1'} + \hat{W}^*  \right]} \right\rbrace \, .
\label{eq:SOMs}
\end{align}

Elementary considerations strictly restrict the possible forms of $ W$.  From Eq.~\eqref{eq:Ls}, meaningful $W$'s need to have real and imaginary parts that are, respectively, odd and even under the exchange of spatial arguments.  It is also natural to impose on $W$ the symmetries of the underlying system dynamics.  Thus, if the dynamics is translationally invariant, then $W$ should depend on the difference of arguments only, $(x_1 - x_{1'})$.  With invariance under inversion, the superpotential needs to be an even function of $(x_1 - x_{1'})$ as well. Consequently, the real part of the superpotential has to vanish.  To suppress off-diagonal elements, the~imaginary part of $ W$ needs to be nonpositive.  Finally, the conservation of particle number (or probability in other contexts) is enforced if the imaginary part vanishes along the $x_1=x_{1'}$ diagonal.  The simplest form of $\text{Im} \, W$, which would be a quadratic one, leads to a Gaussian suppression of the off-diagonal elements. That form has frequently appeared in the theory of decoherence~\cite{omnes02,vacchini05}.  Other forms of $ W$-type functions have been arrived at in studying a subsystem coupled to a chaotic environment~\cite{kusnezov99}.

With our goals in mind, we chose not to suppress elements close to the diagonal for the matrix and to suppress strongly those sufficiently far away.  We have further refrained from changing the suppression abruptly with location of the matrix element, to avoid any significant diffractive effects induced in the matrix.  We have adopted piecewise parabolic changes for the potential,  in-between the diagonal and the suppressed regions, through the parameterization
\begin{align}
 W(x_r) =
\left\{
\begin{array}{@{} c  @{\, , \quad} l @{}}
  0 & 0 \leq |x_r| \leq x_0 \, , \\
  -i  W_0 \, {2 } \big[ |x_r| - x_0 \big]^2/{d_0^2}  & x_0 \leq  |x_r|  \leq x_0 + {d_0}/{2} \, , \\[.3ex]
  -i W_0 \Big( 1 - {2} \big[ |x_r| - (x_0 + d_0) \big]^2/{d_0^2} \Big) & x_0 + {d_0}/{2} \leq  |x_r|  \leq x_0+ d_0 \, , \\[.3ex]
  -i W_0 &  x_0  + d_0 \leq |x_r| \leq L \, .
\end{array}
\right.
\label{eq:cutoff}
\end{align}
In employing FFT within the SOM for our system, we had to make our system periodic.  To retain the periodicity for the evolved matrix elements, we further needed to enforce periodicity on the superpotential.  Thus, with the periodicity of our system being~$2L$, the~superpotential has also been built, for $|x_r| > L$, to satisfy $ W (x_r + 2L) =  W ( x_r)$.

\begin{figure}
  \includegraphics[height=.5\textwidth]{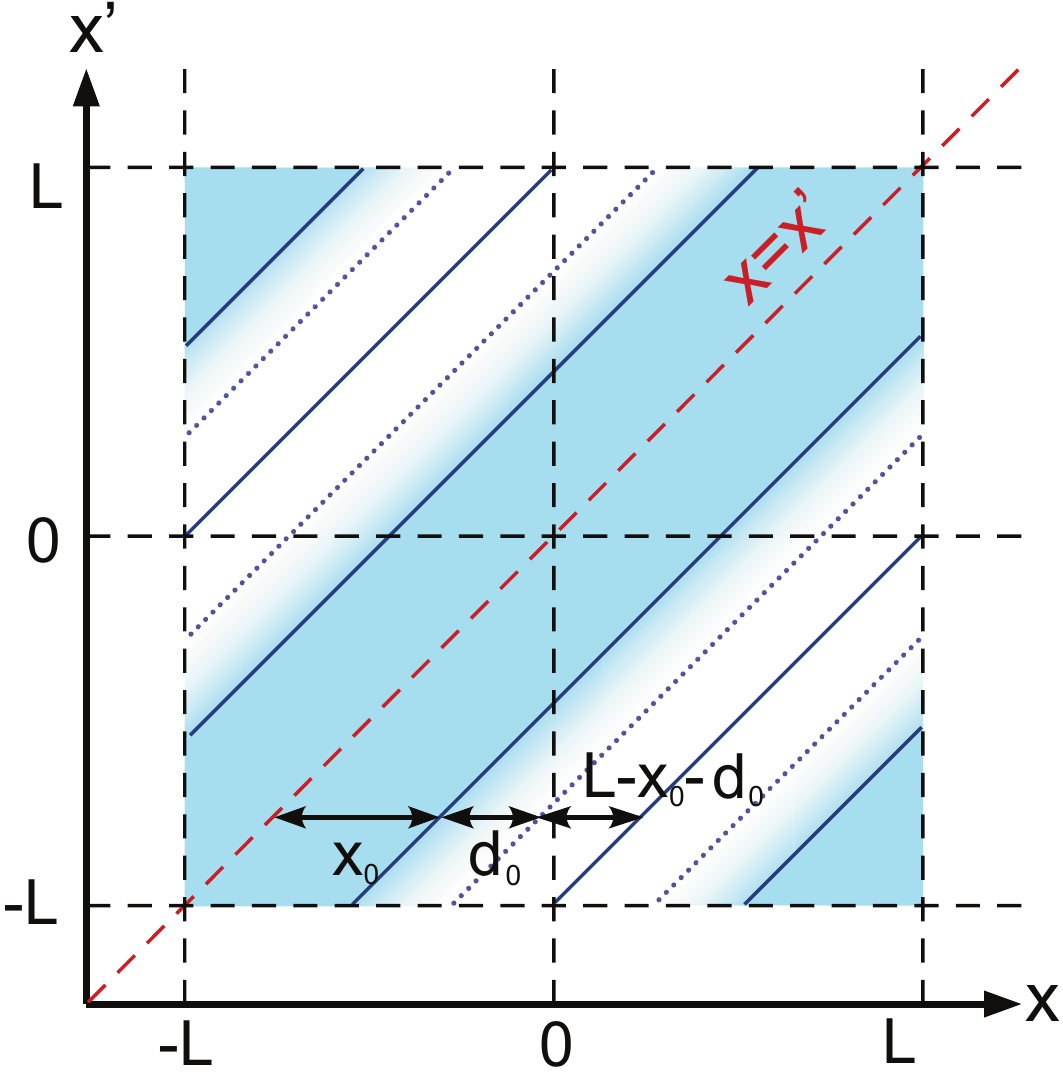}
  \caption{(Color online)
  Structure of the superpotential $W$, and of the corresponding suppression of matrix elements, within the $(x,x')$ plane.  In a band around the diagonal, $|x-x'| \le x_0$, darkened in the figure, the superpotential vanishes and the elements are not suppressed.  At separations $x_0 + d_0 \le |x-x'| \le L$, shown as white, the superpotential is strong and the matrix elements are strongly suppressed. The superpotential and suppression change gradually in-between.  The~periodicity of the system gives rise to repeated bands of directly unaffected and suppressed elements in the corners of the computational space.}
  \label{fig:cutfield}
\end{figure}

The superpotential and the associated suppression pattern of the off-diagonal elements are illustrated in Fig.~\ref{fig:cutfield}.  Note that the periodicity enforces the lack of suppression for the elements next to the corners at $(-L,L)$ and $(L,-L)$.  If elements were suppressed there, fragments moving towards the edges of the calculational region would encounter stronger suppression of off-diagonal elements than fragments in the middle of the region.  Effectively, the fragments would be interacting with the edges, leading to a break of translational invariance.

The three parameters $W_0$, $x_0$ and $d_0$ control the strength and shape of the superpotential. Specifically, $x_0$ regulates the size of the band around the $x=x'$ axis where matrix elements stay intact.  The fraction of elements that get suppressed within our $(x,x')$ plane is $\chi = 1 - x_0/L$.  The steepness in the change of $W$ with position is regulated by $d_0$.  A too small value of $d_0$ may induce spurious numerical oscillations in the density matrix, beyond those already inherent in the matrix.  Finally, $W_0$ regulates the pace at which the elements get suppressed.  Over a time interval of $\Delta t$, the suppression factor for elements away from the diagonal, at $|x_r| > x_0 + d_0$, is $\text{e}^{-2 W_0 \Delta t/\hbar}$.  Unless otherwise specified, we employ a rather large, on nuclear scale, strength $W_0 = 1000 \, \text{MeV}$ and a moderate $d_0 = 2 \, \text{fm}$.  We have confirmed that, within a considerable range of values of $W_0$ and $d_0$, the results of the calculations are fairly similar.

In the reminder of this Section, we shall attempt to quantify the practical importance of the far-away off-diagonal elements of the density matrix within the mean-field evolution of 1D slabs.  To this end, we shall reexamine the three collisions discussed before, representing fusion, break-up and fragmentation.  We shall follow the system evolution, with a~progressively narrower range of retained matrix elements, \emph{i.e.}\ smaller $x_0$, while focussing on the local density and other characteristics of the system associated with the region close to the $x=x'$ diagonal.  Our basic interest is in how far we may be able to push, if at all, the~element suppression without significantly affecting the region close to the diagonal.

Before we proceed, it may be useful to cast the differential equation we solve,
\begin{align}
i \hbar \, \frac{\partial}{\partial t} \, \hat \rho = \left( \hat{\mathcal L} + 2 i \, \text{Im} \, \hat W \right) \, \hat \rho \, ,
\label{eq:EOMLW}
\end{align}
in one more integral form, for the sake of understanding the nature of the forthcoming solutions:
\begin{align}
\rho(x, x'; t) =  - \frac{i}{ \hbar} \int_{t_1}^t \text{d}t' \, \text{e}^{\frac{2}{\hbar} \,  \text{Im} \, W(x,x') \, (t-t')} \, \left( \hat{\mathcal L} \hat{\rho} \right) (x,x';t')
 + \text{e}^{\frac{2}{\hbar} \,  \text{Im} \, W(x,x') \, (t-t_1)} \, \rho (x, x'; t_1) \, .
\label{eq:rhoWrho}
\end{align}
In the above, we have made use of our conclusions on $W$ to separate formally the time evolution driven by $ \hat{\mathcal L}$ from that associated to $\textrm{Im} W$.  The time $t_1$ precedes $t$.  When we do not suppress any elements, or else in the region where $W$ happens to vanish, the exponential factors containing $W$ become equal to~1.  Otherwise, Eq.~\eqref{eq:rhoWrho} demonstrates that large values of~$W$ act to quickly eliminate the off-diagonal elements of $\rho$.  On the other hand, if $\rho$ was not going to have significant values in the first place, where $W$ was acting, the presence of~$W$ would not change the situation much for the density matrix.  Finally, for any $\rho$ and at least for finite times, in the regions that are away from the suppression, the density matrix is going to evolve in a manner consistent with the absence of any suppression, \emph{i.e.} following the evolution driven by $ \hat{\mathcal L}$.

\begin{figure}
  \includegraphics[height=.4\textheight]{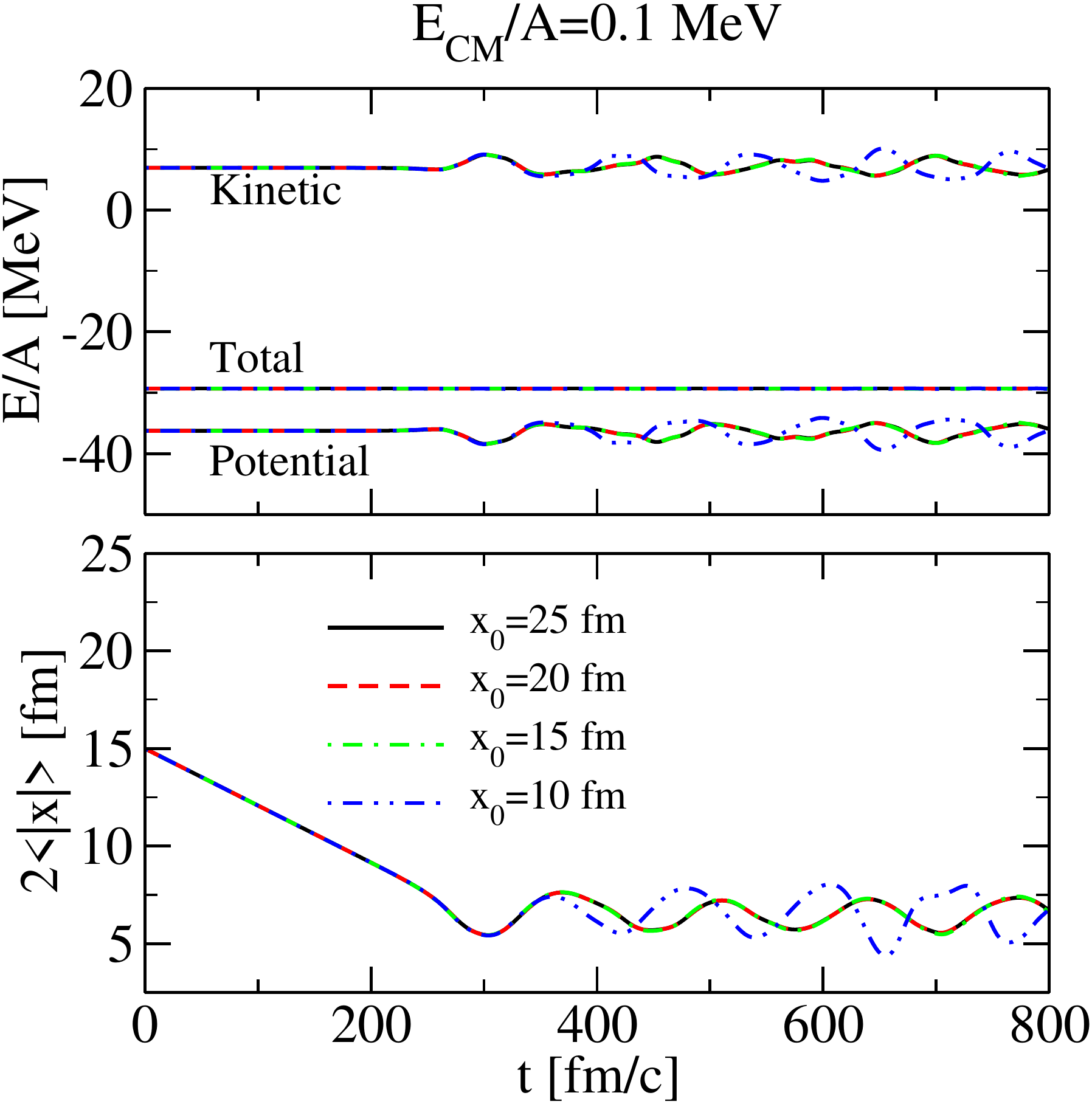}
  \caption{(Color online)
  Time evolution of the energy per particle (upper panel) and of the system extent (lower panel) in a collision at ${E_{CM}}/{A}=0.1 \, \text{MeV}$ for different values of the cutoff parameter $x_0$ associated to the suppression of off-diagonal elements in the density matrix.}
  \label{fig:enerfrag_fus}
\end{figure}

Let us start the practical investigations of the suppression with the low-energy fusion-reaction at ${E_{CM}}/{A}=0.1 \, \text{MeV}$ and consider global quantities for that reaction.  The top panel of Fig.~\ref{fig:enerfrag_fus} shows the net energy of the system as well as the different contributions to this energy, as a function of time, for different values of the cutoff~$x_0$.  The~value of $x_0 = L = 25 \, \text{fm}$ represents the standard evolution, without any suppression.  As is apparent in~Fig.~\ref{fig:enerfrag_fus}, the net energy remains very well conserved throughout the evolution, even for the relatively low cutoff value of $x_0 =10 \, \text{fm}$.  Moreover, the contributions to the energy do not appear to depend on the cutoff for $x_0 \ge 15 \, \text{fm}$.  As to the $x_0 = 10 \, \text{fm}$ case, the~kinetic and potential energies evolve in the same manner as for the original evolution up to $t \sim 370 \, \text{fm}/c$.  Thereafter, though, differences start to emerge.  The bottom panel of~Fig.~\ref{fig:enerfrag_fus} shows the evolution of the extent of the system for different~$x_0$.  As with the energy breakdown, the size appears independent of the cutoff, when $x_0 \ge 15 \, \text{fm}$.  In the $x_0 = 10 \, \text{fm}$ case, like it has been discussed for the energy, the size evolves in the same manner as without any element suppression, until $t \sim 370 \, \text{fm}/c$ and only then differences emerge.  In the lower panel, we also note that the differences begin to emerge only as the system begins to recontract following its initial expansion.  In essence, the excessive, for this energy, suppression of matrix elements reduces the period of collective oscillations for the system.  As~time progresses, the~accumulated difference in phase of the collective oscillations between the original and the $x_0 =10 \, \text{fm}$ evolutions leads to a beating for the difference in size between the two evolutions.

\begin{figure}
  \includegraphics[width=.8\textwidth]{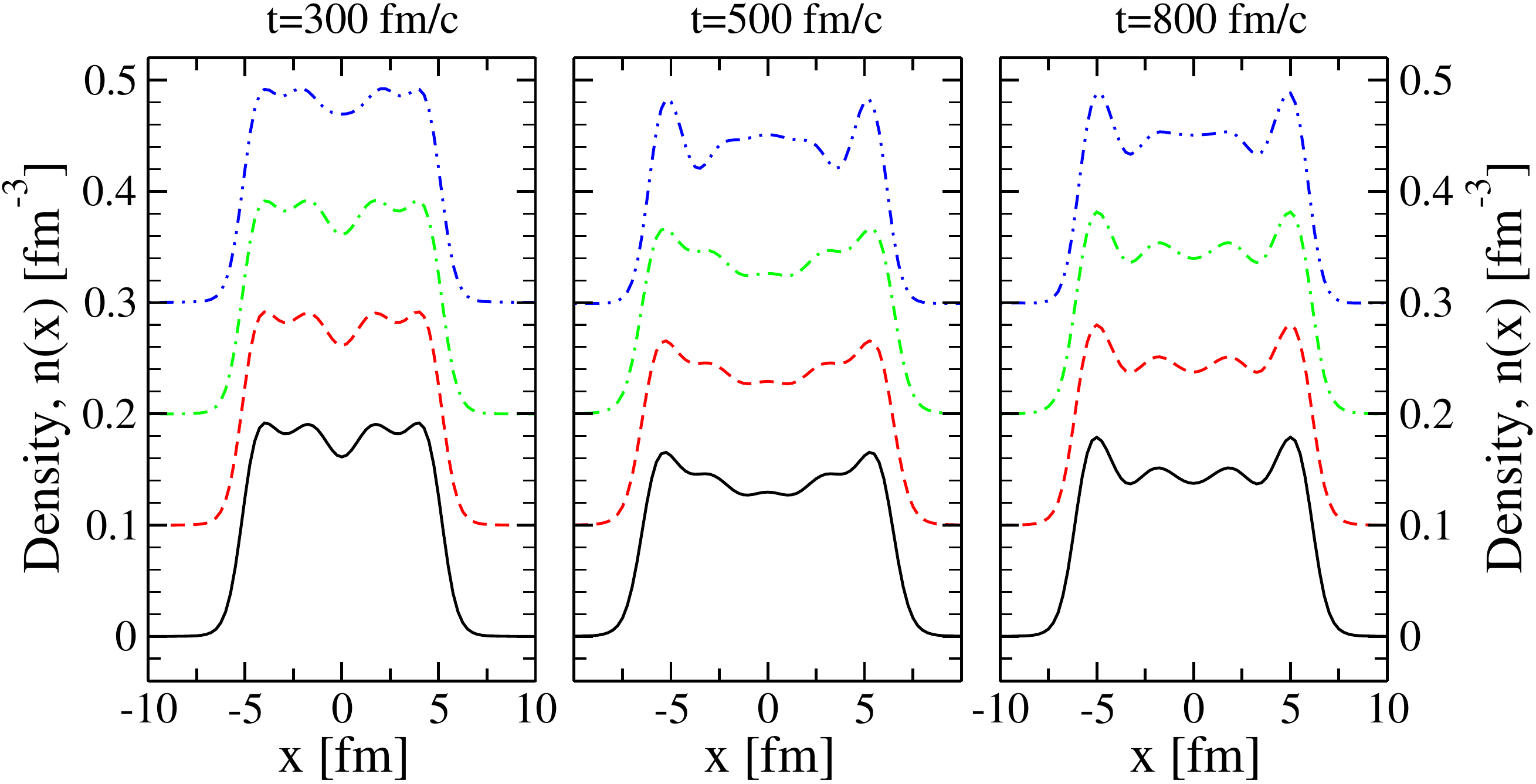}
  \caption{(Color online)
  Density profiles in the $E_{CM}/A = 0.1 \, \text{MeV}$ collision, at sample times, for different values of the cutoff parameter~$x_0$. The profiles are staggered by $0.1 \, \text{fm}^{-3}$ to provide better insight into details.  From bottom to top, the lines are for the standard evolution, and then $x_0 = 20, 15$ and $10 \, \text{fm}$, respectively.
  }
  \label{fig:dencut_fus}
\end{figure}

Figure~\ref{fig:dencut_fus} next shows the density in the ${E_{CM}}/{A}=0.1 \, \text{MeV}$ collision, at sample times, for different values of the cutoff parameter~$x_0$.  In this more differential representation of the collision, again no differences can be observed between the standard evolution and the evolutions with $x_0 \ge 15 \, \text{fm}$.  For $x_0 = 10 \, \text{fm}$, the central dip in the density is slightly filled at $t=300 \, \text{fm}/c$, compared to the standard evolution.  The difference at $500 \, \text{fm}/c$, between the $x_0 = 10 \, \text{fm}$ and the standard evolution, may be attributed to the different phase in collective oscillation.  At $t=800 \, \text{fm}/c$, the densities for those two evolutions are similar as the oscillations within those evolutions progress through the analogous intermediate state, cf.~Fig.~\ref{fig:enerfrag_fus}.

Overall, we find that we can suppress up to about 50\% of all matrix elements for our calculational region, cf.\ also Fig.~\ref{fig:2dfus}, without affecting any essential features of the fusion reaction.  When we suppress more, the system appears to be affected quantitatively at the later stages of its evolution.  However, the system still evolves through similar stages as without element suppression, it just arrives at those stages at modified times.  Interestingly, the conservation of net energy survives to a surprising degree, even when quite aggressive suppressions of the off-diagonal elements are introduced into the evolution.

To an extent, we may find the above results quite understandable.  If we look at Fig.~\ref{fig:2dfus}, we observe that significant matrix elements extend from the diagonal only by $|x-x'|$ about equal to the thickness of an $A=16$ system, or $\sim 13\, \text{fm}$ after Eq.~\eqref{eq:lA} (cf.~also~Fig.~\ref{fig:enerfrag_fus}, accounting that the thickness of a uniform system is equal to $4 \langle |x| \rangle$).  Thus, if we choose a large cutoff $x_0 \ge 15 \, \text{fm}$, we never suppress any significant values of matrix elements.  A~more abrupt suppression might give rise to stronger kinetic energy effects, even for small suppressed elements, but we have prevented that through the graduation of the suppression, \emph{i.e.} through a finite $d_0$.  In the case of $x_0 = 10 \, \text{fm}$, we do not suppress any significant elements before the slab make contact either, cf.~Fig.~\ref{fig:2dfus}. It is only after the slabs fuse and the orbitals manage to spread themselves over the extent of the compound slab, by $t \sim 300 \, \text{fm}/c$, that the $x_0 = 10 \, \text{fm}$ suppression really begins to affect the matrix.  However, phase correlations should only start to matter upon the return of the waves incident on the opposing boundary of the compound slab, hence the delay, until the system recontracts, for significant changes in the global features to emerge.

\begin{figure}
  \includegraphics[width=.8\textwidth]{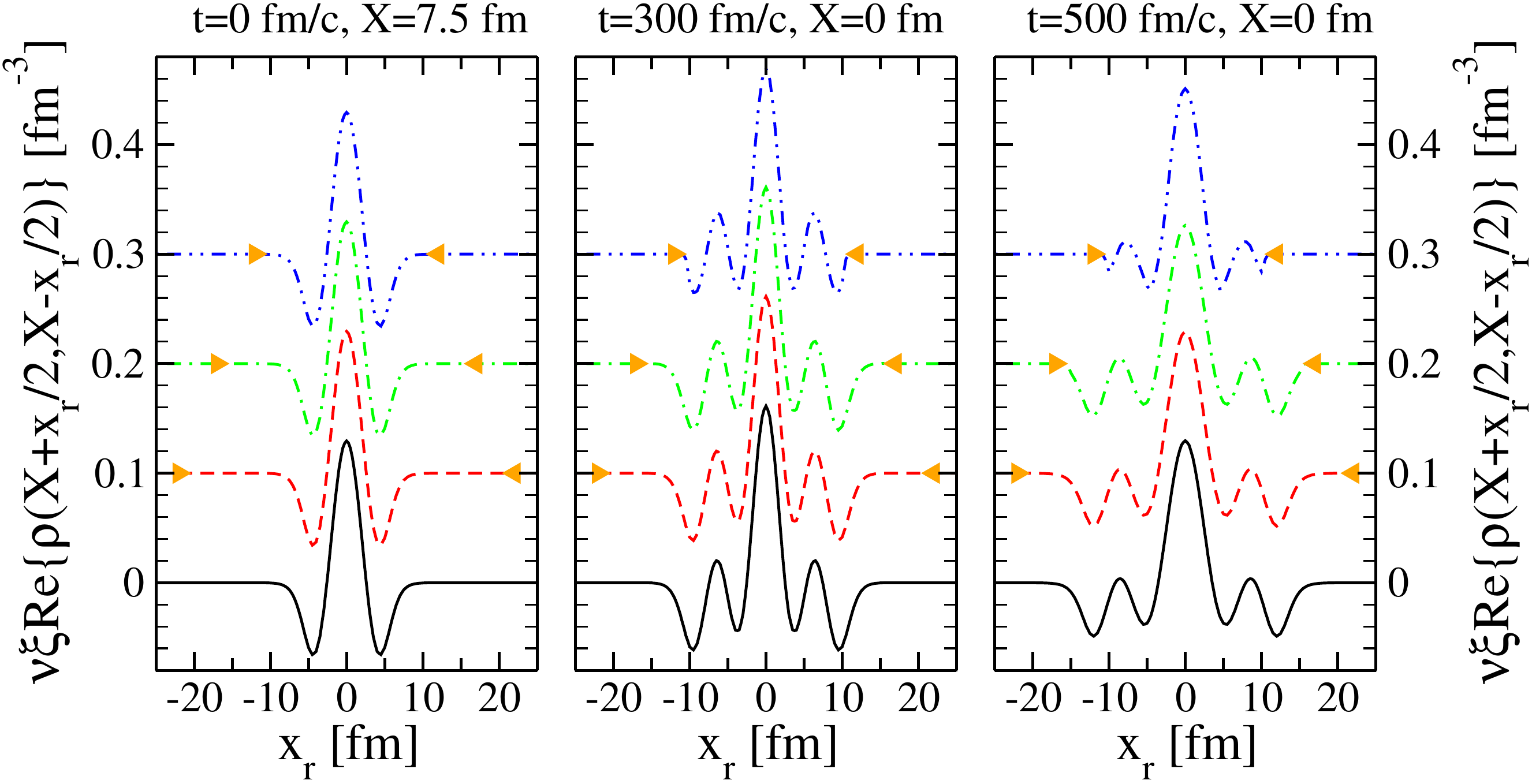}
  \caption{(Color online)
  Real part of the scaled density matrix $\xi \nu \, \rho(x,x';t)$ for the slab collision at $E_{CM}/A = 0.1 \, \text{MeV}$, along a line $x + x' = 2X = const$, perpendicular to the diagonal $x=x'$ in the $(x,x')$ plane, for different values of the cutoff parameter~$x_0$. The results are shown against the coordinate difference $x-x' = x_r$, at sample values of $X$ and $t$.  The results for different $x_0$ are staggered by $0.1 \, \text{fm}^{-3}$ to provide better insight into details.  From bottom to top, the results are for the standard evolution, and then $x_0 = 20, 15$ and $10 \, \text{fm}$, respectively.  The~locations where element suppression sets in, $|x_r| \ge x_0$, are marked with triangles. }
  \label{fig:denoff_cut_fus}
\end{figure}

Figure~\ref{fig:denoff_cut_fus} provides additional insight into the off-diagonal structure of the matrix in the context of element suppression.  The figure displays values of the matrix along lines perpendicular to the $x=x'$ diagonal, at different $x_0$.  For the display, we~have chosen the initial time $t=0$ and the cut at $X=(x+x')/2 = 7.5 \, \text{fm}$, across the center of the right slab, and, further, two later times and a cut across the system center at $X=0$.  With the $t=0$ density matrix extending up to about $|x_r| \sim 7 \, \text{fm}$, neither of the suppressions in the figure affects that matrix.  At $t= 300 \, \text{fm}/c$, the matrix has spread out in transverse directions up to $|x_r| \sim 10 \, \text{fm}$.  The $x_0 = 10 \, \text{fm}$ cut starts barely to affect the matrix, while the $x_0 \ge 15 \, \text{fm}$ cuts leave the matrix intact.  At $t = 500 \, \text{fm}/c$, the undisturbed matrix is spread out a bit farther than at $t=300 \, \text{fm}/c$.  The $x_0 > 15 \, \text{fm}$ cuts still leave the matrix intact.  On the other hand, the effect of matrix suppression for $x_0 = 10 \, \text{fm}$ has by now propagated from $|x_r| \sim 10 \, \text{fm}$ down to the vicinity of the axis $x_r \sim 0$, affecting in particular the density. Remarkably, though, the overall structure of the matrix at $|x_r| \lesssim x_0$ remains pretty unchanged by that last significant cut.

\begin{figure}
  \includegraphics[height=.4\textheight]{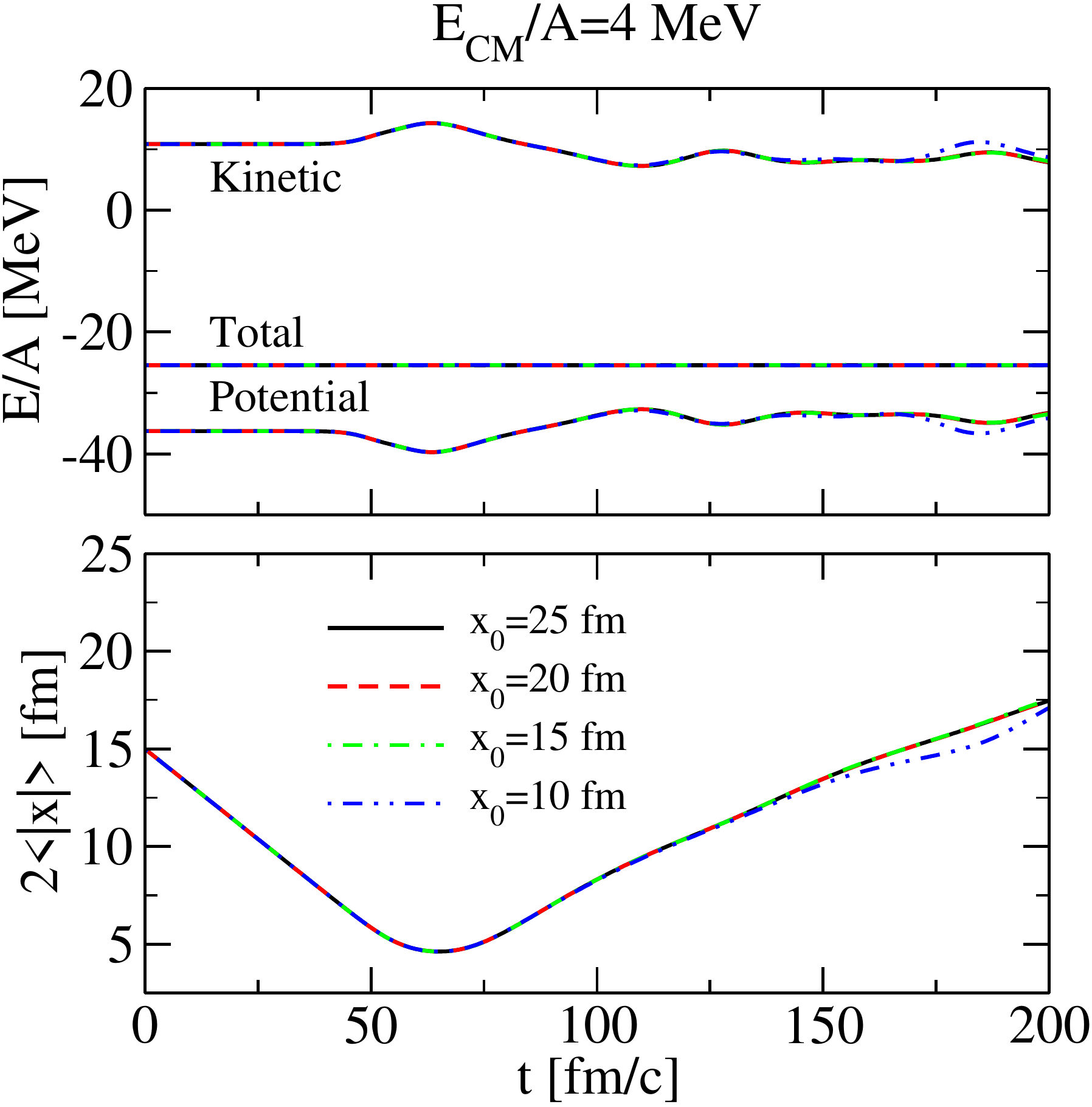}
  \caption{(Color online)
  The same as Fig.~\ref{fig:enerfrag_fus} but for a collision at $E_{CM}/A = 4 \, \text{MeV}$.}
  \label{fig:enerfrag_fis}
\end{figure}

In the preceding Section we have seen that the support of the density matrix spreads out over the whole calculational $(x,x')$ area at higher energies, rather than staying compact as in the low-energy fusion-reactions.  Given our low-energy experience in this Section, we might then expect that even modest suppressions of the elements in the higher-energy reactions could produce noticeable changes late in the system evolution compared to the standard evolution.  To~test this, Fig.~\ref{fig:enerfrag_fis} shows the evolution of the energy components and of the system extent for the reaction at $E_{CM}/A = 4 \, \text{MeV}$, when utilizing different cutoff parameters~$x_0$ in the suppression of matrix elements.  Consistently with our earlier experience, we may note an excellent energy conservation in the presence of element suppression.  However, we may also note that the energy breakdown and the system extent turn out to be fairly independent of the cuts applied to the elements, contradicting the naive expectations.  Only in the case of $x_0 = 10 \, \text{fm}$, we see small $\lesssim 5 \%$ changes in the energy breakdown and system extent in the late $t \gtrsim 150 \, \text{fm}/c$ stages of the reaction, compared to the unperturbed evolution.  We expected that, with the increase in collision energies, the global features of the reactions would be more sensitive to cuts. Instead, Fig.~\ref{fig:enerfrag_fis} shows that these features are far less sensitive to the off-diagonal suppression.

\begin{figure}
  \includegraphics[width=.8\textwidth]{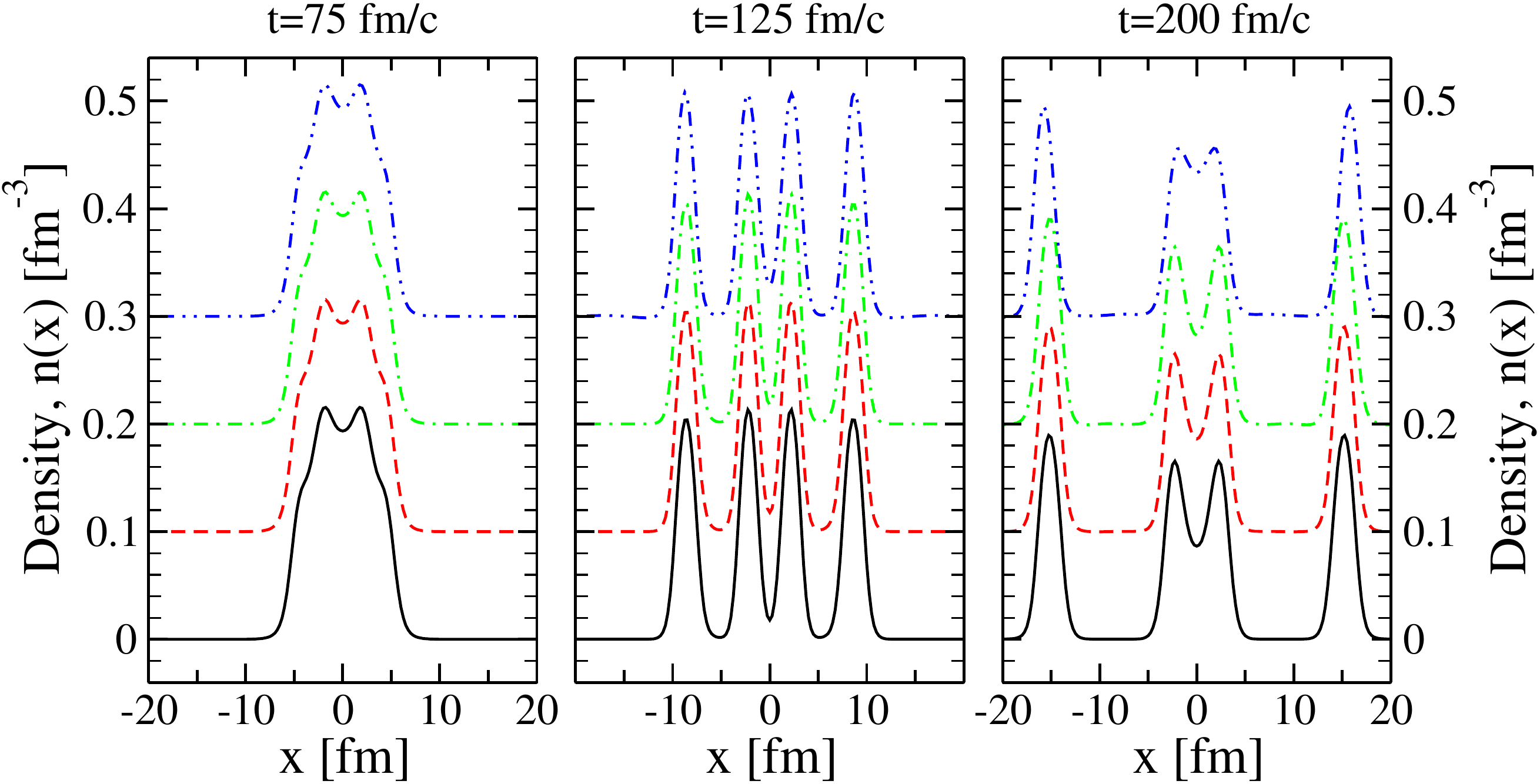}
  \caption{(Color online)
  The same as Fig.~\ref{fig:dencut_fus} but for a collision at $E_{CM}/A=4 \, \text{MeV}$.}
  \label{fig:dencut_fis}
\end{figure}

Figure~\ref{fig:dencut_fis} shows the system density in the $E_{CM}/A = 4 \, \text{MeV}$ reaction, at sample times, for different cutoff parameters~$x_0$.  As has been discussed in Sec.~\ref{sec:offdiag}, at this particular collision energy the system breaks into three fragments, one central with mass $A \sim 8$, and two moving forward and backward in the~CM.  We~can observe in Fig.~\ref{fig:dencut_fis} that the cuts with $x_0 \ge 15 \, \text{fm}$ have no practical effect on the evolution of the density.  The $x_0 = 10 \, \text{fm}$ does not modify the evolution throughout the formation and breakup of the compound slab.  Only at  later stages of the reaction ($t>150 \, \text{fm}/c$), slight differences emerge compared to the standard evolution.  Thus, at $t= 200 \, \text{fm}/c$, the leading fragments are ahead by $\sim 1 \, \text{fm}$ compared to the standard evolution.  This indicates that those fragments emerge at slightly higher velocity from the central region, for $x_0 = 10 \, \text{fm}$.  On the other hand, the central fragment at $t= 200 \, \text{fm}/c$, for $x_0 = 10 \, \text{fm}$, appears shifted in its phase of oscillation compared to the standard evolution.  This seems similar to the situation with the fused slab at $E_{CM}/A = 0.1 \, \text{MeV}$. One might, however, wonder how an $x_0 = 10 \, \text{fm}$ cut can affect an $A \simeq 8$ slab, of much smaller size.  After all, such a cut has no noticeable effect on the ground-state initial $A=8$ slabs.  The~essential issue here is that the central $A \simeq 8$ fragment is actually reconstituted from a highly expanded structure, cf.\ central panel in Fig.~\ref{fig:dencut_fis}, that manages to expand perpendicular to the $x=x'$ axis before recontraction, and it is therefore sensitive to the suppression in the off-diagonal direction.

Additional insight into the element suppression and its consequences at $E_{CM}/A = 4 \, \text{MeV}$ is provided by Fig.~\ref{fig:dencut_fis}, analogous to Fig.~\ref{fig:dencut_fus} for this collision.  At the earliest of the times represented in Fig.~\ref{fig:dencut_fis}, $t=75 \, \text{fm}/c$, the single-particle orbitals and, correspondingly, the~density matrix in the transverse direction are relatively compact.  Only the $x_0 = 10 \, \text{fm}$ cut affects the matrix in a significant manner.  However, the effect of the suppression, if any, would take a while to propagate to the vicinity of the $x=x'$ axis.  The panel for $t=125 \, \text{fm}/c$ shows a cut along the line $X=(x+x')/2 = 5.46 \, \text{fm}$, traversing the patches within the density matrix that represent a~correlation between the central fragment and the leading fragment moving to the right.  The superpotentials with $x_0 \ge 15 \, \text{fm}$ have had no practical effect on those patches, but the potential with $x_0=10 \, \text{fm}$ begins to absorb them.  The panel for $t=200 \, \text{fm}/c$, representing the cut at $X=8.74 \, \text{fm}$, illustrates the ensuing development for those patches.  The $x_0 = 20 \, \text{fm}$ superpotential has had, by that time, still no impact on the patches, while the $x_0 = 15 \, \text{fm}$ superpotential begins to absorb them.  Finally, the $x_0 = 10 \, \text{fm}$ superpotential has, at this point, completely absorbed the off-diagonal patches.  The suppression of the patches, however, appears to be of no consequence for the development of the essential aspects of the system, cf. Fig.~\ref{fig:enerfrag_fis}.  This can be understood in the following terms. When the system expands, the orbitals fragment into individual pieces which take off with different velocities, forming fragments that move away from each other.  With the fragments separating, the correlation patches in the density matrix move away from the $x=x'$ diagonal.  If those patches never move back to the vicinity of the axis, where they could contribute to the observables (momentum will be addressed later in the paper), it becomes irrelevant whether they are allowed to propagate forever or get suppressed.  If the region far-away from the diagonal becomes the region of 'no-return', it ceases to matter whether the dynamics in this region is followed or not, in order to find the essential observables.

\begin{figure}
  \includegraphics[height=.4\textheight]{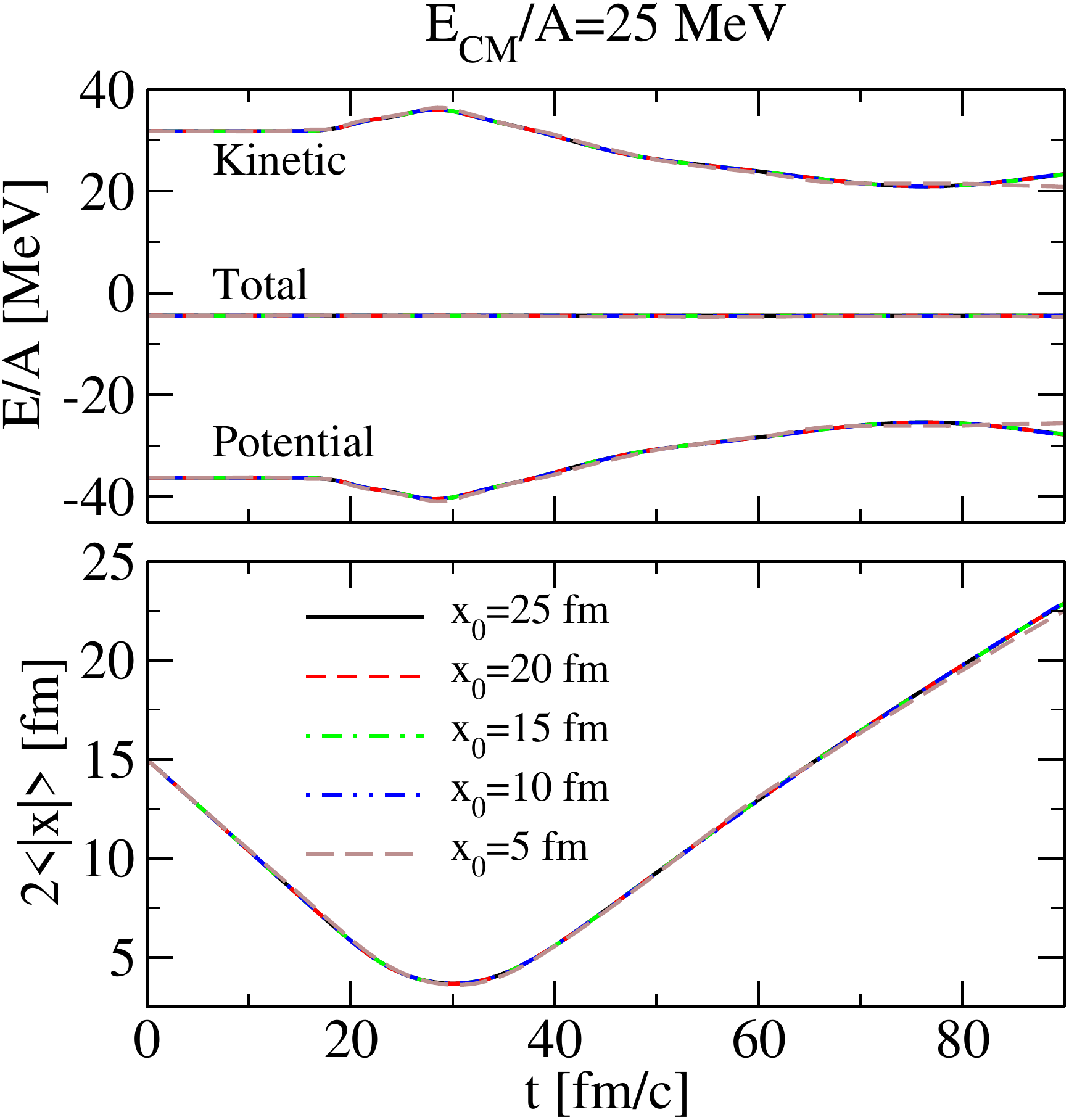}
  \caption{(Color online)
  The same as Fig.~\ref{fig:enerfrag_fus} but for a collision at $E_{CM}/A=25 \, \text{MeV}$.}
  \label{fig:enerfrag_mult}
\end{figure}

A similar situation is found at even higher energies when we examine the impact of element suppression in the fragmentation reaction at $E_{CM}/A = 25 \, \text{MeV}$.  For this particular reaction, besides the results analogous to those for lower-energy reactions, we shall also show the consequences of the suppression with $x_0 = 5 \, \text{fm}$, that directly affects 80\% of the matrix elements within our calculational area.  This specific case will give, in particular, an~opportunity to discuss some practical issues involved in the element suppression.  Figure~\ref{fig:enerfrag_mult} shows first the evolution of the energy components and of the system extent for the different suppression cuts.  No dependence on the parameter $x_0$ can be seen for either quantity if $x_0 \ge 10 \, \text{fm}$.  For $x_0 = 5 \, \text{fm}$, only a slight deviation from the standard evolution may be seen within the energy breakdown at $t > 80 \, \text{fm}/c$ and virtually no deviation is found within the system extent.

\begin{figure}
  \includegraphics[width=.8\textwidth]{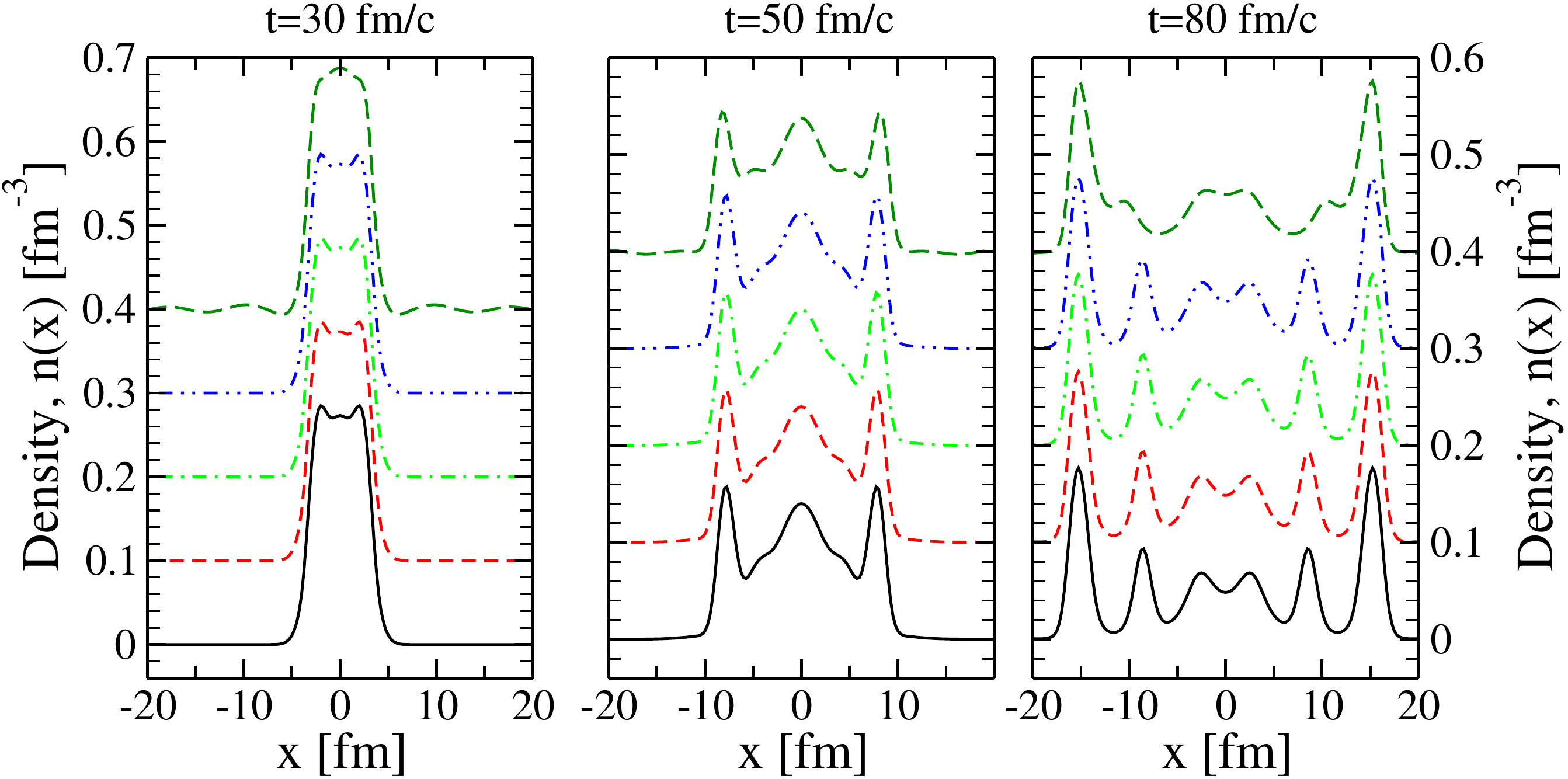}
  \caption{(Color online)
  The same as Fig.~\ref{fig:dencut_fus} but for a collision at $E_{CM}/A=25 \, \text{MeV}$. Note the difference in vertical scales for the left panel and the two remaining panels.}
  \label{fig:dencut_mult}
\end{figure}

Figure~\ref{fig:dencut_mult}, with sample densities, provides more insight into the role of element suppression at~$E_{CM}/A = 25 \, \text{MeV}$.  It is apparent in the figure that superpotentials with $x_0 \ge 10 \, \text{fm}$ have little practical effect on the evolution of density.  The likely reason is that, at this energy and within the followed evolution, the system rapidly breaks up into small fragments and no stable structures form, that might be extended in the $(x-x')$ direction.  On the other hand, it is evident in~Fig.~\ref{fig:dencut_mult}, that the superpotential with $x_0 = 5 \, \text{fm}$ affects the evolution of density in a nontrivial manner.  At the late stage ($t=80 \, \text{fm}/c$ in Fig.~\ref{fig:dencut_mult}), the $x_0 = 5 \, \text{fm}$ system appears to fragment into 3 rather than 5 fragments as in the evolution without any suppression.  Some differences in the evolutions, precursing this outcome, can be already seen in the central panel of Fig.~\ref{fig:dencut_mult}, for $t=50 \, \text{fm}/c$.  Around maximal compression at $t = 30 \, \text{fm}$, the central dip, present in the standard density, is filled for the density with the~$x_0 = 5 \, \text{fm}$ suppression.  A new aspect of the changes in the density with the strong element suppression may also be seen in the $t = 30 \, \text{fm}/c$, and, to some extent, in the $t=50 \, \text{fm}/c$ panel.  For $x_0 = 5 \, \text{fm}$, small-amplitude oscillations extend away from the central fused slab and take the density to negative values.  These illustrate the fact the density matrix, a positive-definite operator, generally loses this property after element suppression.  Indeed, while unitarity is preserved, \emph{i.e.}~the integral along the diagonal does not change, the individual values of the density are modified and there is no restriction on their values. We may encounter negative density values for less drastic suppressions as well, but these are usually quantitatively negligible. Generally, the appearance of those values will depend on the shape of the element suppression.  Note that for drastic suppressions, such as $x_0 = 5 \, \text{fm}$ discussed here, the low-density oscillations may become visible, even though they still make negligible contributions to the global quantities such as the system extent or breakdown of energy into different components.  When evaluating potential energy, we~have, so far, adopted a convention that we replace the argument of the energy by zero when the density is negative.  In~relation to the above, we have also found that the population of single-particle states differs from the mean-field values once the off-diagonal suppression is introduced.  In place of the usual $N_\alpha=1$ and $0$ values, fractional occupations emerge, with a distribution that could be interpreted in terms of a temperature.  The emergence of fractional occupations can be expected, since the element suppression effectively decouples the portions of single-particle wavefunctions that have moved out with different fragments.  However, for very strong suppressions, such as $x_0 = 5 \, \text{fm}$ here, we can also encounter small negative occupations $N_\alpha$ as well as occupations slightly exceeding~$1$.

Figure~\ref{fig:denoff_cut_mult} provides next an insight into the impact of the cuts on the real part of the density matrix within the $(x,x')$ plane in the $E_{CM}/A = 25 \, \text{MeV}$ reaction at $t=80 \, \text{fm}/c$.  The~matrix without any element suppression is similarly represented in the right panel of Fig.~\ref{fig:2dmult}.  While the superpotentials with $x_0 \ge 10 \, \text{fm}$ remove matrix elements quite effectively there where they are applied away from the diagonal, they leave the vicinity of the $x=x'$ axis practically unchanged.  Even the $x_0 = 5 \, \text{fm}$ superpotential leaves the vicinity of the axis fairly similar to that of the standard evolution.  Furthermore, Fig.~\ref{fig:denoff_cut_mult} shows values of the real part of the density matrix, along the lines perpendicular to the $x=x'$ diagonal, at sample times in the $E_{CM}/A = 25 \, \text{MeV}$ reaction, for different suppression parameters~$x_0$.  As~for the previously investigated reactions, the cuts play a more prominent role later, rather than earlier in the reaction.  While the $x_0 \ge 10 \, \text{fm}$ cuts suppress the density matrix at the later stages of the reaction, they leave, as seen here again, the immediate vicinity of the $x=x'$ axis unchanged. On the other hand, the $x_0 = 5 \, \text{fm}$ cut eventually induces changes in the structure of the matrix in the immediate vicinity of the axis, specifically at $t=80 \, \text{fm}/c$. Note that the different frequency of oscillations may be interpreted in terms of a different momentum per nucleon for a given piece of matter at a specific location.

Overall, we have seen that extensive suppressions of far-away elements of the density matrix may be carried out without affecting central features of the reaction dynamics.  The~degree to which the suppressions may be applied depends on the energy of the reaction and on the size of stable structures that are formed in the dynamical evolution.  The higher the energy and the smaller the objects, the more compact the region around the diagonal of the density matrix that needs to be retained.  While entanglement of internal wavefunctions between fragments emerging from a reaction leads to correlation patches that move away from the diagonal (potentially up to infinity for infinite times), one does not need to follow the evolution of these patches if the fragments are never to meet again. As an additional test for the importance of the correlations, besides what has been discussed, we have followed the development of reactions allowing the fragments to leave the periodic computational interval.  The fragment that leaves the computational interval enters the next one and/or enters the original interval back from the other side, depending on the interpretation.  With this, we allowed the correlated fragments to collide.  Within the $(x,x')$ plane, the correlation patch may be seen as traveling to the diagonal within the next computational box and/or as reentering the original box from another side.  On~the diagonal, the patches encounter the parent density pulses undergoing a collision.  We~have carried those calculations with and without element suppression.  In such calculations, any level of element suppression eliminates the patches that travel from one diagonal to another.  Contrary to naive expectations, though, we did not find any significant differences in the evolution of those overextended collisions, whether or not we had suppressed the elements and the correlation patches traveling across the diagonals.  Quite likely, the relative insensitivity to the interference in the collisions of the emerging fragments can be associated with the large relative momenta between fragments.  There is, though, one aspect of the calculations that we found to be highly sensitive to any suppression of the structures in far-away elements. This is the time-reversibility of the dynamics, and we shall discuss that aspect of the dynamics in the following section.

\section{Time-reversibility of the dynamics}
\label{sec:rev}

Generally, the  time evolution of the density matrix in the mean-field approximation obeys time-reversal symmetry in the microscopic sense \cite{ring}, \emph{i.e.}\ the evolution of a system, following the mean-field dynamics, is describable in terms of the mean-field dynamics when ran backwards in time.  Within the symmetry, if we evolved a system past a collision, following the mean-field dynamics, we should be able to reverse the direction of evolution and reach the initial state.  We have, like others \cite{ring}, exploited that symmetry to test our numerical code, and found the symmetry to be very well satisfied at the numerical level.  Erasing of elements in the density matrix is, though, a generally irreversible process.  However, the degree to which this irreversibility is important clearly depends on the circumstances.  Thus, if the erased elements are small (or even zero) to start with, the erasing should presumably not be very important.  Further, we have seen that even removal of significant elements may not affect much a system under expansion.  In the following, we examine in practice the level of irreversibility induced by suppressing the elements of the density matrix, taking as an example the case of the parameter $x_0=10 \, \text{fm}$.  This particular suppression has left the central features of the $E_{CM} = 25 \, \text{MeV}$ reaction unchanged and has affected the lower energy reactions at a semi-quantitative but not qualitative level.

Handling of the imaginary superpotential in the time-reversed case requires some elementary care.  At the formal level, the change of the direction of time in the equation of motion, Eq.~\eqref{eq:EOMLW}, may be accomplished by taking a complex conjugate of the sides of that equation.  For the conjugate equation, the forward evolution in time will be progressing through stages in reverse order to those in the original forward evolution.  In the reversed evolution, the states will be described in terms of the complex conjugate matrix.  In taking the complex conjugate of the equation, the superpotential will be conjugated as well and, as a purely imaginary object, it will change its sign.  With this change in sign, the superpotential will lead to an artificial increase in the magnitude of matrix elements found beyond the cutoff, rather than to a decrease.  However, since we are aiming at the reduction of the calculational space, we are interested in erasing the information no matter in which direction the evolution progresses.  Consequently, we need to change the sign of the superpotential by hand when reversing the time evolution.  Note, besides, that a strong potential which increases off-diagonal elements arbitrarily would give rise to a numerically unstable evolution.

\begin{figure}
  \includegraphics[width=.5\textwidth]{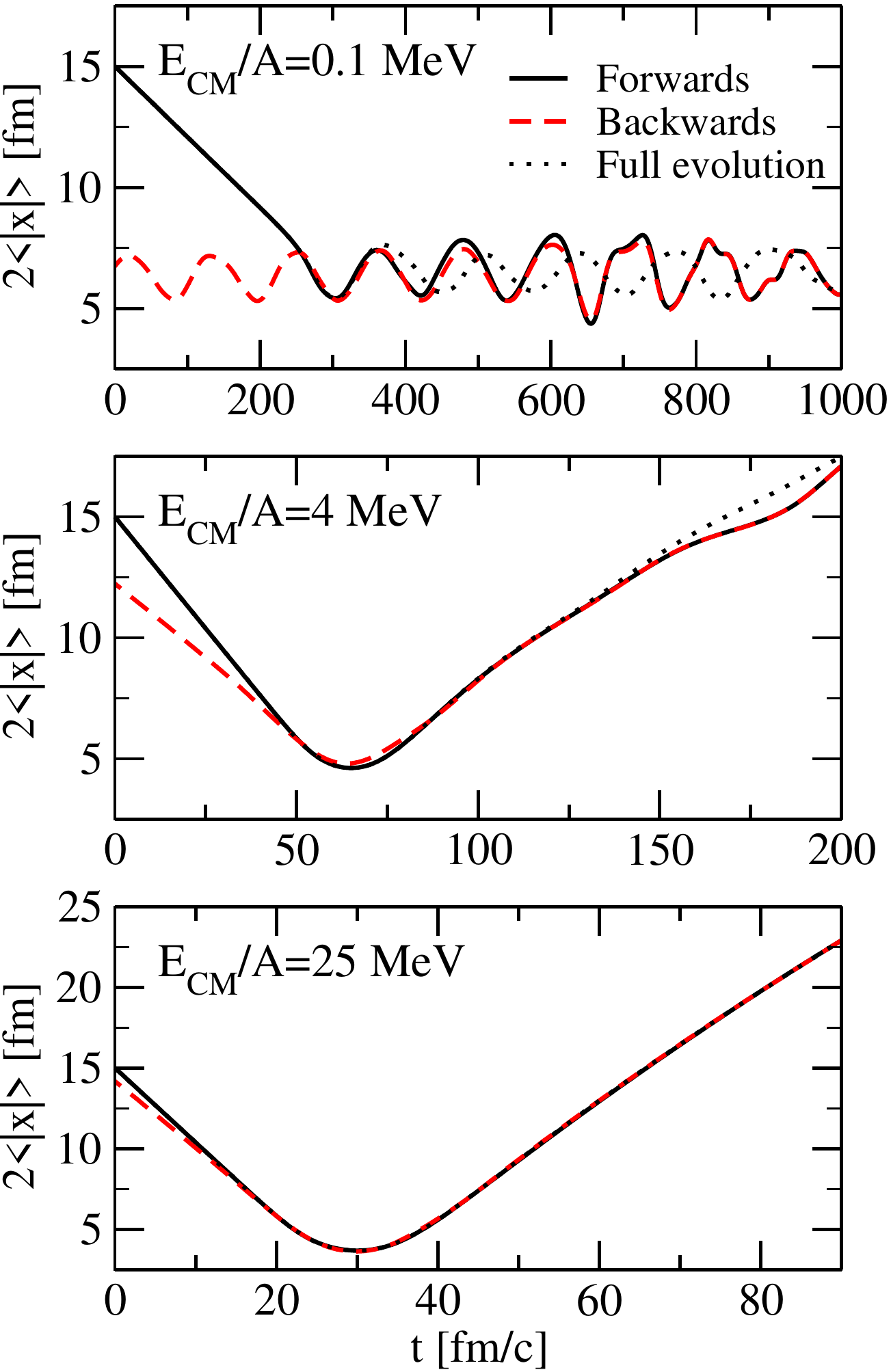}
  \caption{(Color online)
  Evolution of the system extent for the collision of two $A=8$ slabs at three energies.  The standard, fully reversible evolution is represented by dotted lines.  The forward evolution, with the $x_0 = 10 \, \text{fm}$ suppression, is represented by solid lines.  The backward evolution, with the suppression, is represented by dashed lines. In each case, the reversal of evolution is applied at the end of the displayed time interval.}
  \label{fig:timerev}
\end{figure}

As a characterization of the system development under time reversal, we show in Fig.~\ref{fig:timerev} the evolution of the system extent for the three collision energies considered before.  Without element suppression, the extent for a reversed evolution retraces the extent for the system evolving in the forward direction.  In Fig.~\ref{fig:timerev}, the forward and backward evolution without element suppression are accounted for by one line.  For the forward evolution with element suppression, at the two lower energies, we can recognize in Fig.~\ref{fig:timerev} the late-time deviations from the standard evolution that were already discussed in the previous section.  At the highest of the represented energies, no late-time deviations are actually visible.  For each of the collision energies, the reversal of the evolution is applied at the end of the displayed evolutions: at $t=1000$, $200$ and $90 \, \text{fm}/c$, respectively, at the energies of $E_{CM} = 0.1$, $4$ and $25 \, \text{MeV}$.  The backward evolution initially follows very well the forward evolution with suppression, especially well at the two higher energies.  Gradually, however, deviations build up, and the system extents evolved in the two ways, with and without suppression, separate at all energies and for every case for times prior to the initial compression.  This is indicative of the fact that the system, at either of the studied energies, does not evolve back into the initial state.  A particularly dramatic situation develops at the lowest of the energies, where the system evolved backward in time stays compact and oscillates in size, rather than separating into the two original fragments, for which system extent would increase with decreasing time.  At the two other energies, we can observe a reduced slope in the dependence of system extent around the initial time.  That indicates a reduced speed and/or a reduced mass for the fragments, compared to the original initial state, both implying a~certain level of internal excitation.  Clearly, with element suppression, the time reversal symmetry is broken at the level of global system characteristics.  The degree of violation of time reversal symmetry, due to element suppression at fixed~$x_0$, depends on the collision energy.  This is similar to the situation with the impact of element suppression on the forward time evolution.

\begin{figure}
  \includegraphics[width=.8\textwidth]{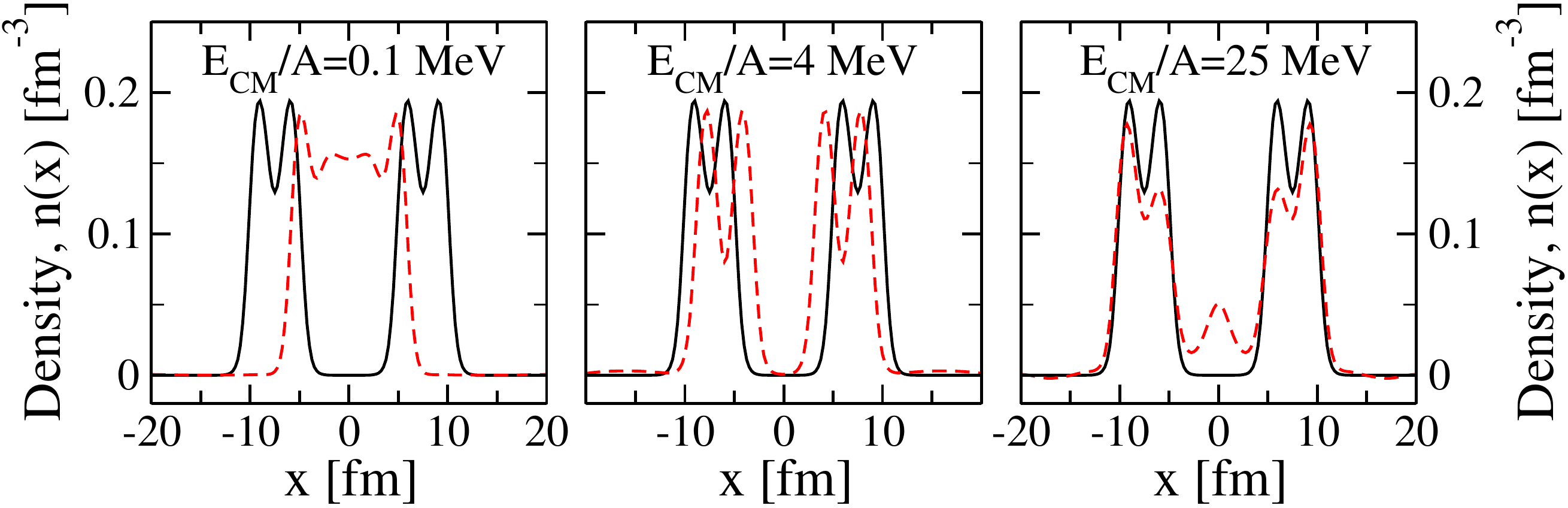}
  \caption{(Color online)
  Comparison of the initial density (solid lines) to the density obtained at $t=0$ when first evolving the system forward and then backward in time, with element suppression in the density matrix, for a collision of two $A=8$ slabs at three energies.}
  \label{fig:timerevden}
\end{figure}

Further insight into the time-reversal violation induced by off-diagonal element suppression is provided by Fig.~\ref{fig:timerevden}, that compares the initial system density to the density at $t=0$ obtained from evolving the system first forward and then backward in time, at the three collision energies.  For the system evolved forward and backward at the lowest energy of $E_{CM} = 0.1 \, \text{MeV}$, with element suppression, we can directly see in Fig.~\ref{fig:timerevden} a fused system at $t=0$, that has remained fused and oscillating since the initial contact.  At $E_{CM} = 4 \, \text{MeV}$, with element suppression, the system evolves back into two highly excited fragments around $t=0$, that are slowed down compared to the original initial state.  The level of excitation is evidenced by the depth of the dips in density for these fragments in the central panel of Fig.~\ref{fig:timerevden}.  Finally, at $E_{CM} = 25 \, \text{MeV}$ the system evolves back into two excited fragments with a low-density neck formed in-between.

At each energy, the system that is evolved back to $t=0$, with suppressed elements in the density matrix, acquires some features typical for the final states at the specific energy, be that fusion, fragment excitation, or formation of a neck. This indicates that the correlation patches within the density matrix that may be far from the $x=x'$ axis late in the system forward evolution, and that move back towards the axis when the time is reversed, are essential in the restoration of the density matrix of the coherent initial state.  When these off-diagonal elements are erased, the initial state coherence is lost and it is not possible to recover a well separated, two-slab system. Note, also, that for the forward time direction, the entanglements are of far less importance because coherence is less of an issue at late stages of nuclear reactions.

One aspect of the collisions that has not been addressed yet is their momentum content.  In~the~next Section, we remedy this by analyzing a mixed, configuration and momentum space, representation of the density matrix. The discussion will also bring in another interpretation for the erasing of the far off-diagonal elements within the spatial representation.

\section{Wigner distribution}
\label{sec:wigner}

As has been discussed in the Introduction, high-energy central reactions are commonly described in the literature in terms of the semiclassical BE. The dynamic quantities in BE-type of approaches are the so-called Wigner functions, distributions of particles jointly in the configuration and momentum space. Wigner distributions have been also employed in the literature in the context of TDHF, in particular to gain guidance in supplementing TDHF with a collision term~\cite{grange81,tang81,kohler80,kohler84,kohler79}.  Within nuclear structure, the Wigner functions represented a departure point for semiclassical approximations~\cite{ring}.  Elsewhere, distributions in the mixed configuration and momentum representations have been found to provide interesting insights into aspects of quantal dynamics~\cite{hillery84}. For 1D systems, as will be evident, the Wigner functions are particularly straightforward to visualize.

The Wigner function $f_W$ results from Fourier transforming the density matrix in the relative coordinate:
\begin{align}
f_W(x,p) = \int \text{d} x_r \, \text{e}^{- \frac{i p x_r}{\hbar}} \, \rho( x + x_r/2, x - x_r/2) \, .
\label{eq:wigt}
\end{align}
The transformation yields a real function $f_W$ that, in the classical limit, can be interpreted as a phase-space distribution~\cite{hillery84}.  This function is, however, generally not positive definite. In~quantum mechanics, it only acquires that property after $f_W$ is averaged by folding it with a sufficiently wide Gaussian in position and momentum. Consistently with its interpretation, projecting the Wigner function onto the $x$-axis yields, up to a factor, the 1D spatial density:
\begin{align}
\int  \frac{\text{d}p}{2 \pi \hbar}  \, f_W(x,p) = \rho( x, x) = \frac{1}{\nu} \, n_1(x) \, .
\end{align}
Similarly, projecting onto the $p$-axis yields, up to a factor, the density in momentum space.

Under proper conditions~\cite{danielewicz84a}, which include meeting in different ways the semiclassical limit, the KB equations yield a BE for $f_W$.  The collision term in BE is associated with the correlation contributions to the KB equations, cf.~the l.h.s.~of Eqs.~\eqref{eq:kb1} and \eqref{eq:kb1p}. This collision term, in particular, includes the effects of statistics.  Aside from being highly exploited in central high-energy nuclear reactions~\cite{bonasera94}, BE with this type of a collision term has been beneficial in other fields~\cite{haug,frensley90}.

When the semiclassical limit is applied on top of the mean-field dynamics, the Vlasov equation for $f_W$ emerges from the equation of motion \eqref{eq:denmf} for the density matrix.  This represents a~reduced form of BE with effects of collisions suppressed. Specifically, given the peaking of the density matrix in the relative coordinate around zero, we can attempt to expand the difference of mean-field potentials on the r.h.s.\ of Eq.~\eqref{eq:denmf}:
\begin{align}
U(x_1) - U(x_{1'}) \approx (x_1 - x_{1'}) \, \left. \frac{\text{d}U}{\text{d} x} \right|_{x = \frac{x_1 + x_{1'}}{2}} \, .
\label{eq:Uexp}
\end{align}
Combining this approximation with a Fourier transformation in the relative coordinate of both sides of Eq.~\eqref{eq:denmf}, yields the Vlasov equation for $f_W$:
\begin{align}
\frac{\partial f_W}{ \partial t} + \frac{p}{m} \, \frac{\partial f_W}{\partial x}
- \frac{\partial U}{\partial x} \, \frac{\partial f_W}{\partial x} = 0 \, .
\label{eq:Vlasov}
\end{align}
The Vlasov aspect of the equation is in the self-consistent evaluation of $U$ from the single-particle density. The equation for the evolution of $f_W$ under the influence of an {\em external} potential $U$ is of an identical form.

Following Eqs.~\eqref{eq:rhonm} and \eqref{eq:3.2}, the width of the main peak around the diagonal of the density matrix for our ground-state slabs is about $\pi \hbar/p_{F1} \approx 3.2 \, \text{fm}$, see also Fig.~\ref{fig:offdenfus}.  For a~general potential~$U$, the validity of Eq.~\eqref{eq:Vlasov} requires a slow variation of the potential over that specific distance.  As the momentum scale is compared with the scale in position and $\hbar$ is involved, this condition is semiclassical in nature.  Notably, for a $U$ that is largely linear in density at low to moderate densities, cf.~Eq.~\eqref{eq:meanf}, the condition of slow variation will not be satisfied for our slabs, given their shape, see~Fig.~\ref{fig:ad_den}.  On the other hand, Eq.~\eqref{eq:Uexp} may yield coarsely correct results.  Otherwise, since the approximation of a linear expansion of $U$ at best addresses the immediate vicinity of the diagonal in the density matrix, the issues of any structures moving away from the diagonal are left out in the Vlasov description.  Note, however, that in the case where $U$ is quadratic in position, Eq.~\eqref{eq:Uexp} is exact.  For such a~potential, it does not matter whether there are any off-diagonal structures present or not.  In particular, the Wigner function for the HO initial state of the adiabatic evolution will represent a static solution to Eq.~\eqref{eq:Vlasov}.

In spite of these limitations, the Vlasov equation \eqref{eq:Vlasov} is still useful as a reference when analyzing collisions in terms of Wigner functions.  The equation can be written in different ways and, in particular, can be cast into the form of a single-particle Liouville equation.  Upon defining the single-particle energy as
\begin{align}
\epsilon (x,p;t) = \frac{p^2}{2 m} + U(x,t) \, ,
\label{eq:epsilon}
\end{align}
the Vlasov equation may be rewritten as
\begin{align}
\frac{\partial f_W}{ \partial t} + \frac{\partial \epsilon}{\partial p} \, \frac{\partial f_W}{\partial x}
- \frac{\partial \epsilon}{\partial x} \, \frac{\partial f_W}{\partial p} & =
\frac{\partial f_W}{ \partial t} + \frac{\partial}{\partial x} \left(
\frac{\partial \epsilon}{\partial p} \, f_W\right)
+ \frac{\partial}{\partial p} \left(-
\frac{\partial \epsilon}{\partial x} \, f_W\right)
\notag \\
& \equiv \frac{\partial f_W}{ \partial t} +  i \hat{\mathcal L} \, f_W = 0 \, .
\label{eq:phLiou}
\end{align}
A combination of the terms on the l.h.s.\ of Eq.~\eqref{eq:phLiou} may be recognized as the Poisson bracket of the Wigner function with the single-particle energy.  Upon rewriting of the l.h.s.\ of the Vlasov equation after the first equality in \eqref{eq:phLiou}, the Vlasov equation acquires the form of a continuity equation in phase space, with
$\left( \frac{\partial \epsilon}{\partial p} , -\frac{\partial \epsilon}{\partial x}  \right)$ representing the local phase-space velocity-vector and $\left( \frac{\partial \epsilon}{\partial p} \, f_W, -\frac{\partial \epsilon}{\partial x} \, f_W \right)$ representing the phase-space flux.  The semiclassical limit of the Liouville operator in phase space is then given by
\begin{align}
i \hat{\mathcal L} = \frac{\partial}{\partial x} \left(
\frac{\partial \epsilon}{\partial p} \, \cdot \right)
- \frac{\partial}{\partial p} \left(
\frac{\partial \epsilon}{\partial x} \, \cdot \right) \, .
\end{align}

If the equations for the phase-trajectories $(x(t),p(t))$ are introduced,
\begin{align}
\frac{\text{d} x}{\text{d} t} = \frac{\partial \epsilon}{\partial p}
\hspace*{2em} \text{and} \hspace*{2em}
\frac{\text{d} p}{\text{d} t} = - \frac{\partial \epsilon}{\partial x} \, ,
\label{eq:CEOM}
\end{align}
the Liouville equation is found to imply that the phase-space density is constant along the trajectories,
\begin{align}
\frac{\text{d}}{\text{d} t} \, f_W (x(t),p(t);t) = 0 \, , \hspace*{1.5em}
\text{i.e.} \hspace*{1.5em} f_W (x(t),p(t);t) = f_W (x(0),p(0);0) \, .
\end{align}
The classical phase space density described by the Vlasov equation is just the number of particles per phase-space volume.  Within the mean-field dynamics, those particles move in time along the trajectories given by \eqref{eq:CEOM}.  The conservation of the density along these trajectories may be associated with the fact that the classical evolution, represented by Eqs.~\eqref{eq:CEOM}, preserves the phase-space volume spanned by the trajectories. The latter follows from the fact that the phase-space divergence of the phase-space velocity vanishes and it may be also determined by examining directly the changes in the phase-space Jacobian:
\begin{align}
\frac{\text{d}}{\text{d} t} \, \frac{\partial(x(t),p(t))}{\partial(x(0),p(0))} = 0 \, , \hspace*{1.5em} \text{hence} \hspace*{1.5em}
\frac{\partial(x(t),p(t))}{\partial(x(0),p(0))} = 1 \, .
\end{align}
The first equality follows from employing the equations of motion, Eq.~\eqref{eq:CEOM}, in calculating the derivative of the Jacobian.  The practical outcome is that both the phase-space distribution and the phase-space itself behave as an incompressible fluid.  In a sense, the distribution may be viewed as a set of markings on locations within a phase-space fluid that evolves with time in an incompressible manner, consistent with Eq.~\eqref{eq:CEOM}.

In the following, we shall examine the Wigner function for the collision of $A=8$ slabs at $E_{CM} = 25 \, \text{MeV}$.  We choose that particular energy because the Wigner function is then  most spread out in momentum space, making many details particularly visible.  With calculations carried out in a periodic box of $-L \le x \le L$, our Wigner function is defined for momenta $n \pi \hbar/2L$, where $n$ is integer.  With $L=25 \, \text{fm}$, this provides a momentum resolution of $\Delta p = \pi \hbar/2L = 12.4 \, \text{MeV}/c$ or, equivalently, a wavevector resolution of $\Delta k = \pi/L = 0.063 \, \text{fm}^{-1}$. Note that the~integration over $x_r$ in \eqref{eq:wigt}, for the periodic box, extends over a distance of~$2L$.

\begin{figure}
  \includegraphics[width=.9\textwidth]{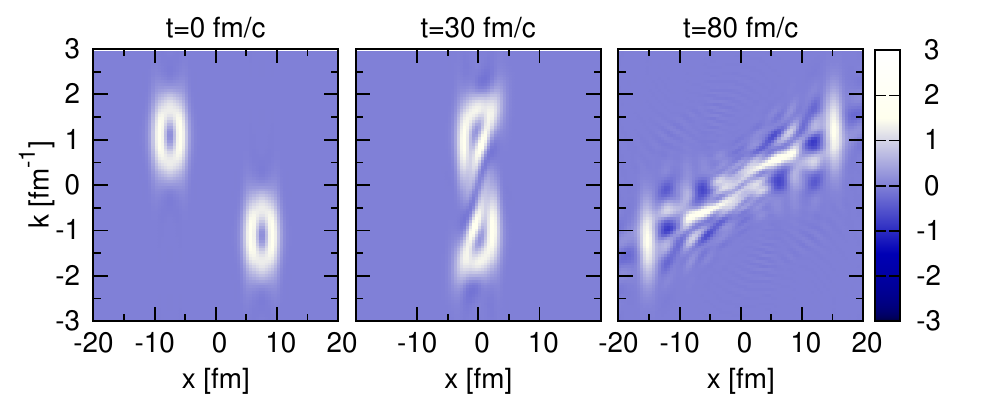}
  \caption{(Color online)
  Intensity plot of the Wigner distribution, $f_W(x,p)$, for the collision of $A=8$ slabs at $E_{CM}/A=25 \, \text{MeV}$.  The left, center and right panels represent, respectively, the sample times of $t=0$, $30$ and $80 \, \text{fm}/c$.  The vertical scale is $k = p/\hbar$.}
  \label{fig:wignermult}
\end{figure}

The left panel in Fig.~\ref{fig:wignermult} shows the Wigner distribution for the initial state of our system at $E_{CM} = 25 \, \text{MeV}$.  With the initial density matrix equal to the sum of density matrices for the two slabs, cf.~Eq.~\eqref{eq:net_rho}, the Wigner function is a sum of the Wigner functions for the two slabs, each represented by a ring structure in Fig.~\ref{fig:wignermult}.  The centers of those rings are displaced from the origin by the displacement of the slabs from $x=0$ (their centroids lie at $\pm x=7.5\ \text{fm}$, cf.~upper left panel of Fig.~\ref{fig:denmat_ea25}) and by $\pm P/\hbar=1.10\ \text{fm}^{-1}$ from $p=0$.  The~latter displacement follows from the fact that multiplication of a~density matrix by a phase-factor $\text{e}^{\pm i P x_r}$, cf.~Eq.~\eqref{eq:boost}, leads, under Fourier transformation, to a~displacement of the Wigner function in momentum by~$\pm P$. Relative to its center, a ring represents the Wigner function of a ground-state slab at rest.

The Wigner function of each slab is basically a sum of the Wigner functions for each filled orbital, cf.~Eqs.~\eqref{eq:rhowf} and \eqref{eq:wigt}.  For orbitals that are parity eigenstates, the orbital Wigner functions are invariant under inversion in $(x,p)$-space.  Given that the density of the ground-state slabs is similar to that in the ground state of the corresponding HO potential, cf.~Fig.~\ref{fig:ad_den}, we expect similarity in the Wigner functions as well.  Classical HO orbits for constant single-particle energy, $\epsilon = p^2/2m + m \, \Omega^2 \, x^2/2$, are ellipses in phase space. They would become circles if the momentum axis was rescaled by a factor of $1/(m \Omega)$ and, consistently with this, the Wigner functions of HO orbitals become isotropic with such a rescaling. They are given by Laguerre polynomials multiplying Gaussians in distance from the origin~\cite{hillery84}.  As~the~lowest orbital is a Gaussian, the corresponding HO Wigner function is also a Gaussian, cf.~Eq.~\eqref{eq:wigt}.  Clearly, however, a~Gaussian does not explain the ring structures in the left panel of Fig.~\ref{fig:wignermult}.  The responsible party for such a structure is actually the second orbital, which is antisymmetric and, thus, has a node at the origin.  For the HO potential, the ring may also be imagined in terms of a collection of classical elliptical orbits.  With time, the particles circle on those orbits in the clockwise direction with a period of $2\pi/\Omega \simeq 92 \, \text{fm}/c $, cf.~\eqref{eq:w0}.  For a mean-field potential, the bounded orbits will not be substantially different and therefore they show an elliptical structure.

With the single-particle energies of Eq.~\eqref{eq:epsilon} being continuous functions of location in phase space, an evolution consistent with the classical limit of Eq.~\eqref{eq:Vlasov} should preserve the topology of the original Wigner function. Consequently, if the original topology is in the form of two rings, those two rings should be recognizable later in the evolution. Large distortions in the shape might appear, but their topology should be preserved in the time evolution. Consequently, the same set of values for the function inside and along the circumference of the rings might be expected.

The next stage of the reaction, represented in the central panel of Fig.~\ref{fig:wignermult}, presents the actual quantum evolution of the Wigner function and corresponds to the stage of overlap and maximal compression of the two slabs.  In spite of the maximal compression, which corresponds to a single compound slab in configuration space, the two rings in the Wigner distribution can be perfectly distinguished. In the initial state, the two rings are separated in phase space and this topology is preserved here, even though the rings overlap in configuration space.  The rings are, however, distorted compared to the initial state.  After the slabs come into contact, the nucleons that get first affected are those that move towards the other slab.  With the potential barrier between the slabs gradually disappearing, the fastest nucleons will move into the other slab.  Slower nucleons, being in the course of getting reflected from the potential barrier of their parent slab, will end up losing their momentum only partially at the merger stage of the slabs, as the barrier disappears.  Finally, the slowest nucleons, moving towards the outer boundary of their parent slab, continue to do so at that stage.  This produces a characteristic  trapezoidal shape for each of the rings in the center panel of Fig.~\ref{fig:wignermult}, representing $t=30 \, \text{fm}/c$.

In spite of the similarities between the initial state and the maximal compression Wigner functions, there are structures in the latter functions that are of a pure quantum nature. One nonclassical aspect of the phase-space distribution is that of the thinning of each ring around a line starting from the side of the ring lowest in momentum magnitude and ending at the ring's center.  The values of the distribution drop there everywhere radially across the ring, in contradiction to what is allowed for a~classical distribution starting from that in the left panel of Fig.~\ref{fig:wignermult}.  Another nonclassical aspect are the values of the distribution right in-between the two rings.  For a~classical evolution, we would expect a valley with vanishing values of the Wigner function.  Instead significant negative values are found there.  The specific valley has been observed before in 1D slab collisions \cite{kohler79,kohler84} and it has been discussed in the context of the lack of collisions in TDHF.  Correlations, if included, would generally act to thermalize the single-particle Wigner function, gradually filling the valley~\cite{grange81}.

The Wigner distribution for the late stage of the reaction, $t = 80 \, \text{fm}/c$, is shown in the third panel of Fig.~\ref{fig:wignermult}.  The distribution is complicated and we may have a hard time recognizing traces of the original rings.  In large part this is due to quantum-mechanical effects whose presence could be anticipated given the substantial off-diagonal structures encountered in the spatial representation of the density matrix.  In spite of those complications, it becomes generally apparent, from the sequence of phase-space images in Fig.~\ref{fig:wignermult}, that the two fragments leading in the configuration space, that emerge from the collision, see also~Fig.~\ref{fig:denmat_ea25}, represent the fastest nucleons from the opposing slabs.  Those nucleons have punched through the matter and have remained the fastest within the system as the collision progressed.  The~original ring structures may be still recognized at $t = 80 \, \text{fm}/c$ in the regions immediately adjacent to the peaks in phase space that represent the leading fragments.  From there on, towards lower momenta, streaks may be observed that can be attributed to the same original slab as the leading fragment.  However, the ring structure cannot be recognized within the respective streak region.  In fact, three or more parallel ridges of greater intensity may be counted there within the remnants of each of the slabs, while the classical dynamics could explain the presence of only two parallel ridges of specific spatial length, formed over a~specific time.  A closer examination reveals many more fainter and more narrowly spaced parallel ridges at $t = 80 \, \text{fm}/c$, down to the momentum resolution from the inverse size of the calculation region.  These clearly represent interference patterns that mar any phase-space structures that could be attributed to a classical dynamics.  Between positive values of the function for the ridges, negative values occur in the valleys.  The~gross pattern of the distribution at $t = 80 \, \text{fm}/c$ is that of a developing Hubble correlation between position and average momentum, cf. Eq.~\eqref{eq:eXxr}.

In the preceding Sections, we have discussed the role of far off-diagonal elements of the density matrix and the effects of a potential suppression of those elements.  We will now discuss the interrelation between the suppression of elements and the momentum space features of the Wigner function.  Clearly, given Eq.~\eqref{eq:wigt}, the Wigner function must be affected by element suppression.  To gain an understanding of the effect, let us denote as $f_W$ the Wigner function for some density matrix, $\rho$.  Let us further denote as $f_{\mathcal P}$ the Wigner function for a matrix obtained from $\rho$ by suppressing the far-away elements according to some profile function, ${\mathcal P}(x_r)$:
\begin{align}
f_{\mathcal P} (x,p) = \int \text{d} x_r \, \text{e}^{- \frac{i p x_r}{\hbar}} \, {\mathcal P}(x_r) \, \rho( x + x_r/2, x - x_r/2) \, ,
\label{eq:wigP}
\end{align}
where ${\mathcal P}(0) = 1$.  In the case that we have discussed in Sec.~\ref{sec:offdiag}, ${\mathcal P}$ is generated by the exponential suppression factor associated to the superoperator field, $W$.  For a sharp cutoff, with matrix elements put to zero at $|x_r| > x_0$, the profile function would be ${\mathcal P}(x_r) = \theta (x_0 - |x_r|)$.  As generally known, a~product in configuration space transforms into a~convolution in the conjugate momentum space.  Upon expressing the density matrix $\rho$ in terms of $f_W$, following Eq.~\eqref{eq:wigt}, we find:
\begin{align}
f_{\mathcal P} (x,p) = \int \text{d} p' \, {\mathcal P}(p - p') \, f_W(x,p') \, ,
\label{eq:fPW}
\end{align}
where
\begin{align}
{\mathcal P} (p) = \frac{1}{2 \pi \hbar} \int \text{d} x_r \, \text{e}^{- \frac{i p x_r}{\hbar}} \, {\mathcal P}(x_r) \, .
\label{eq:Pp}
\end{align}
Given that ${\mathcal P}(x_r=0) = 1$, we find that
\begin{align}
\int \text{d} p \, {\mathcal P} (p) = 1 \, ,
\end{align}
and we therefore observe in Eq.~\eqref{eq:fPW} that the Wigner function $f_{\mathcal P}$ amounts to an average out in momentum of the function $f_W$.  The weight function in averaging is the Fourier transform of the profile function.  In the case of the exemplary sharp cutoff profile above, the weight function is, after \eqref{eq:Pp}, ${\mathcal P}(p)= \frac{1}{\pi p} \sin{\frac{px_0}{\hbar}}$, with the range of averaging in momentum of $\sim  \pi \hbar/(2 x_0)$.  The general conclusion that follows from this investigation is that the suppression of far-away elements should lead to a blurring of finer momentum-structures within the Wigner distribution.

\begin{figure}
  \includegraphics[width=.9\textwidth]{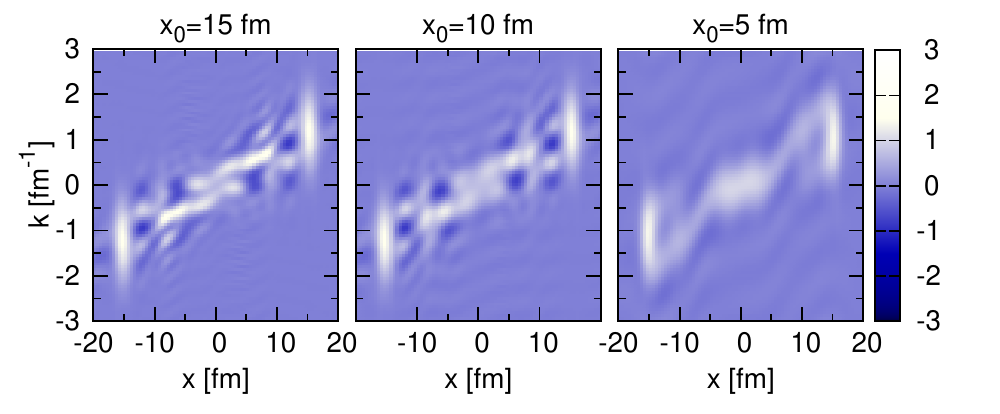}
  \caption{(Color online) Intensity plot of the Wigner distribution, $f_W$, at $t=80 \, \text{fm}/c$, for $E_{CM}/A = 25 \, \text{MeV}$ $A=8$ slab collisions, from calculations with different element suppressions.  The left, center and right panels represent, respectively, calculations with the cut-off parameter value of $x_0 = 15$, $10$ and $5 \, \text{fm}$.  The vertical scale is $k=p/\hbar$.}
  \label{fig:wignercut}
\end{figure}

\begin{figure}
  \includegraphics[width=.9\textwidth]{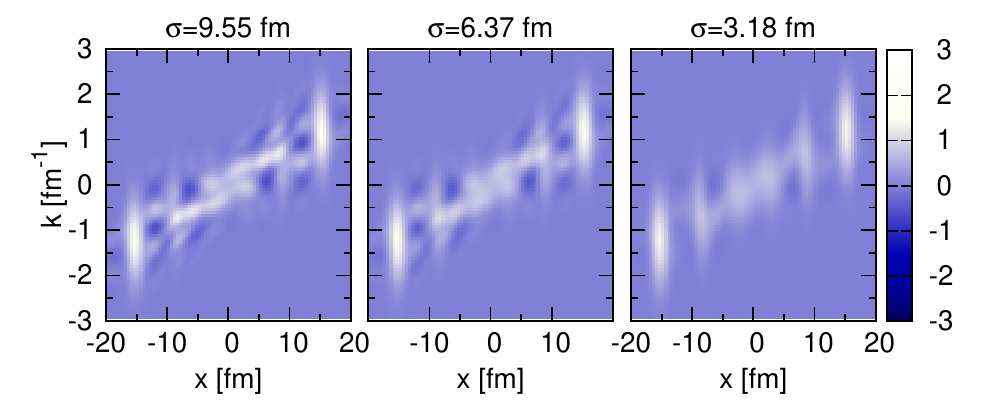}
  \caption{(Color online)
  Intensity plot of the Wigner function, $f_{\mathcal P}$, obtained by convoluting the function $f_W$ from the standard calculation of the $E_{CM}/A = 25 \, \text{MeV}$ collision, for $t=80 \, \text{fm}/c$, with a Gaussian [see Eq.~\eqref{eq:Gaussian}].  The width of the Gaussian for the three displayed panels has been taken in proportion to the cutoff, $x_0$, in the three respective panels of Fig.~\ref{fig:wignercut}, cf.~Eq.~\eqref{eq:sigx}. With this, there is a match between the range of Gaussian averaging and the expected range of blurring of the Wigner function due to the suppression of elements in the evolution of the density matrix.}
  \label{fig:wignergauss}
\end{figure}

The above expectation is tested in Fig.~\ref{fig:wignercut}, that shows the Wigner distribution at the late stage, $t = 80 \, \text{fm}/c$, of the $E_{CM}/A = 25 \, \text{MeV}$ collision, from calculations with different cutoff parameter $x_0$ in the superpotential~$W$ of Eq.~\eqref{eq:cutoff}.  At this stage of the collision, the~Wigner function is particularly nuanced.  If we compare the left panel in Fig.~\ref{fig:wignercut} to the right panel in Fig.~\ref{fig:wignermult}, representing the standard evolution, we can see that even the cutoff at $x_0 = 15 \, \text{fm}$ significantly blurs the Wigner function.  As the cut-off in $W$ is reduced, from the left towards the right panel in Fig.~\ref{fig:wignercut}, the Wigner function gets progressively more blurred, with the finer interference patterns disappearing.  The patterns that remain may be actually attributed to the abruptness in suppressing elements as a function of $x_r$.  For comparison, we further show in Fig.~\ref{fig:wignergauss} the Wigner function from the standard evolution, \emph{i.e.}\ that represented in Fig.~\ref{fig:wignermult}, but now with the function directly blurred at $t=80 \, \text{fm}/c$ by folding it in momentum with a~Gaussian profile function of a prescribed width.  Specifically, we make the momentum range for the profile function,
\begin{align}
{\mathcal P} (p) = \frac{\sigma}{\sqrt{2 \pi} \hbar} \, \exp{\left( - \frac{\sigma^2 \, p^2}{2 \hbar^2} \right)} \, ,
\label{eq:Gaussian}
\end{align}
about the same as that for the weight from a sharp cutoff at $x_0$, with
\begin{align}
\sigma = \frac{2 x_0}{\pi} \, .
\label{eq:sigx}
\end{align}
In comparing Figs.~\ref{fig:wignercut} and \ref{fig:wignergauss}, we see a good degree of agreement between the corresponding blurred distributions, for the two larger values of $x_0$, in the regions of significant values of the distribution.  For $x_0 = 5 \, \text{fm}$, the agreement is qualitative.  Note that the convolution in momentum with Eq.~\eqref{eq:Gaussian} does not change the density for the standard evolution and the densities for evolutions with and without cutoffs have been already compared in Fig.~\ref{fig:dencut_mult}.

Several conclusions, pertaining to the single-particle quantities and their evolution, may be drawn from the above considerations.  Suppressing the far-off diagonal elements in the density matrix is equivalent to smoothing the Wigner function in momentum.  The more severe the suppression of elements, the greater is the level of smoothing of the Wigner function.  This equivalence is important for possible extensions, where correlations are incorporated into the single-particle approach using the KB equations.  In that case, NGFs with different time-arguments need to be considered, see Sec.~\ref{sec:KB}.  The smoothing of Wigner distributions for the purpose of discarding redundant information might be generalized in a more straightforward manner in the correlated case than the suppression of off-diagonal matrix elements. We can further observe that, for moderate levels of smoothing, it is sufficient to follow only the self-consistent evolution of the smoothed Wigner distribution, to arrive at a faithful information on the smoothed distribution in the future.  Finally, it is important to note that, in the region of significant values of the Wigner distribution, the particular details of smoothing, other than its overall range, are likely to be irrelevant, as seen by comparing Figs.~\ref{fig:wignercut} and \ref{fig:wignergauss}.

\section{Conclusions}
\label{sec:conclusions}

In spite of their considerable general potential in describing quantal time-dependent systems, NGF techniques have been, so far, underutilized in nuclear physics.  Specific potential applications within nuclear physics include the description of large-amplitude collective motion and of central nuclear collisions.  The techniques could yield a seamless description for these areas and for nuclear structure~\cite{dickhoff}.  The scarce use of these techniques to study nuclear reactions in practice could be attributed to the anticipated large computational effort.  In this paper, we have investigated whether all the information that would be naturally followed within the NGF approach, actually needs to be maintained in describing nuclear reactions.  This is the major source of computational effort within the NGF method and suitable information suppression techniques would therefore have the potential to make such calculations feasible.

For this purpose, we have studied 1D slab collisions within the mean-field approximation.  While our starting point has been equivalent to the TDHF approach, the manner of discarding of information pursued here could not have been investigated within an approach relying on wavefunctions only.  We have derived correspondences between 3D and 1D systems that help to comprehend the features of the latter.  In solving the single-particle evolution, we have demonstrated in practice the possibility of arriving at a mean-field ground-state, through an adiabatic transformation of the interaction.  We hope to be able to arrive similarly at the ground state for correlated dynamics.  Following our general goal, within the mean-field approximation, we have examined the structure of single-particle density matrices for collisions.  The matrices are generally peaked along the diagonal in their spatial arguments.  For the ground state, the width of the peak of significant values of the matrix elements around the diagonal is proportional to the inverse of the Fermi wavevector.  Otherwise, initial density matrices are confined within square-like structures in the $(x,x')$ plane, with the side of their compact support equal to the spatial extent of the respective slab.

In the dynamical evolution associated with a reaction, significant values of far off-diagonal elements may appear due to the entanglement of single-particle wavefunctions for nucleons that have the possibility of moving out with different momenta and different fragments from the reaction. Applying an absorptive superpotential within an evolution of the density matrix, we have demonstrated that the entanglement patches in the far-off diagonal regions of the matrix have little effect on the evolution close to the diagonal.  This can be attributed to the fact that the emerging fragments have little chance to encounter each other again in an~expanding system, particularly at low relative velocities where phases in the wavefunctions, recorded in the patches, might matter.  Within the structure of the density matrix, those entanglement patches generally travel away from the diagonal of the matrix, never to come back. Even when the element suppression with the superpotential is severe and the density matrices of the individual fragments are modified, we find that the intrinsic collective motions for the fragments are only affected at a semiquantitative level. Also, we find that the energy conservation is quite robust when off-diagonal elements of the density matrix are suppressed.  These findings bode well for the possibility of carrying out realistic NGF calculations of nuclear collisions, where the discarding of information, such as in elements far away from the diagonal of function arguments, becomes an absolute necessity.  On the other hand, we find that maintaining the far off-diagonal elements is important for preserving time-reversibility of the mean-field evolution.  Under time reversal, the correlation patches within the density matrix travel back towards the diagonal of the matrix, where they contribute to the restoration of original density matrix for the initial state.  If~the~correlation patches are removed, the state restored in the course of backward evolution acquires features that are actually characteristic for final states of the collisions within the specific energy regime.

As is well known, Fourier transformation of the density matrix in relative coordinates yields the Wigner function.  The semiclassical Vlasov equation for the Wigner function results under the assumption that the density matrix is peaked  sharply enough around the diagonal within its arguments so that the mean-field in the equation of motion can be expanded in the distance away from the diagonal.  In aiming at suppressing the off-diagonal elements of the density matrix, we do not insist on the validity of the semiclassical expansion.  Rather, we aim at a situation where the element suppression is dialed to achieve a~compromise between the computational effort to be undertaken and the detail needed in the description of physical processes.  The proximity of a system to the semiclassical limit would then just act to improve the chances for reaching a satisfactory compromise.  The suppression of the far-away elements for the matrix, in the spatial representation, is equivalent to smoothing of the matrix within the momentum variable in the Wigner representation.  Such a smoothing procedure removes fine fringe patterns in the structure of the Wigner function.  The equivalence between off-diagonal element suppression and momentum space smoothing is also important for the way in which discarding of information needs to be generalized to the case of NGF with correlations. In that case, the smoothing might be a~more natural generalization when NGF with different time arguments are considered.

To our knowledge, this has been the first time when the off-diagonal elements in the density matrix have been investigated within the context of nuclear collisions.  The original scope of this investigation has been modest, aiming at a reduction of the computational effort within the NGF approach, with the testing ground being the 1D mean-field dynamics.  In the end, however, the investigation touches upon some general aspects of quantal many-body dynamics.  Possibly the most important of those is the growing redundancy of information within a system undergoing expansion.

\section{Acknowledgments}
This work was supported by the National Science Foundation, under Grant No.\ PHY-0800026 and under the Research Experiences for Undergraduates Program, by STFC, under Grant ST/F012012, and by a Marie-Curie Intra-European Fellowship within the 7$^\text{th}$ European Community Framework Programme.

\appendix
\section{3D interpretation of 1D densities}
\label{appendix1}

Results of 1D calculations may be interpreted in terms of 3D matter, when assuming that the 3D matter is uniform in two directions $y$ and $z$, while generally nonuniform in the remaining $x$ direction.  In the two uniform directions, the matter is described in terms of a~set of plane-wave wavefunctions that multiply the wavefunctions labeled with $\alpha$, describing variation in the $x$ direction and incorporating spin and isospin degrees of freedom.  The~transverse set is assumed frozen, independent of $\alpha$ or time.  Under these assumptions, the density of 3D matter, $n_3(x,t) \equiv n(x,t)$, becomes proportional to the density, $n_1(x,t)$, of 1D matter calculated from the 1D density matrix or its wavefunctions $\phi_\alpha(x,t)$,
\begin{align}
n(x,t) = \xi \, n_1(x,t) \, .
\label{eq:xiscale}
\end{align}

The scaling factor $\xi$ may be established following the requirement that the energy of uniform matter needs to minimize at the normal density $n_0$.  In the derivation, we shall invoke the Hugenholtz-van Hove theorem, both for the standard isotropic 3D matter and for the 1D matter to be interpreted in the 3D manner. The Hugenholtz-van Hove theorem states that the energy per nucleon and the Fermi energy should coincide when the system is in its ground state.  In the mean-field approximation, both the energy per nucleon and the Fermi energy consist of kinetic and mean-field terms.  The Fermi energy at normal density is
\begin{align}
\varepsilon_{F1,3} = \frac{p_{F1,3}^2}{2m} + U(n_0) \, ,
\end{align}
where the Fermi momentum, $p_F$, is either that of 1D matter or that of the standard 3D matter.  The total energy per nucleon is
\begin{align}
e_3 = \frac{3}{5} \, \frac{p_{F3}^2}{2m} + e_U (n_0) \, ,
\end{align}
for the 3D matter and
\begin{align}
e_1 = \frac{1}{3} \, \frac{p_{F1}^2}{2m} + e_U (n_0) \, ,
\end{align}
for the 1D matter.  Here, $e_U$ is the contribution to the energy associated with the mean-field.  In the 1D case we consistently include only the kinetic-energy contribution associated with the $x$ direction.  The Hugenholtz-van Hove theorem states that
\begin{align}
\varepsilon_{F1} & = e_1 \, , \\
\varepsilon_{F3} & = e_3 \, .
\end{align}
Upon subtracting the equations side-by-side, the mean-field contributions drop out and we find the relation
\begin{align}
p_{F1} = \sqrt{\frac{3}{5}} \, p_{F3} \, ,
\label{eq:pF13}
\end{align}
at normal density.  We further have
\begin{align}
\label{eq:npF1}
n_1 = \frac{\nu \, p_{F1}}{\pi \hbar} \, , \\
n_0 = \frac{\nu \, p_{F3}^3}{6 \pi^2 \hbar^3} \, .
\label{eq:npF3}
\end{align}
Upon combining Eqs.~\eqref{eq:xiscale}, \eqref{eq:pF13}, \eqref{eq:npF1} and \eqref{eq:npF3}, we find
\begin{align}
\xi = \sqrt{\frac{5}{3}} \left( \frac{\pi \, n_0^2}{6 \, \nu^2} \right)^{1/3} \simeq 1.04 \left( \frac{n_0}{\nu} \right)^{2/3} \, .
\end{align}
Thanks to Eq.~\eqref{eq:xiscale} and the considerations that follow~\eqref{eq:xiscale}, 3D mean-field parametrizations from the literature can be immediately adapted for 1D.

The considerations above can further provide an estimate for the thickness of a slab for a given~$A$ in the 1D interpretation:
\begin{align}
\ell \simeq \frac{A}{n_1} = \frac{\xi A}{n_0} = \sqrt{\frac{5}{3}} \left( \frac{\pi}{6 \nu^2 \, n_0}     \right)^{1/3} \, A \simeq 0.80 \, \text{fm} \, A \, .
\label{eq:lA}
\end{align}
In the above, we have employed $n_0 \simeq 0.16 \, \text{fm}^{-3}$.  Otherwise, when we quote a mass number $A$ in the 1D interpretation, this represents the number of nucleons in the 3D interpretation within a transverse area of $1/\xi$.

\section{Oscillator frequency for 1D slabs}
\label{appendix2}

To minimize the adjustments that the system needs to go through during the adiabatic switching on of the interactions, one needs to choose suitable precursor HO states.  These states are characterized by two parameters: the total number of filled shells, $N_s$, and the oscillator frequency,~$\Omega$.  The first parameter determines the number of nucleons $A$ in the 1D interpretation of the results and, further, the number of nucleons per unit area in the 3D interpretation, see Sec.~\ref{sec:MF} and Appendix \ref{appendix1}.  For spin-isospin saturated systems, studied here, $A=\nu \, N_s$, with $\nu=4$.  To achieve a density in slab interior which is close to the normal density for uniform matter, the frequency $\Omega$ needs to be adequately correlated with~$A$.

\begin{figure}
  \begin{center}
    \includegraphics[width=.75\textwidth]{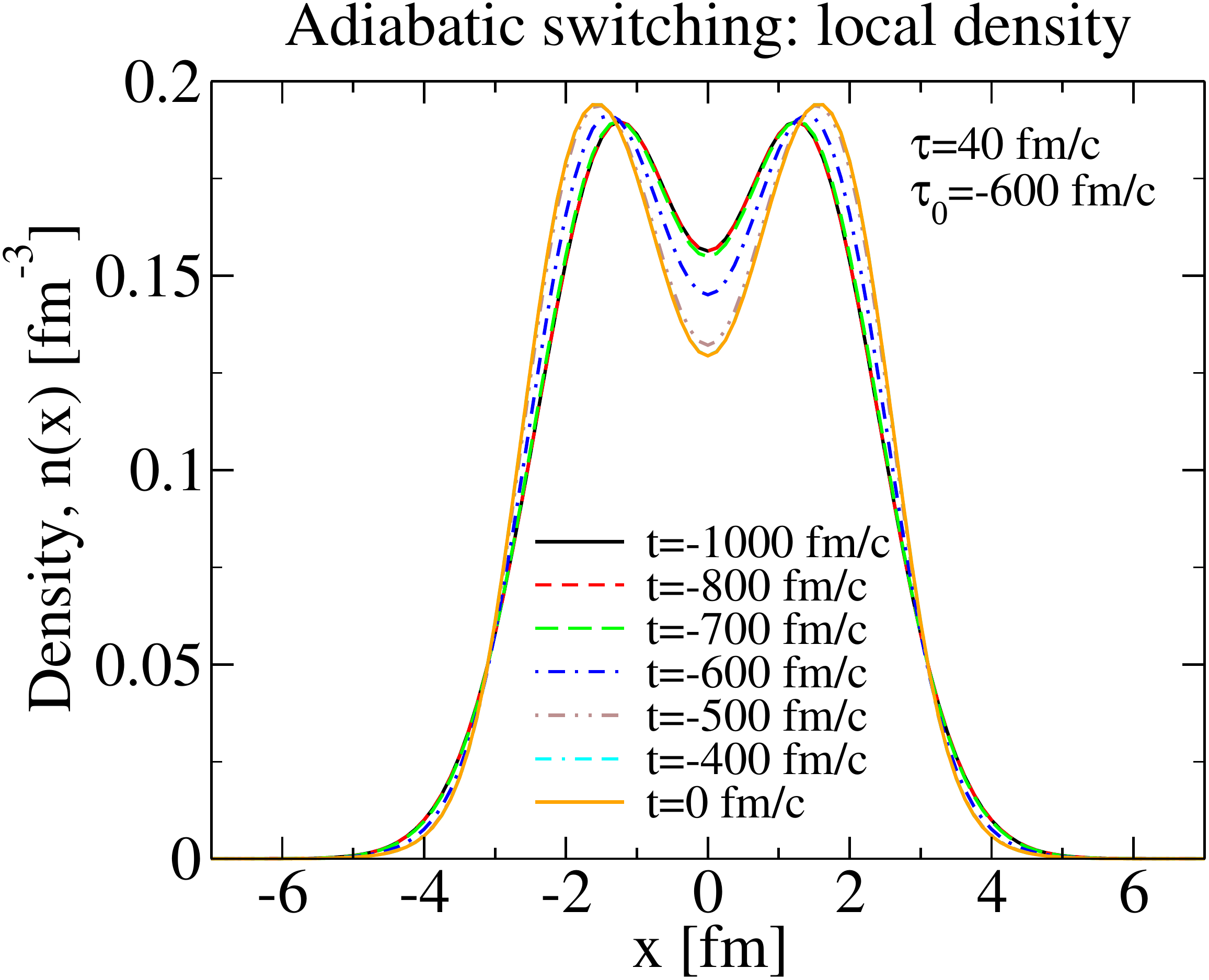}
    \caption{(Color online) Evolution of the density profile, when starting from an HO configuration with $N_s=2$ filled shells (or $A=8$ in 1D interpretation) at time $t=-1000\, \text{fm}/c$ and transforming the single-particle potential to the mean-field form, according to Eq.~\eqref{eq:Ut} with Eq.~\eqref{eq:Fft} and~\eqref{eq:ft}, $\tau = 40 \, \text{fm}/c$ and $\tau_0 = -600 \, \text{fm}/c$.}
    \label{fig:ad_den}
  \end{center}
\end{figure}

To establish what $\Omega$ should be used for a given $A$, we shall follow a procedure analogous to that employed in the literature for the 3D HO.  The well-known 3D result of $\hbar \Omega = 41 A^{-1/3} \, \text{MeV}$ \cite{ring} is found by combining the virial theorem for HO with an expression for the mean square radius of a uniform sphere.  Correspondingly, we start out from the virial theorem for a single-particle state $\alpha$ of the 1D HO:
\begin{align}
\label{eq:1shell}
\frac{1}{2} \, m \, \Omega^2 \, \left\langle x^2 \right\rangle_\alpha =
\frac{\hbar \Omega}{2} \, \left(n_\alpha + \frac{1}{2} \right) \, ,
\end{align}
where $\left\langle x^2 \right\rangle_\alpha$ is the mean square position within the state $\alpha$ and $n_\alpha$ is the oscillator quantum number.  If $N_s$ states are filled, the corresponding quantum number values are $n_\alpha = 0,1, \cdots , N_s - 1$.  The mean square position within a slab is
\begin{align}
\frac{\nu}{A} \sum_\alpha \left\langle x^2 \right\rangle_\alpha = \left\langle x^2 \right\rangle \, .
\end{align}
Upon summing up both sides of Eq.~\eqref{eq:1shell} over the $N_s$ shells, we find
\begin{align}
\hbar \Omega =
\frac{\hbar^2 \, \nu \, N_s^2}{2 m \, \langle x^2 \rangle \, A } =
\frac{\hbar^2 \, A}{2 m \, \nu \, \langle x^2 \rangle}
 \, ,
\end{align}
where we have employed $A=\nu \, N_s$.  For a uniform slab of total thickness $\ell$, we would have had $\langle x^2 \rangle = \ell^2/12$.  If we aim, for large $A$, for a slab of average density $A/\ell = n_1$, we need to choose an HO frequency of
\begin{align}
\hbar \Omega = \frac{6 \hbar^2 \, n_1^2}{m \, \nu A} = \frac{6 \hbar^2 \, n_0^2}{m \, \xi^2 \, \nu A} \sim \frac{108 \, \text{MeV}}{A} \, ,
\label{eq:w0}
\end{align}
where we have made used of the connection between 1D and 3D density derived in Appendix~\ref{appendix1}. The utility of the derived relation, for adiabatic switching, can be tested \emph{a~posteriori} by comparing the density profiles of the  predecessor HO state and that of the corresponding self-consistent slab.  In the example provided in Fig.~\ref{fig:ad_den}, the profiles of both the precursor HO slab and the self-consistent mean-field one are very close.  We have observed a similar agreement over a wide range of particle numbers $A$, which suggests that Eq.~\eqref{eq:w0} indeed yields a good choice of the frequency for the starting state.


\bibliographystyle{apsrev}
\bibliography{tdgf}

\begin{thebibliography}{55}
\expandafter\ifx\csname natexlab\endcsname\relax\def\natexlab#1{#1}\fi
\expandafter\ifx\csname bibnamefont\endcsname\relax
  \def\bibnamefont#1{#1}\fi
\expandafter\ifx\csname bibfnamefont\endcsname\relax
  \def\bibfnamefont#1{#1}\fi
\expandafter\ifx\csname citenamefont\endcsname\relax
  \def\citenamefont#1{#1}\fi
\expandafter\ifx\csname url\endcsname\relax
  \def\url#1{\texttt{#1}}\fi
\expandafter\ifx\csname urlprefix\endcsname\relax\def\urlprefix{URL }\fi
\providecommand{\bibinfo}[2]{#2}
\providecommand{\eprint}[2][]{\url{#2}}

\bibitem[{\citenamefont{Kadanoff and Baym}(1962)}]{kadanoff}
\bibinfo{author}{\bibfnamefont{L.~P.} \bibnamefont{Kadanoff}} \bibnamefont{and}
  \bibinfo{author}{\bibfnamefont{G.}~\bibnamefont{Baym}},
  \emph{\bibinfo{title}{Quantum Statistical Mechanics}}
  (\bibinfo{publisher}{Benjamin}, \bibinfo{address}{New York},
  \bibinfo{year}{1962}).

\bibitem[{\citenamefont{Danielewicz}(1984{\natexlab{a}})}]{danielewicz84a}
\bibinfo{author}{\bibfnamefont{P.}~\bibnamefont{Danielewicz}},
  \bibinfo{journal}{Ann. Phys. (N.Y.)} \textbf{\bibinfo{volume}{152}},
  \bibinfo{pages}{239} (\bibinfo{year}{1984}{\natexlab{a}}).

\bibitem[{\citenamefont{Botermans and Malfliet}(1990)}]{botermans90}
\bibinfo{author}{\bibfnamefont{W.}~\bibnamefont{Botermans}} \bibnamefont{and}
  \bibinfo{author}{\bibfnamefont{R.}~\bibnamefont{Malfliet}},
  \bibinfo{journal}{Phys. Rept.} \textbf{\bibinfo{volume}{198}},
  \bibinfo{pages}{115} (\bibinfo{year}{1990}).

\bibitem[{\citenamefont{K\"ohler}(1995)}]{kohler95}
\bibinfo{author}{\bibfnamefont{H.~S.} \bibnamefont{K\"ohler}},
  \bibinfo{journal}{Phys. Rev. C} \textbf{\bibinfo{volume}{51}},
  \bibinfo{pages}{3232} (\bibinfo{year}{1995}).

\bibitem[{\citenamefont{Haug and Jauho}(1996)}]{haug}
\bibinfo{author}{\bibfnamefont{H.}~\bibnamefont{Haug}} \bibnamefont{and}
  \bibinfo{author}{\bibfnamefont{A.-P.} \bibnamefont{Jauho}},
  \emph{\bibinfo{title}{Quantum Kinetics in Transport and Optics of
  Semiconductors}} (\bibinfo{publisher}{Springer}, \bibinfo{address}{Berlin},
  \bibinfo{year}{1996}).

\bibitem[{\citenamefont{Gasenzer et~al.}(2005)\citenamefont{Gasenzer, Berges,
  Schmidt, and Seco}}]{gasenzer-2005-72}
\bibinfo{author}{\bibfnamefont{T.}~\bibnamefont{Gasenzer}},
  \bibinfo{author}{\bibfnamefont{J.}~\bibnamefont{Berges}},
  \bibinfo{author}{\bibfnamefont{M.~G.} \bibnamefont{Schmidt}},
  \bibnamefont{and} \bibinfo{author}{\bibfnamefont{M.}~\bibnamefont{Seco}},
  \bibinfo{journal}{Phys. Rev. A} \textbf{\bibinfo{volume}{72}},
  \bibinfo{pages}{063604} (\bibinfo{year}{2005}), \eprint{cond-mat/0507480}.

\bibitem[{\citenamefont{Mart\'inez et~al.}(2007)\citenamefont{Mart\'inez,
  Bescond, Barker, Svizhenko, Anatram, Millar, and Asenov}}]{negf07}
\bibinfo{author}{\bibfnamefont{A.}~\bibnamefont{Mart\'inez}},
  \bibinfo{author}{\bibfnamefont{M.}~\bibnamefont{Bescond}},
  \bibinfo{author}{\bibfnamefont{J.~R.} \bibnamefont{Barker}},
  \bibinfo{author}{\bibfnamefont{A.}~\bibnamefont{Svizhenko}},
  \bibinfo{author}{\bibfnamefont{M.~P.} \bibnamefont{Anatram}},
  \bibinfo{author}{\bibfnamefont{C.}~\bibnamefont{Millar}}, \bibnamefont{and}
  \bibinfo{author}{\bibfnamefont{A.}~\bibnamefont{Asenov}},
  \bibinfo{journal}{IEEE Trans. Electron Devices}
  \textbf{\bibinfo{volume}{54}}, \bibinfo{pages}{2213} (\bibinfo{year}{2007}).

\bibitem[{\citenamefont{Dahlen and {van Leeuwen}}(2007)}]{dahlen07}
\bibinfo{author}{\bibfnamefont{N.~E.} \bibnamefont{Dahlen}} \bibnamefont{and}
  \bibinfo{author}{\bibfnamefont{R.}~\bibnamefont{{van Leeuwen}}},
  \bibinfo{journal}{Phys. Rev. Lett.} \textbf{\bibinfo{volume}{98}},
  \bibinfo{pages}{153004} (\bibinfo{year}{2007}), \eprint{cond-mat/0703411}.

\bibitem[{\citenamefont{Danielewicz and Bertsch}(1991)}]{Danielewicz:1991dh}
\bibinfo{author}{\bibfnamefont{P.}~\bibnamefont{Danielewicz}} \bibnamefont{and}
  \bibinfo{author}{\bibfnamefont{G.~F.} \bibnamefont{Bertsch}},
  \bibinfo{journal}{Nucl. Phys. A} \textbf{\bibinfo{volume}{533}},
  \bibinfo{pages}{712} (\bibinfo{year}{1991}).

\bibitem[{\citenamefont{Ivanov et~al.}(2000)\citenamefont{Ivanov, Knoll, and
  Voskresensky}}]{Ivanov:1999tj}
\bibinfo{author}{\bibfnamefont{Y.~B.} \bibnamefont{Ivanov}},
  \bibinfo{author}{\bibfnamefont{J.}~\bibnamefont{Knoll}}, \bibnamefont{and}
  \bibinfo{author}{\bibfnamefont{D.~N.} \bibnamefont{Voskresensky}},
  \bibinfo{journal}{Nucl. Phys. A} \textbf{\bibinfo{volume}{672}},
  \bibinfo{pages}{313} (\bibinfo{year}{2000}), \eprint{nucl-th/9905028}.

\bibitem[{\citenamefont{Effenberger and Mosel}(1999)}]{Effenberger:1999uv}
\bibinfo{author}{\bibfnamefont{M.}~\bibnamefont{Effenberger}} \bibnamefont{and}
  \bibinfo{author}{\bibfnamefont{U.}~\bibnamefont{Mosel}},
  \bibinfo{journal}{Phys. Rev. C} \textbf{\bibinfo{volume}{60}},
  \bibinfo{pages}{051901} (\bibinfo{year}{1999}), \eprint{nucl-th/9906085}.

\bibitem[{\citenamefont{Cassing and Juchem}(2000)}]{Cassing:1999wx}
\bibinfo{author}{\bibfnamefont{W.}~\bibnamefont{Cassing}} \bibnamefont{and}
  \bibinfo{author}{\bibfnamefont{S.}~\bibnamefont{Juchem}},
  \bibinfo{journal}{Nucl. Phys. A} \textbf{\bibinfo{volume}{665}},
  \bibinfo{pages}{377} (\bibinfo{year}{2000}), \eprint{nucl-th/9903070}.

\bibitem[{\citenamefont{Danielewicz}(1984{\natexlab{b}})}]{danielewicz84b}
\bibinfo{author}{\bibfnamefont{P.}~\bibnamefont{Danielewicz}},
  \bibinfo{journal}{Ann. Phys. (N.Y.)} \textbf{\bibinfo{volume}{152}},
  \bibinfo{pages}{305} (\bibinfo{year}{1984}{\natexlab{b}}).

\bibitem[{\citenamefont{Tohyama}(1987)}]{tohyama87}
\bibinfo{author}{\bibfnamefont{M.}~\bibnamefont{Tohyama}},
  \bibinfo{journal}{Phys. Rev. C} \textbf{\bibinfo{volume}{36}},
  \bibinfo{pages}{187} (\bibinfo{year}{1987}).

\bibitem[{\citenamefont{K\"ohler}(1996)}]{PhysRevE.53.3145}
\bibinfo{author}{\bibfnamefont{H.~S.} \bibnamefont{K\"ohler}},
  \bibinfo{journal}{Phys. Rev. E} \textbf{\bibinfo{volume}{53}},
  \bibinfo{pages}{3145} (\bibinfo{year}{1996}).

\bibitem[{\citenamefont{Bo\ifmmode~\dot{z}\else
  \.{z}\fi{}ek}(1997)}]{PhysRevC.56.1452}
\bibinfo{author}{\bibfnamefont{P.}~\bibnamefont{Bo\ifmmode~\dot{z}\else
  \.{z}\fi{}ek}}, \bibinfo{journal}{Phys. Rev. C}
  \textbf{\bibinfo{volume}{56}}, \bibinfo{pages}{1452} (\bibinfo{year}{1997}),
  \eprint{nucl-th/9704007}.

\bibitem[{\citenamefont{Juchem et~al.}(2004)\citenamefont{Juchem, Cassing, and
  Greiner}}]{PhysRevD.69.025006}
\bibinfo{author}{\bibfnamefont{S.}~\bibnamefont{Juchem}},
  \bibinfo{author}{\bibfnamefont{W.}~\bibnamefont{Cassing}}, \bibnamefont{and}
  \bibinfo{author}{\bibfnamefont{C.}~\bibnamefont{Greiner}},
  \bibinfo{journal}{Phys. Rev. D} \textbf{\bibinfo{volume}{69}},
  \bibinfo{pages}{025006} (\bibinfo{year}{2004}).

\bibitem[{\citenamefont{Lindner and Muller}(2008)}]{Lindner:2007am}
\bibinfo{author}{\bibfnamefont{M.}~\bibnamefont{Lindner}} \bibnamefont{and}
  \bibinfo{author}{\bibfnamefont{M.~M.} \bibnamefont{Muller}},
  \bibinfo{journal}{Phys. Rev. D} \textbf{\bibinfo{volume}{77}},
  \bibinfo{pages}{025027} (\bibinfo{year}{2008}), \eprint{0710.2917}.

\bibitem[{\citenamefont{Dickhoff and {van Neck}}(2005)}]{dickhoff}
\bibinfo{author}{\bibfnamefont{W.}~\bibnamefont{Dickhoff}} \bibnamefont{and}
  \bibinfo{author}{\bibfnamefont{D.}~\bibnamefont{{van Neck}}},
  \emph{\bibinfo{title}{Many-Body Theory Exposed!}} (\bibinfo{publisher}{World
  Scientific}, \bibinfo{address}{London}, \bibinfo{year}{2005}).

\bibitem[{\citenamefont{Reinhard and Suraud}(1992)}]{Reinhard199298}
\bibinfo{author}{\bibfnamefont{P.~G.} \bibnamefont{Reinhard}} \bibnamefont{and}
  \bibinfo{author}{\bibfnamefont{E.}~\bibnamefont{Suraud}},
  \bibinfo{journal}{Ann. Phys. (N.Y.)} \textbf{\bibinfo{volume}{216}},
  \bibinfo{pages}{98 } (\bibinfo{year}{1992}).

\bibitem[{\citenamefont{Greiner and Leupold}(1998)}]{greiner98}
\bibinfo{author}{\bibfnamefont{G.}~\bibnamefont{Greiner}} \bibnamefont{and}
  \bibinfo{author}{\bibfnamefont{S.}~\bibnamefont{Leupold}},
  \bibinfo{journal}{Ann. Phys. (N.Y.)} \textbf{\bibinfo{volume}{270}},
  \bibinfo{pages}{328 } (\bibinfo{year}{1998}).

\bibitem[{\citenamefont{Bonche et~al.}(1976)\citenamefont{Bonche, Koonin, and
  Negele}}]{bonche76}
\bibinfo{author}{\bibfnamefont{P.}~\bibnamefont{Bonche}},
  \bibinfo{author}{\bibfnamefont{S.}~\bibnamefont{Koonin}}, \bibnamefont{and}
  \bibinfo{author}{\bibfnamefont{J.~W.} \bibnamefont{Negele}},
  \bibinfo{journal}{Phys. Rev. C} \textbf{\bibinfo{volume}{13}},
  \bibinfo{pages}{1226} (\bibinfo{year}{1976}).

\bibitem[{\citenamefont{Negele}(1982)}]{tdhf}
\bibinfo{author}{\bibfnamefont{J.~W.} \bibnamefont{Negele}},
  \bibinfo{journal}{Rev. Mod. Phys.} \textbf{\bibinfo{volume}{54}},
  \bibinfo{pages}{913} (\bibinfo{year}{1982}).

\bibitem[{\citenamefont{Umar and Oberacker}(2006)}]{Umar:2006ub}
\bibinfo{author}{\bibfnamefont{A.~S.} \bibnamefont{Umar}} \bibnamefont{and}
  \bibinfo{author}{\bibfnamefont{V.~E.} \bibnamefont{Oberacker}},
  \bibinfo{journal}{Phys. Rev. C} \textbf{\bibinfo{volume}{73}},
  \bibinfo{pages}{054607} (\bibinfo{year}{2006}), \eprint{nucl-th/0603038}.

\bibitem[{\citenamefont{Bertsch and Das~Gupta}(1988)}]{Bertsch:1988ik}
\bibinfo{author}{\bibfnamefont{G.~F.} \bibnamefont{Bertsch}} \bibnamefont{and}
  \bibinfo{author}{\bibfnamefont{S.}~\bibnamefont{Das~Gupta}},
  \bibinfo{journal}{Phys. Rept.} \textbf{\bibinfo{volume}{160}},
  \bibinfo{pages}{189} (\bibinfo{year}{1988}).

\bibitem[{\citenamefont{Bonasera et~al.}(1994)\citenamefont{Bonasera,
  Gulminelli, and Molitoris}}]{bonasera94}
\bibinfo{author}{\bibfnamefont{A.}~\bibnamefont{Bonasera}},
  \bibinfo{author}{\bibfnamefont{F.}~\bibnamefont{Gulminelli}},
  \bibnamefont{and}
  \bibinfo{author}{\bibfnamefont{J.}~\bibnamefont{Molitoris}},
  \bibinfo{journal}{Phys. Rep.} \textbf{\bibinfo{volume}{243}},
  \bibinfo{pages}{1} (\bibinfo{year}{1994}).

\bibitem[{\citenamefont{Aichelin and Stoecker}(1986)}]{Aichelin:1986wa}
\bibinfo{author}{\bibfnamefont{J.}~\bibnamefont{Aichelin}} \bibnamefont{and}
  \bibinfo{author}{\bibfnamefont{H.}~\bibnamefont{Stoecker}},
  \bibinfo{journal}{Phys. Lett. B} \textbf{\bibinfo{volume}{176}},
  \bibinfo{pages}{14} (\bibinfo{year}{1986}).

\bibitem[{\citenamefont{Ono et~al.}(1992)\citenamefont{Ono, Horiuchi, Maruyama,
  and Ohnishi}}]{Ono:1991uz}
\bibinfo{author}{\bibfnamefont{A.}~\bibnamefont{Ono}},
  \bibinfo{author}{\bibfnamefont{H.}~\bibnamefont{Horiuchi}},
  \bibinfo{author}{\bibfnamefont{T.}~\bibnamefont{Maruyama}}, \bibnamefont{and}
  \bibinfo{author}{\bibfnamefont{A.}~\bibnamefont{Ohnishi}},
  \bibinfo{journal}{Phys. Rev. Lett.} \textbf{\bibinfo{volume}{68}},
  \bibinfo{pages}{2898} (\bibinfo{year}{1992}).

\bibitem[{\citenamefont{Ring and Schuck}(1980)}]{ring}
\bibinfo{author}{\bibfnamefont{P.}~\bibnamefont{Ring}} \bibnamefont{and}
  \bibinfo{author}{\bibfnamefont{P.}~\bibnamefont{Schuck}},
  \emph{\bibinfo{title}{The Nuclear Many-Body Problem}}
  (\bibinfo{publisher}{Springer}, \bibinfo{address}{Berlin},
  \bibinfo{year}{1980}).

\bibitem[{\citenamefont{Wong}(1982)}]{wong82}
\bibinfo{author}{\bibfnamefont{C.~Y.} \bibnamefont{Wong}},
  \bibinfo{journal}{Phys. Rev. C} \textbf{\bibinfo{volume}{25}},
  \bibinfo{pages}{1460} (\bibinfo{year}{1982}).

\bibitem[{\citenamefont{Lacroix et~al.}(2004)\citenamefont{Lacroix, Ayik, and
  Chomaz}}]{lacroix04}
\bibinfo{author}{\bibfnamefont{D.}~\bibnamefont{Lacroix}},
  \bibinfo{author}{\bibfnamefont{S.}~\bibnamefont{Ayik}}, \bibnamefont{and}
  \bibinfo{author}{\bibfnamefont{P.}~\bibnamefont{Chomaz}},
  \bibinfo{journal}{Prog. Part. Nucl. Phys.} \textbf{\bibinfo{volume}{52}},
  \bibinfo{pages}{497} (\bibinfo{year}{2004}).

\bibitem[{\citenamefont{Simenel and Chomaz}(2003)}]{simenel03}
\bibinfo{author}{\bibfnamefont{C.}~\bibnamefont{Simenel}} \bibnamefont{and}
  \bibinfo{author}{\bibfnamefont{P.}~\bibnamefont{Chomaz}},
  \bibinfo{journal}{Phys. Rev. C} \textbf{\bibinfo{volume}{68}},
  \bibinfo{pages}{024302} (\bibinfo{year}{2003}), \eprint{nucl-th/0209069}.

\bibitem[{\citenamefont{Flocard et~al.}(1978)\citenamefont{Flocard, Konnin, and
  Weiss}}]{flocard78}
\bibinfo{author}{\bibfnamefont{H.}~\bibnamefont{Flocard}},
  \bibinfo{author}{\bibfnamefont{S.~E.} \bibnamefont{Konnin}},
  \bibnamefont{and} \bibinfo{author}{\bibfnamefont{M.~S.} \bibnamefont{Weiss}},
  \bibinfo{journal}{Phys. Rev. C} \textbf{\bibinfo{volume}{17}},
  \bibinfo{pages}{1682} (\bibinfo{year}{1978}).

\bibitem[{\citenamefont{Tang et~al.}(1981)\citenamefont{Tang, Dasso, Esbensen,
  Broglia, and Winther}}]{tang81}
\bibinfo{author}{\bibfnamefont{H.~H.~K.} \bibnamefont{Tang}},
  \bibinfo{author}{\bibfnamefont{C.~H.} \bibnamefont{Dasso}},
  \bibinfo{author}{\bibfnamefont{H.}~\bibnamefont{Esbensen}},
  \bibinfo{author}{\bibfnamefont{R.~A.} \bibnamefont{Broglia}},
  \bibnamefont{and} \bibinfo{author}{\bibfnamefont{A.}~\bibnamefont{Winther}},
  \bibinfo{journal}{Phys. Lett. B} \textbf{\bibinfo{volume}{101}},
  \bibinfo{pages}{10} (\bibinfo{year}{1981}).

\bibitem[{\citenamefont{K{\"o}hler}(1980)}]{kohler80}
\bibinfo{author}{\bibfnamefont{H.~S.} \bibnamefont{K{\"o}hler}},
  \bibinfo{journal}{Nucl. Phys. A} \textbf{\bibinfo{volume}{343}},
  \bibinfo{pages}{315} (\bibinfo{year}{1980}).

\bibitem[{\citenamefont{K{\"o}hler and Nilsson}(1984)}]{kohler84}
\bibinfo{author}{\bibfnamefont{H.~S.} \bibnamefont{K{\"o}hler}}
  \bibnamefont{and} \bibinfo{author}{\bibfnamefont{B.~S.}
  \bibnamefont{Nilsson}}, \bibinfo{journal}{Nucl. Phys. A}
  \textbf{\bibinfo{volume}{417}}, \bibinfo{pages}{541} (\bibinfo{year}{1984}).

\bibitem[{\citenamefont{Grang{\'e} et~al.}(1981)\citenamefont{Grang{\'e},
  Richert, Wolschi, and Weidenm{\"u}ller}}]{grange81}
\bibinfo{author}{\bibfnamefont{P.}~\bibnamefont{Grang{\'e}}},
  \bibinfo{author}{\bibfnamefont{J.}~\bibnamefont{Richert}},
  \bibinfo{author}{\bibfnamefont{G.}~\bibnamefont{Wolschi}}, \bibnamefont{and}
  \bibinfo{author}{\bibfnamefont{H.~A.} \bibnamefont{Weidenm{\"u}ller}},
  \bibinfo{journal}{Nucl. Phys. A} \textbf{\bibinfo{volume}{356}},
  \bibinfo{pages}{260} (\bibinfo{year}{1981}).

\bibitem[{\citenamefont{Tohyama}(1985)}]{tohyama85}
\bibinfo{author}{\bibfnamefont{M.}~\bibnamefont{Tohyama}},
  \bibinfo{journal}{Phys. Lett. B} \textbf{\bibinfo{volume}{163}},
  \bibinfo{pages}{14} (\bibinfo{year}{1985}).

\bibitem[{\citenamefont{K{\"o}hler and Flocard}(1979)}]{kohler79}
\bibinfo{author}{\bibfnamefont{H.~S.} \bibnamefont{K{\"o}hler}}
  \bibnamefont{and} \bibinfo{author}{\bibfnamefont{H.}~\bibnamefont{Flocard}},
  \bibinfo{journal}{Nucl. Phys. A} \textbf{\bibinfo{volume}{323}},
  \bibinfo{pages}{189} (\bibinfo{year}{1979}).

\bibitem[{\citenamefont{Baym}(1962)}]{baym62}
\bibinfo{author}{\bibfnamefont{G.}~\bibnamefont{Baym}}, \bibinfo{journal}{Phys.
  Rev.} \textbf{\bibinfo{volume}{127}}, \bibinfo{pages}{1391}
  (\bibinfo{year}{1962}).

\bibitem[{\citenamefont{Kwong et~al.}(1998)\citenamefont{Kwong, Bonitz, Binder,
  and K\"ohler}}]{kwong98}
\bibinfo{author}{\bibfnamefont{N.~H.} \bibnamefont{Kwong}},
  \bibinfo{author}{\bibfnamefont{M.}~\bibnamefont{Bonitz}},
  \bibinfo{author}{\bibfnamefont{R.}~\bibnamefont{Binder}}, \bibnamefont{and}
  \bibinfo{author}{\bibfnamefont{H.~S.} \bibnamefont{K\"ohler}},
  \bibinfo{journal}{Phys. Stat. Sol. B} \textbf{\bibinfo{volume}{206}},
  \bibinfo{pages}{197} (\bibinfo{year}{1998}).

\bibitem[{\citenamefont{Balzer et~al.}(2009)\citenamefont{Balzer, Bonitz, van
  Leeuwen, Dahlen, and Stan}}]{balzer09}
\bibinfo{author}{\bibfnamefont{K.}~\bibnamefont{Balzer}},
  \bibinfo{author}{\bibfnamefont{M.}~\bibnamefont{Bonitz}},
  \bibinfo{author}{\bibfnamefont{R.}~\bibnamefont{van Leeuwen}},
  \bibinfo{author}{\bibfnamefont{N.~E.} \bibnamefont{Dahlen}},
  \bibnamefont{and} \bibinfo{author}{\bibfnamefont{A.}~\bibnamefont{Stan}},
  \bibinfo{journal}{Phys. Rev. B} \textbf{\bibinfo{volume}{79}},
  \bibinfo{pages}{245306} (\bibinfo{year}{2009}), \eprint{arXiv:0810.2425}.

\bibitem[{\citenamefont{K\"ohler et~al.}(1999)\citenamefont{K\"ohler, Kwong,
  and Yousif}}]{kohler99}
\bibinfo{author}{\bibfnamefont{H.~S.} \bibnamefont{K\"ohler}},
  \bibinfo{author}{\bibfnamefont{N.~H.} \bibnamefont{Kwong}}, \bibnamefont{and}
  \bibinfo{author}{\bibfnamefont{H.~A.} \bibnamefont{Yousif}},
  \bibinfo{journal}{Comp. Phys. Comm.} \textbf{\bibinfo{volume}{123}},
  \bibinfo{pages}{123} (\bibinfo{year}{1999}).

\bibitem[{\citenamefont{Feit et~al.}(1982)\citenamefont{Feit, Fleck, and
  Steiger}}]{feit82}
\bibinfo{author}{\bibfnamefont{M.~D.} \bibnamefont{Feit}},
  \bibinfo{author}{\bibfnamefont{J.~A.} \bibnamefont{Fleck}}, \bibnamefont{and}
  \bibinfo{author}{\bibfnamefont{A.}~\bibnamefont{Steiger}},
  \bibinfo{journal}{J. Comp. Phys.} \textbf{\bibinfo{volume}{47}},
  \bibinfo{pages}{412} (\bibinfo{year}{1982}).

\bibitem[{\citenamefont{Messiah}(1999)}]{messiah}
\bibinfo{author}{\bibfnamefont{A.}~\bibnamefont{Messiah}},
  \emph{\bibinfo{title}{Quantum Mechanics}} (\bibinfo{publisher}{Dover
  Publications}, \bibinfo{address}{NY}, \bibinfo{year}{1999}).

\bibitem[{\citenamefont{Morawetz and Kohler}(1999)}]{morawetz99}
\bibinfo{author}{\bibfnamefont{K.}~\bibnamefont{Morawetz}} \bibnamefont{and}
  \bibinfo{author}{\bibfnamefont{H.}~\bibnamefont{Kohler}},
  \bibinfo{journal}{Eur. Phys. J. A} \textbf{\bibinfo{volume}{4}},
  \bibinfo{pages}{291} (\bibinfo{year}{1999}), \eprint{nucl-th/9802082}.

\bibitem[{\citenamefont{Thouless and Valatin}(1962)}]{thouless62}
\bibinfo{author}{\bibfnamefont{D.~J.} \bibnamefont{Thouless}} \bibnamefont{and}
  \bibinfo{author}{\bibfnamefont{J.~G.} \bibnamefont{Valatin}},
  \bibinfo{journal}{Nucl. Phys.} \textbf{\bibinfo{volume}{31}},
  \bibinfo{pages}{211} (\bibinfo{year}{1962}).

\bibitem[{\citenamefont{Simenel et~al.}()\citenamefont{Simenel, Avez, and
  Lacroix}}]{simenel08}
\bibinfo{author}{\bibfnamefont{C.}~\bibnamefont{Simenel}},
  \bibinfo{author}{\bibfnamefont{B.}~\bibnamefont{Avez}}, \bibnamefont{and}
  \bibinfo{author}{\bibfnamefont{D.}~\bibnamefont{Lacroix}},
  \bibinfo{note}{{L}ect. Notes Int. Joliot-Curie School, Maubuisson - France,
  2007, arXiv:0806.2714}.

\bibitem[{\citenamefont{M{\'a}rk}(1997)}]{Geza97}
\bibinfo{author}{\bibfnamefont{G.}~\bibnamefont{M{\'a}rk}},
  \bibinfo{journal}{Eur. J. Phys.} \textbf{\bibinfo{volume}{18}},
  \bibinfo{pages}{247} (\bibinfo{year}{1997}).

\bibitem[{\citenamefont{Frensley}(1990)}]{frensley90}
\bibinfo{author}{\bibfnamefont{W.~R.} \bibnamefont{Frensley}},
  \bibinfo{journal}{Rev. Mod. Phys.} \textbf{\bibinfo{volume}{62}},
  \bibinfo{pages}{745} (\bibinfo{year}{1990}).

\bibitem[{\citenamefont{Breuer and Petruccione}(2002)}]{breuer}
\bibinfo{author}{\bibfnamefont{H.~P.} \bibnamefont{Breuer}} \bibnamefont{and}
  \bibinfo{author}{\bibfnamefont{F.}~\bibnamefont{Petruccione}},
  \emph{\bibinfo{title}{The Theory of Open Quantum Systems}}
  (\bibinfo{publisher}{Oxford University Press}, \bibinfo{address}{Oxford, UK},
  \bibinfo{year}{2002}), ISBN \bibinfo{isbn}{0198520638}.

\bibitem[{\citenamefont{Omn\`es}(2002)}]{omnes02}
\bibinfo{author}{\bibfnamefont{R.}~\bibnamefont{Omn\`es}},
  \bibinfo{journal}{Phys. Rev. A} \textbf{\bibinfo{volume}{65}},
  \bibinfo{pages}{052119} (\bibinfo{year}{2002}), \eprint{quant-ph/0304100}.

\bibitem[{\citenamefont{Vacchini}(2005)}]{vacchini05}
\bibinfo{author}{\bibfnamefont{B.}~\bibnamefont{Vacchini}},
  \bibinfo{journal}{Phys. Rev. Lett.} \textbf{\bibinfo{volume}{95}},
  \bibinfo{pages}{230402} (\bibinfo{year}{2005}), \eprint{quant-ph/0510127}.

\bibitem[{\citenamefont{Kusnezov et~al.}(1999)\citenamefont{Kusnezov, Bulgac,
  and Dang}}]{kusnezov99}
\bibinfo{author}{\bibfnamefont{D.}~\bibnamefont{Kusnezov}},
  \bibinfo{author}{\bibfnamefont{A.}~\bibnamefont{Bulgac}}, \bibnamefont{and}
  \bibinfo{author}{\bibfnamefont{G.~D.} \bibnamefont{Dang}},
  \bibinfo{journal}{Phys. Rev. Lett.} \textbf{\bibinfo{volume}{82}},
  \bibinfo{pages}{1136} (\bibinfo{year}{1999}), \eprint{chao-dyn/9901002}.

\bibitem[{\citenamefont{Hillery et~al.}(1984)\citenamefont{Hillery, O'Connel,
  Scully, and Wigner}}]{hillery84}
\bibinfo{author}{\bibfnamefont{M.}~\bibnamefont{Hillery}},
  \bibinfo{author}{\bibfnamefont{R.~F.} \bibnamefont{O'Connel}},
  \bibinfo{author}{\bibfnamefont{M.~O.} \bibnamefont{Scully}},
  \bibnamefont{and} \bibinfo{author}{\bibfnamefont{E.~P.}
  \bibnamefont{Wigner}}, \bibinfo{journal}{Phys. Rep.}
  \textbf{\bibinfo{volume}{106}}, \bibinfo{pages}{121} (\bibinfo{year}{1984}).

\end{thebibliography}

\clearpage

\begin{figure}
  \includegraphics[width=.8\textwidth]{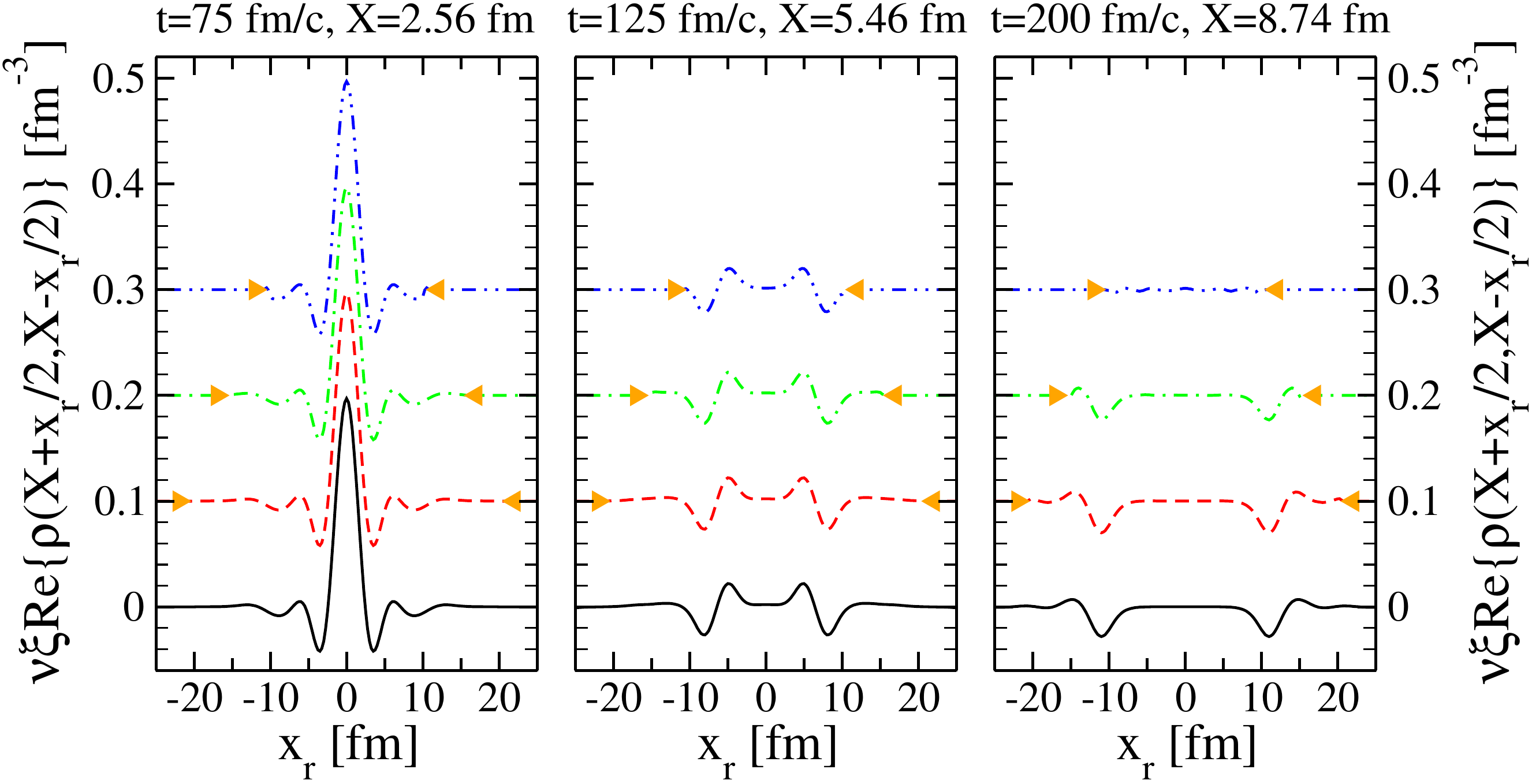}
  \caption{(Color online)
  The same as Fig.~\ref{fig:denoff_cut_fus} but for a collision at $E_{CM}/A=4 \, \text{MeV}$.}
  \label{fig:denoff_cut_fis}
\end{figure}

\begin{figure}
  \includegraphics[width=.9\textwidth]{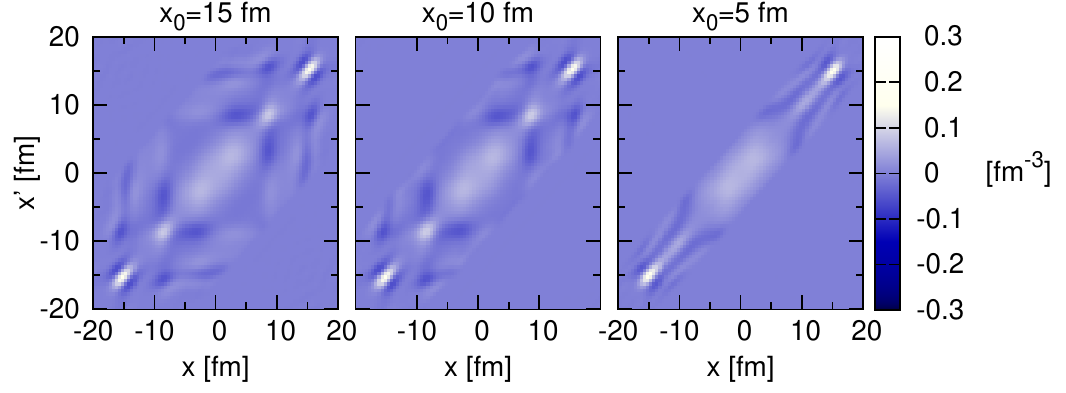}
  \caption{(Color online)
  Intensity plots representing the real part of the scaled density matrix, $\xi \nu \, \rho(x,x';t)$, for a collision at $E_{CM}/A=25 \, \text{MeV}$ at $t=80 \, \text{fm}/c$ in calculations with  different off-diagonal cuts,~$x_0$.  The corresponding intensity plot for a calculation without element suppression may be found in Fig.~\ref{fig:2dmult}.}
  \label{fig:2dcut}
\end{figure}

\begin{figure}
  \includegraphics[width=.8\textwidth]{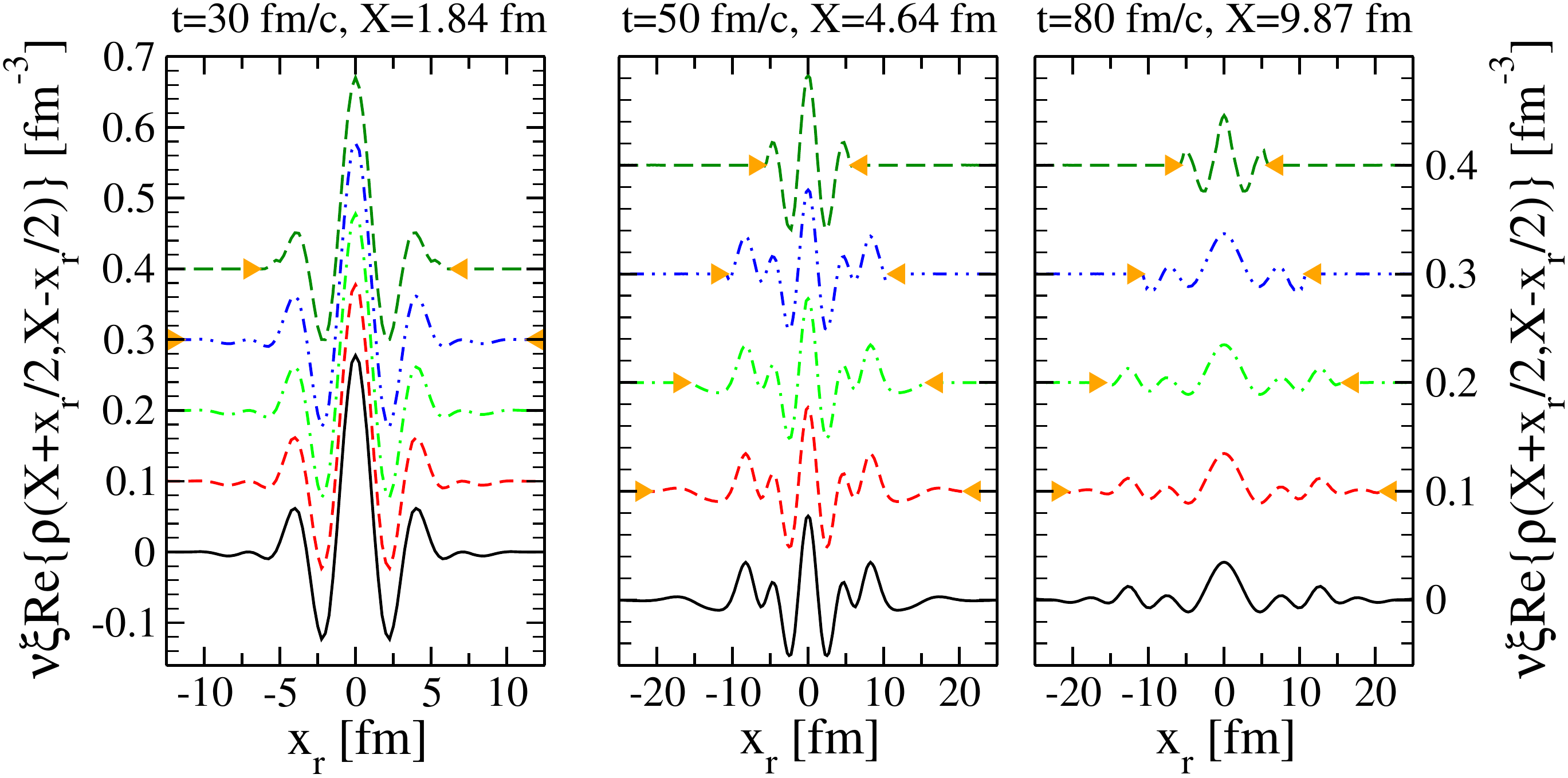}
  \caption{(Color online)
The same as Fig.~\ref{fig:denoff_cut_fis} but for a collision at $E_{CM}/A=25$ MeV. Note the differences in the vertical and horizontal scales for the left panel and the two remaining panels.
  }
  \label{fig:denoff_cut_mult}
\end{figure}

\afterpage{\clearpage}

\end{document}